\documentclass[twocolumn, twoside, superscriptaddress, nofootinbib, floatfix, aps, prc]{revtex4-1}

\usepackage[utf8]{inputenc} % allow utf-8 input
\usepackage[T1]{fontenc}    % use 8-bit T1 fonts
\usepackage{hyperref}       % hyperlinks
\usepackage{url}            % simple URL typesetting
\usepackage{booktabs}       % professional-quality tables
\usepackage{amsfonts}       % blackboard math symbols
\usepackage{nicefrac}       % compact symbols for 1/2, etc.
\usepackage{microtype}      % microtypography
\usepackage{graphicx}
\usepackage{amsmath}
\usepackage{float}
\usepackage[table]{xcolor}
\usepackage{color}
\usepackage{colortbl}
\usepackage{multirow}
\usepackage{tikz}
\usepackage{soul}
\usepackage{bm}
\usepackage{xfrac}
\usepackage{setspace}
\usetikzlibrary{arrows.meta,
                positioning,
                quotes
                }

\DeclareMathOperator{\sgn}{sgn}

\definecolor{midgreen}{rgb}{0.52, 0.73, 0.4}
\newcommand{\code}[1]{\textsc{#1}}
\def\TRENTo{{\sc t\kern-.05em \lower.5ex\hbox{r}\kern-.025em e\kern-.05em n\kern-.05em t\kern-.09em}o}
\def\iccing{{\sc i\kern-.05em c\kern-.05em c\kern-.05em i\kern-.05em n\kern-.05em g\kern-.05em}}
\def\ccake{{\sc c\kern-.05em c\kern-.05em a\kern-.05em k\kern-.05em e\kern-.05em}}
\def\vUSPhydro{v-{\sc u\kern-.05em s\kern-.05em p\kern-.05em}hydro}
\graphicspath{ {figures/} }
\newcommand{\KOMPOST}{\code{kømpøst}}
\newcommand{\vect}[1]{\vec{#1}}
\def\l{\left}
\def\r{\right}

\begin{document}
%==========================================================================
\title{BSQ Conserved Charges in Relativistic Viscous Hydrodynamics solved with Smoothed Particle Hydrodynamics}

\author{Christopher Plumberg}
\affiliation{Illinois Center for Advanced Studies of the Universe, Department of Physics, University of Illinois at Urbana-Champaign, Urbana, IL 61801, USA}
\affiliation{Natural Science Division, Pepperdine University, Malibu, CA 90263, USA}
\author{Dekrayat Almaalol}
\affiliation{Illinois Center for Advanced Studies of the Universe, Department of Physics, University of Illinois at Urbana-Champaign, Urbana, IL 61801, USA}
\author{Travis Dore}
\affiliation{Illinois Center for Advanced Studies of the Universe, Department of Physics, University of Illinois at Urbana-Champaign, Urbana, IL 61801, USA}
\author{D\'ebora Mroczek}
\affiliation{Illinois Center for Advanced Studies of the Universe, Department of Physics, University of Illinois at Urbana-Champaign, Urbana, IL 61801, USA}
\author{Jordi Salinas San Mart\'in}
\affiliation{Illinois Center for Advanced Studies of the Universe, Department of Physics, University of Illinois at Urbana-Champaign, Urbana, IL 61801, USA}
\author{Willian M.~Serenone}
\affiliation{Illinois Center for Advanced Studies of the Universe, Department of Physics, University of Illinois at Urbana-Champaign, Urbana, IL 61801, USA}
\author{Lydia Spychalla}
\affiliation{Illinois Center for Advanced Studies of the Universe, Department of Physics, University of Illinois at Urbana-Champaign, Urbana, IL 61801, USA}
\author{Patrick Carzon}
\affiliation{Illinois Center for Advanced Studies of the Universe, Department of Physics, University of Illinois at Urbana-Champaign, Urbana, IL 61801, USA}
\affiliation{Department of Mathematics and Physical Sciences, Franciscan University of Steubenville, Steubenville, OH 43952, USA}
\author{Matt Sievert}
\affiliation{Department of Physics, New Mexico State University, Las Cruces, NM 88003, USA}
\author{Fernando G. Gardim}
\affiliation{Instituto de Ci\^encia e Tecnologia, Universidade Federal de Alfenas, 37715-400 Po\c cos de Caldas, MG, Brazil
}
\affiliation{Illinois Center for Advanced Studies of the Universe, Department of Physics, University of Illinois at Urbana-Champaign, Urbana, IL 61801, USA}
\author{Jacquelyn Noronha-Hostler}
\affiliation{Illinois Center for Advanced Studies of the Universe, Department of Physics, University of Illinois at Urbana-Champaign, Urbana, IL 61801, USA}
%=============================================================================
\begin{abstract}
%=============================================================================
Conservation laws play a crucial role in the modeling of heavy-ion collisions, including the those for charges such as baryon number (B), strangeness (S), and electric charge (Q).
In this study, we present a new 2+1 relativistic viscous hydrodynamic code called \ccake{} which uses the Smoothed Particle Hydrodynamics (SPH) formalism to locally conserve BSQ charges, together with an extended description of the multi-dimensional equation of state (EoS) obtained from lattice Quantum Chromodynamics.
Initial conditions for \ccake{} are supplied by the \iccing{} model, which samples gluon splittings into quark anti-quark pairs to generate the initial BSQ charge distributions.
We study correlations between the BSQ charges and find that local BSQ fluctuations remain finite during the evolution, with corresponding chemical potentials of ($\sim100$--$200 \,\rm MeV$) at freeze-out.
We find that our framework produces reasonable multiplicities of identified particles and that \iccing{} has no significant effect on the collective flow of all charged particles nor of identified particles when only one particle of interest is considered.
However, we show specifically for Pb\texttt{+}Pb collisions at the LHC $\sqrt{s_{NN}}=5.02$ TeV that \iccing{} does have an effect on collective flow of identified particles if two particles of interest are considered. 
\end{abstract}
%==============================================================================
%\end{frontmatter}
\maketitle
%%%%%%%%%%%%%%%%%%%%%%%%%%%%%%%%%%%%%%%%%%%%%%%%%
%%%%%%%%%%%%%%%   INTRODU CTION   %%%%%%%%%%%%%%%%
%%%%%%%%%%%%%%%%%%%%%%%%%%%%%%%%%%%%%%%%%%%%%%%%%
\section{Introduction}
\label{sec:Introduction}
%==============================================================================
Relativistic heavy-ion collisions provide a unique laboratory for exploring the dynamics and properties of nuclear matter at extreme temperatures and densities.
In such environments, Quantum Chromodynamics (QCD) predicts that nuclear matter will transition into a deconfined state which is commonly known as quark-gluon plasma (QGP).
The QGP can be extremely well described by relativistic viscous hydrodynamics \cite{Heinz:2013th,Luzum:2013yya,DerradideSouza:2015kpt}.

The dynamical evolution of the QGP is dictated  by the interactions of its constituents and conservation laws.
From QCD, we know that quark flavor is conserved in every fundamental interaction such that each quark flavor becomes a conserved charge in the system.
In the context of heavy-ion collisions, this implies the existence of conserved charges such as Baryon number (B), Strangeness (S), and Electric charge (Q).%
\footnote{Note that heavier quarks such as the charm may also be relevant, \cite{Capellino:2023cxe} although they are produced in such low abundances that they are extremely unlikely to reach chemical equilibrium with the rest of the system \cite{Moore:2004tg}.}
Based on the initial conditions of heavy-ion collisions that depend on the BSQ charges of the colliding ions, we expect strangeness neutrality to be preserved globally where the average strangeness density $\rho_S$ is zero $\langle \rho_S\rangle=0$ (no hypernuclei are used) and the  electric charge to baryon number ratio to be conserved and dictated by the choice of ions.
For example, given an ion that is collided in heavy-ion collisions we can define the constraint $\langle \rho_Q\rangle = (Z/A) \langle \rho_B\rangle$\footnote{Here the angled brackets $\langle \ldots \rangle$ to denote a suitably defined ensemble average, such as an average over many collision events.} where $Z$ is the proton number, $A$ is the number of protons plus neutrons, $ \rho_Q$ is the  electric charge density, and $ \rho_B$ is the baryon density.  For typical ions used in heavy-ion collisions such as gold and lead, then  $Z/A\sim 0.4$.

While these constraints must hold \textit{globally}, they do not need to hold \textit{locally}.
For instances, the initial state starts with zero strangeness but due to gluon splittings into quark anti-quark pairs ($q\bar{q}$) then fluctuations in strangeness are generated such that net-strangeness ($\Delta N_S=N_S+N_{\bar{S}}=0$) 
is still zero but the strangeness number $N_S$ and 
anti-Strangeness number $N_{\bar{S}}$ are both non-zero such 
that $N_S=-N_{\bar{S}}\neq 0$.
In fact, the existence of the multitude of strange mesons and baryons in the final state at the Large Hadron Collider (LHC) \cite{ALICE:2016fzo} and the Relativistic Heavy-Ion Collider (RHIC) where $\Delta N_B\sim\Delta N_S\sim\Delta N_Q\sim 0$ are clear indicators that fluctuations of all the conserved charges (BSQ) are relevant.

Meanwhile, the field has also determined that partonic degrees of freedom (either quarks and/or gluons) play a role in the geometry of the initial state, especially in small systems \cite{Schenke:2014zha,Mantysaari:2016ykx,Moreland:2018gsh}. 
Even when these sub-nucleonic fluctuations have been included in the initial state, the BSQ charge fluctuations of these partons have been ignored at LHC energies.\footnote{To be clear, BQ fluctuations have been considered previously at the hadronic level \cite{Shen:2017bsr,Schafer:2021csj,Garcia-Montero:2023gex,Pihan:2023dsb} just not at the quark level. } 
At the LHC and top RHIC energies, the initial state reaches low-x such that gluons are the dominate degrees of freedom \cite{H1:2009pze,Bailey:2020ooq}. 
Thus, for experimental observables only sensitive to the energy density fluctuations it is a safe assumption that gluonic (or nucleonic) degrees-of-freedom are the most important in the initial state.
However, for experimental observables sensitive to hadrons that carry conserved charges, especially multiple conserved charges (e.g. Omega baryons), one could ask if they are sensitive to the influence of gluons splitting into $q\bar{q}$ pairs in the initial state. 

The consideration of $q\bar{q}$ pairs in the initial state, allows for locally distributed, event-by-event fluctuations of  BSQ conserved charges to be transported within the QGP during its evolution.
In fact, large pockets of a specific BSQ charge (or anti-charge) may exist that could alter the azimuthal anisotropies of identified particles \cite{Martinez:2018ygo,Martinez:2019jbu,Carzon:2019qja}.

Accounting for such local charge distributions and their effects requires the ability to enforce the conservation laws locally in hydrodynamic simulations.
The purpose of this paper is to develop and present a new code designed to do this.
%%%%%%%%%%%%%%%%%%%%%%%%%%%%%%%%%%%%%%%%%%%%%%%%%%%%%%%%%%%%%%%%%%%%%%%%%%%
\subsection{Modelling relativistic viscous hydrodynamics with BSQ charges}
%%%%%%%%%%%%%%%%%%%%%%%%%%%%%%%%%%%%%%%%%%%%%%%%%%%%%%%%%%%%%%%%%%%%%%%%%%
The evolution of the QGP in heavy-ion collisions at the LHC and top energies RHIC has been modeled using relativistic viscous hydrodynamics in both 2+1D and 3+1D \cite{Niemi:2015voa,Noronha-Hostler:2015uye,Eskola:2017bup,Giacalone:2017dud,JETSCAPE:2020shq,Bernhard:2019bmu,Nijs:2020roc,Denicol:2018wdp,Schafer:2021csj}.
Many models organize the collision evolution into four stages: initial state, pre-equilibrium, hydrodynamics, and a hadronic afterburner.
In order to accurately take into account local charge fluctuations, charge conservation must be enforced at all stages of the evolution.
For the initial state, whether the initial conditions incorporate baryon stopping at low beam energies \cite{Shen:2017bsr} or gluon splitting into quark/anti-quark pairs at high energies \cite{Martinez:2018ygo,Martinez:2019jbu,Carzon:2019qja}, global charge conservation must be respected during the sampling process.
Furthermore, all three charges must also be conserved locally as they propagate within any pre-equilibrium phase (see, e.g., \cite{Carzon:2023zfp}).
Likewise, hadronic afterburners that incorporate BSQ charges already exist \cite{Bass:1998ca,Weil:2016zrk}, although the freeze-out sampling to connect hydrodynamics to these afterburners is nontrivial \cite{Oliinychenko:2019zfk,Oliinychenko:2020cmr}.

The hydrodynamic phase presents its own challenges once conserved charges are included.
One major complication pertains to the QCD equation of state (EoS) at finite density, which depends not only on temperature $T$, but also on the baryon chemical potential $\mu_B$, strangeness chemical potential $\mu_S$, and electric charge chemical potential $\mu_Q$ \cite{Monnai:2019hkn,Noronha-Hostler:2019ayj,Bellwied:2019pxh,Monnai:2021kgu,Karthein:2021nxe,Aryal:2020ocm}.
The EoS, which is needed to close the hydrodynamic equations of motion, is therefore four-dimensional (4D) once BSQ conservation is enforced.
Another significant challenge arises because the out-of-equilibrium hydrodynamic expansion must incorporate the 3 conserved (BSQ) currents with a $3\times 3$ diffusion matrix \cite{Greif:2017byw,Fotakis:2019nbq}.
Furthermore, second-order relativistic viscous hydrodynamic equations of motion with BSQ diffusion acquire coupling terms between shear and bulk viscosity and also the BSQ currents \cite{Denicol:2012cn,Almaalol:2022pjc}.
These new terms carry their own transport coefficients which must first be derived in some theoretical framework (e.g., kinetic theory \cite{Fotakis:2022usk} or holography \cite{Rougemont:2017tlu,Grefa:2022sav}).
They also require new thermodynamic derivatives that are not needed for vanishing chemical potentials.
All of this in turn generates new equations of motion which must be solved during the system's evolution, in addition to those which reflect the local conservation of the charges themselves.
Modeling hydrodynamics with conserved charges is a highly involved task which requires development on several fronts in order to reach the levels of sophistication already available in hydrodynamics at zero density.

Recent work has begun to explore the baryon-rich QGP through relativistic hydrodynamic simulations with one conserved charge B \cite{Denicol:2018wdp,Du:2019obx}, two conserved charges BS \cite{Fotakis:2019nbq}, or three conserved charges BSQ \cite{Schafer:2021csj}.
These initial results have found a sensitivity to enforcing strangeness neutrality in the EoS \cite{Monnai:2021kgu}, a correlation between the strangeness density and baryon density when both are allowed to fluctuate \cite{Fotakis:2019nbq}, and a deviation from ideal isentropic trajectories due to the influence of diffusive effects \cite{Du:2021zqz}.
In addition, many interesting questions remain, such as the mapping of initial to final state at the beam energy scan (see discussions in \cite{Martinez:2019jbu,Carzon:2019qja}) and the influence that a potential critical point may have on the dynamics \cite{Rajagopal:2019xwg,Dore:2020jye,Dore:2022qyz,Du:2021zqz}.
Incorporating conserved charges into hydrodynamic models of QGP thus opens up a wide range of potential applications.

In this paper, we make an important step-forward by developing an open-source 2+1 BSQ relativistic viscous hydrodynamic code called \ccake{} (Conserved ChArges HydrodynamiK Evolution).
We use the Lagrangian numerical method called Smoothed Particle Hydrodynamics (SPH) based on the original \vUSPhydro{} code \cite{Noronha-Hostler:2013gga,Noronha-Hostler:2014dqa}.
Shear and bulk viscous effects are implemented according to the new Israel-Stewart-DNMR $\mathcal{O}(\text{Re}^{-1})^2$ formalism \cite{Almaalol:2022pjc}.
Only a subset of the second-order terms in this formalism are included in this study, but we plan to include more in subsequent studies.
In the charge sector, we implement only ideal BSQ currents; diffusive corrections will likewise be added in future work.
We have checked that \ccake{} passes all known benchmarks for hydrodynamic simulations of nuclear collisions, including the Gubser tests with shear viscosity \cite{Marrochio:2013wla} and with ideal BSQ currents \cite{Denicol:2018wdp}.
Furthermore, the initial conditions for \ccake{} are obtained from \iccing{} \cite{Martinez:2019jbu,Carzon:2019qja}, which uses gluon splittings into quark/anti-quark pairs to generate local distributions of energy density and the BSQ charge densities.
This enables us to study fluctuations of BSQ conserved charges with a realistic 4D EoS based on lattice QCD \cite{Noronha-Hostler:2019ayj} in an expanding QGP medium for the first time.
Additionally, we explore the BSQ density profiles and correlations between BSQ charges over time, collective flow of identified particles, and the effects on the system's passage through the QCD phase diagram over time.
%%%%%%%%%%%%%%%%%%%%%%%%%%%%%%%%%%%%%%%%%%%%%%%%%%%%%%%%%%%
\subsection{Structure of the paper}
%%%%%%%%%%%%%%%%%%%%%%%%%%%%%%%%%%%%%%%%%%%%%%%%%%%%%%%%%%%
The paper is organized as follows.
In Sec.\ \ref{sec:BSQframework} we provide the details of our new BSQ framework.  The main components of our framework include:
  the \TRENTo{}\texttt{+}\iccing{} initial state (Sec.\ \ref{sec:ICCING});
  the equations of motion and the SPH formalism used to solve (Secs.\ \ref{sec:Hydro} and \ref{sec:SPHformalism});
  the transport coefficients (Sec.\ \ref{sec:TransportCoefficients});
  our implementation of the 4D EoS and how it is connected to the hydrodynamic evolution (Secs.\ \ref{sec:EoS} and \ref{sec:TimeDerivativesOfThermodynamicVariables});
  benchmark tests of \ccake{}, including the Gubser test, time checks, and energy loss (Sec.\ \ref{sec:BenchmarkTest});
  and our freeze-out procedure for ideal BSQ currents, together with several changes that were needed to our original \vUSPhydro{} algorithm to handle BSQ currents (Secs.\ \ref{sec:Freeze-out}, \ref{sec:constant-efo}, and \ref{sec:CF_SPH}).
In Sec.\ \ref{sec:Results}, we use our new BSQ framework to observe the time variation of the local BSQ densities in Sec.\ \ref{sec:BSQdyn}, the trajectories through the QCD phase diagram for an \iccing{} event in Sec.\ \ref{sec:Trajectory} and the remaining fluctuations in BSQ chemical potentials at freeze-out in Sec.\ \ref{sec:BSQDensityProfiles}.
In Sec.\ \ref{sec:densityT} we demonstrate the very non-linear scaling between $\rho_B$ vs $\mu_B$ at finite temperatures in order to better understand the range of densities. 
In Sec.\ \ref{sec:h0} we calculate experimental observables for a single event and understand the influence of \iccing{} smoothed out to different scales.
In Sec.\ \ref{sec:resultsII}, we proceed to calculate the multiplicity and flow harmonics of charged particles and identified particles, which are shown in Secs.\ \ref{sec:AnisotropyOfBSQDensities} and \ref{sec:flowPID}.
Concluding remarks are offered in Sec.\ \ref{sec:Conclusions}, along with a discussion of future work.

%%%%%%%%%%%%%%%%%%%%%%%%%%%%%%%%%%%%%%%%%%%%%%%%%
%%%%%%%%%%%%%%%   BSQ FRAMEWORK   %%%%%%%%%%%%%%%
%%%%%%%%%%%%%%%%%%%%%%%%%%%%%%%%%%%%%%%%%%%%%%%%%
\section{BSQ framework}
\label{sec:BSQframework}
%==============================================================================
The BSQ framework presented here was based upon the original 2+1D relativistic viscous hydrodynamic code, \vUSPhydro{} \cite{Noronha-Hostler:2013gga,Noronha-Hostler:2014dqa}, that has been used extensively at $\mu_B=0$ to make predictions across beam energies \cite{Noronha-Hostler:2015uye,Gardim:2016nrr,Prado:2016szr,Giacalone:2017uqx,Alba:2017hhe,Giacalone:2017dud,Sievert:2019zjr,Katz:2019fkc,Katz:2019qwv,Rao:2019vgy,Carzon:2020xwp,Barbosa:2021ccw}. A few highlights include the first study of a deformed $^{129}$Xe in heavy-ion collisions \cite{Giacalone:2016mdr}, the first  study of the influence of event-by-event fluctuations with realistic hydrodynamic backgrounds on the $R_{AA}\times v_2$ puzzle \cite{Noronha-Hostler:2016eow,Betz:2016ayq},  predictions for D mesons across the proposed system size scan at the LHC \cite{Katz:2019qwv}, and the inclusion of a full particle list motivated by lattice QCD comparisons \cite{Alba:2017hhe}.

Below we detail each separate component of our BSQ framework.  Significant upgrades were required for this work and, therefore, we have decided to rename the hydrodynamic model as \ccake{} to avoid confusion with the previous version and also to provide credit to the enormous effort of the authors here, most of whom were not involved with the original \vUSPhydro{} code development.  Furthermore, this paper is the first time that we are aware of that a Smoothed Particle Hydrodynamic code of a relativistic viscous fluid with multiple conserved charges has been developed. With this paper, we are also releasing an open-source version of \ccake{} at \cite{ccakesite}.
%
%%%%%%%%%%%%%%%%%%%%%%%%%%%%%%%%%%%%%%%%%%%%%%%%%
%%%%%%%%%%%   ICCING INITIAL STATE   %%%%%%%%%%%%
%%%%%%%%%%%%%%%%%%%%%%%%%%%%%%%%%%%%%%%%%%%%%%%%%
\subsection{\texorpdfstring{\iccing{}}{ICCING} Initial State}
\label{sec:ICCING}
%=============================================================================

To run a BSQ hydrodynamic model one requires an initial condition that initializes not just the energy-momentum tensor $T^{\mu\nu}$ but also each respective charge current $N_X^{\mu}$ where $X=B,S,Q$. 
Thus, one requires both an initial distribution of BSQ charge density but also the contribution to the out-of-equilibrium charge current. 
At this time, we consider \iccing{} initial conditions  \cite{Martinez:2019jbu,Carzon:2019qja} that produce BSQ charge fluctuations at vanishing net-baryon densities.  
\iccing{} reads in a generic initial condition, assuming that the energy density distribution only includes gluons. 
It then samples the probability that each gluon produces a quark/anti-quark pair.  
Since each flavor or quark carries certain quantum numbers, e.g., a strange quark is characterized by having $B=\text{\sfrac{1}{3}}$, $S=-1$, $Q=-\text{\sfrac{1}{3}}$, by producing a quark distribution we are, in fact, constructing initial conditions with a BSQ density field. 
However, \iccing{} ensures that the net densities are always zero since quarks are always produced in pairs.
\begin{figure}
    \includegraphics[width=\linewidth]{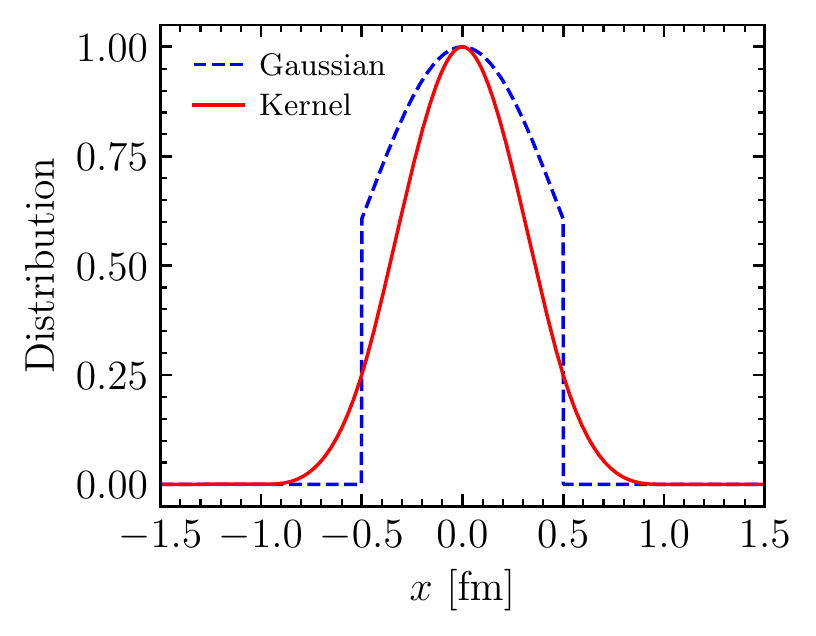}

    \caption{(color online) Comparison of Gaussian (dashed blue) and kernel (solid red) quark/anti-quark density profiles. The cut-off for the Gaussian profile is $0.5$ fm while the kernel function approaches zero smoothly at $1$ fm.
    }
    \label{fig:KernelVsGaussian}
\end{figure}
There are subtleties in connecting the \iccing{} model to \ccake{}, specifically concerning how the charge density is distributed in the initial state. We found that three changes were required in order to couple \iccing{} to \ccake{}:
\begin{itemize}
    \item The underlying distribution function to describe the size of a gluon had to be changed from a Gaussian distribution to a cubic spline kernel.
    \item The normalization of the initial entropy density from \TRENTo{} \cite{Moreland:2014oya} had to be altered in order to reproduce final multiplicities from the LHC.
    \item A perturbative cut-off had to be enforced (see \cite{Carzon:2023zfp}).
\end{itemize}
We discuss each of these changes in greater detail below.

%%%%%%%%%%%%%%%%%%%%%%%%%%%%%%%%%%%%%%%%%%%%%%%%%
%%%%%%%%%%   DISTRIBUTION FUNCTION   %%%%%%%%%%%%
%%%%%%%%%%%%%%%%%%%%%%%%%%%%%%%%%%%%%%%%%%%%%%%%%
\subsubsection{Distribution function}

In the original version of \iccing{} from \cite{Carzon:2019qja}, the profile used to distribute the charge density of a quark or anti-quark was chosen to be a 2D Gaussian with the radius $r$ of the distributed charge being the width of the Gaussian,
\begin{align} \label{eq:GaussianQuarkDensity}
Gaussian(x,y) = \frac{Q}{2\pi r^2 \tau_0} \, \exp\left[ -\frac{x^2 + y^2}{2 r^2} \right] \Theta(r_g-r),
\end{align}
where $r^2=x^2+y^2$, $Q$ is the total charge of the quark, and the factors $1 / 2\pi r^2$ and $1 / \tau_0$ come from the normalization of the 2D Gaussian and conversion from a 2D density per unit rapidity to a 3D density per unit volume, respectively. 
This profile was chosen due to its simplicity.
However, an effect of using a Gaussian profile for the quark/anti-quark charge density is that at the edge of the radius of the $q\bar{q}$ pair is a large gradient in the charge density which can be problematic for hydrodynamics. 
To address this and provide a generalization of the quark/anti-quark profile, we added a new class in \iccing{} called \texttt{Mask} in \texttt{Mask.h} which implements a kernel function option. 
The motivation for including a kernel function version of the quark/anti-quark density profile is that it smoothly approaches zero at the radius of the gluon, $r_g$, whereas the Gaussian has infinite tails and, thus, requires a cut-off. 
Here we use the same kernel function originally from \cite{Monaghan:1992rr} (although other kernels would be possible, see \cite{1985A&A...149..135M}) as what is used in SPH (described in subsequent sections) that uses a cubic spline in 2+1 dimensions:
\begin{equation}\label{eq:KernelQuarkDensity}
    W\left(\frac{|\vect{r}|}{h};h\right)=\frac{10}{7\pi h^2}\times f\left(\frac{|\vect{r}|}{h}\right),
\end{equation}
where
\begin{equation}\label{eq:CubicSpline}
    f(\xi) = \begin{cases}
1-\frac{3}{2}\xi^2+\frac{3}{4}\xi^3 & \text{if}\quad 0 \leq \xi <1 \\
\frac{1}{4}\left(2-\xi\right)^3 & \text{if}\quad 1 \leq \xi <2 \\
0 & \text{otherwise},
    \end{cases}
\end{equation}
where $h$ is the ``smoothing scale", $|\vect{r}|$ is the transverse radius, and $\xi$ is just the variable fed into $f$ where we have defined $\xi=|{\vect{r}}|/h$.  For \iccing{} we define the radius of the gluon to be $r_g=2h$ because in our defined kernel function the contribution of the gluon is exactly $0$ beyond $2h$.

A one-dimensional comparison between the two profiles is presented in Fig.~\ref{fig:KernelVsGaussian}. 
The radius at which the Gaussian profile is cut-off is at $r_g=0.5$ fm and consistent with the analysis in Ref.~\cite{Carzon:2020xwp}. 
The radius of the kernel function profile is chosen to be $r_g=1$ fm here so that it matches closely to the Gaussian profile and effectively just adds tails to the distribution.
We have checked the the energy density profile remains nearly identical for both the Kernel and Gaussian distribution, similar to what was previously shown in \cite{Carzon:2020xwp}.

\begin{figure*}
    \centering
\includegraphics[keepaspectratio, width=\linewidth]{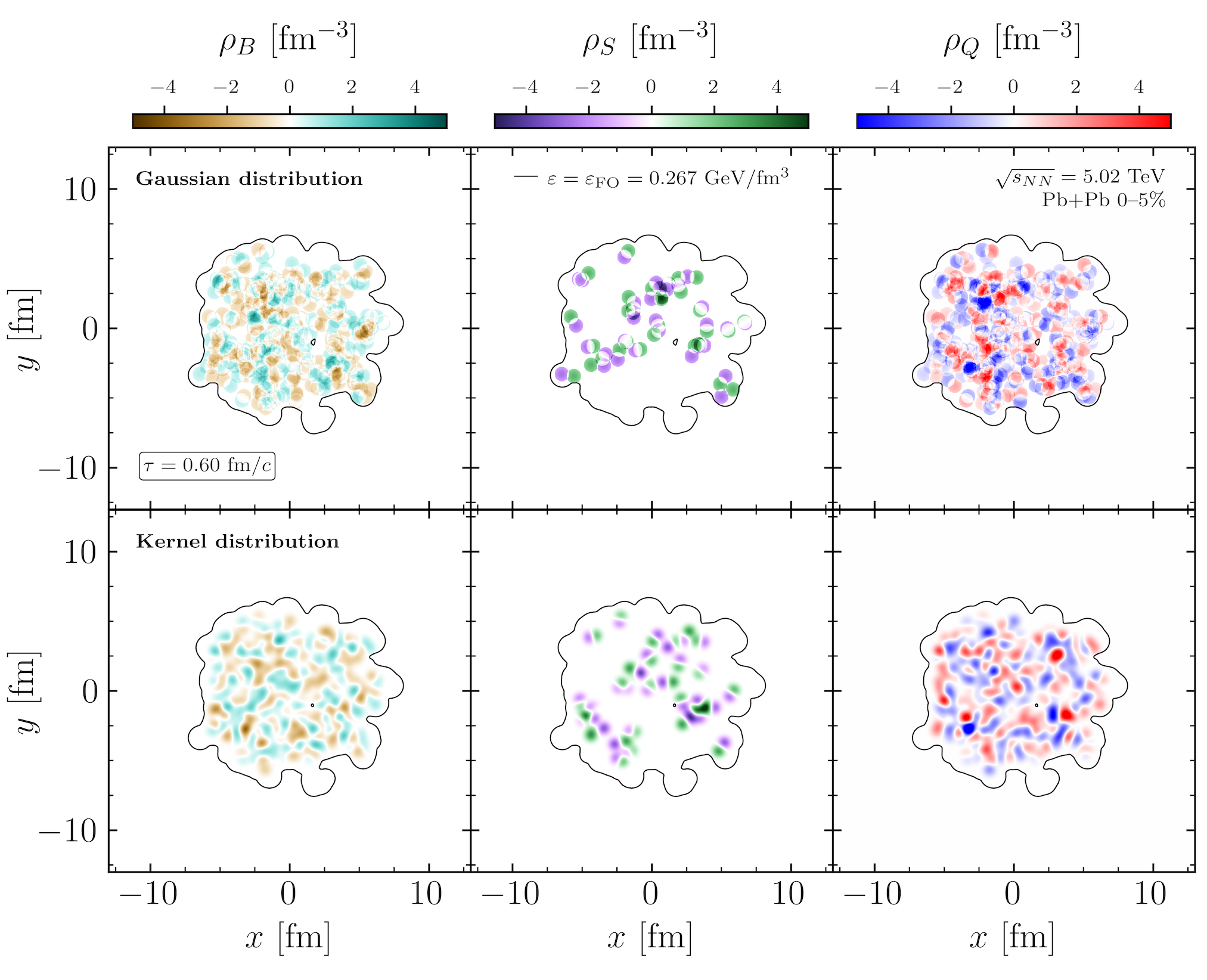}
    \caption{(color online) From left to right: baryon, strangeness, and electric charge density profiles after the \iccing{} procedure for a single event. From top to bottom, Gaussian and cubic spline kernel distributions. In all panels, a contour of constant energy density, where $\varepsilon=\varepsilon_\text{FO}$, is displayed as a reference. 
    }
    \label{fig:GausKern}
\end{figure*}

In Fig.\ \ref{fig:GausKern} we compare the  baryon density $\rho_B$, strangeness density $\rho_S$, and electric charge density $\rho_Q$ distributions for both the Gaussian distribution (top) and the cubic spline kernel (bottom) for a single \iccing{} event corresponding to a central collision. 
We find that the cubic spline leads to a smoother distribution with fewer edge effects around the quark/anti-quark pairs.  
%The energy density profiles look nearly identical and the shape of the charge distributions are unaffected by our choice.  
Moreover, it is clear that small scale structure is smoother with the cubic spline. Thus,  hydrodynamics is able to handle the cubic spline kernel better than the sharp cut-offs from the Gaussian distribution.

%%%%%%%%%%%%%%%%%%%%%%%%%%%%%%%%%%%%%%%%%%%%%%%%%
%%%%%%%%%%   NORMALIZATION CONSTANT   %%%%%%%%%%%
%%%%%%%%%%%%%%%%%%%%%%%%%%%%%%%%%%%%%%%%%%%%%%%%%
\subsubsection{Normalization constant}\label{sec:norm}

All heavy-ion collision simulations that rely on relativistic hydrodynamics include a free parameter which sets the overall normalization of the initial conditions, such that
\begin{eqnarray}
    S &=& \mathcal{A}\, \tau_0 \int_{V} \tilde{s}(\vect{r}, \tau_0) \, d\vect{r}\, \\
    &=& \mathcal{N}  \int_{V} \tilde{s}(\vect{r}, \tau_0) \, d\vect{r},
\end{eqnarray}
where $S$ here is the total entropy,  $\tilde{s}(\vect{r}, \tau_0)$ is the un-normalized entropy density field at the initial time $\tau_0$ (then the normalized entropy density is denoted as $s(\vect{r}, \tau_0)$), $\mathcal{A}$ is the normalization constant without the inclusion of $\tau_0$, and $\mathcal{N}$ [fm] is the normalization constant which absorbs the $\tau_0$ term, i.e., $\mathcal{N}=\tau_0\mathcal{A}$.  One could apply the normalization constant to the energy density instead of the entropy.  However, we choose to use entropy here since it most naturally connects to the final multiplicity. 

From initial state models, one only obtains $\tilde{s}(\vect{r}, \tau_0)$ (either in 2 or 3 dimensions, depending on the model).  Then one simply adjusts $\mathcal{N}$ to reproduce experimental data of the total multiplicity (typically in central collisions).  Various studies \cite{Blaizot:1987hu,Blaizot:1987cc,Kolb:2001qz,Gardim:2019xjs,Sievert:2019zjr,Carzon:2021tif} (this is by no means a complete list) have found a rather direct connection between the total initial  entropy and the final multiplicity, even in the case of viscous systems where entropy is produced throughout the hydrodynamic expansion.  Thus, initial total entropy is often used as a proxy for multiplicity.  
Previous versions of \vUSPhydro{} use $\mathcal{N}=119$ for Pb\texttt{+}Pb $\sqrt{s_{NN}} = 5.02 $ TeV and the standard parameters from \cite{Alba:2017hhe}. 

Because \iccing{} requires the correct energy density scale to perform sampling, one must determine $\mathcal{N}$ \emph{before} running \iccing{}.  Initial papers with \iccing{} \cite{Martinez:2019jbu,Carzon:2019qja,Carzon:2023zfp} assumed that $\mathcal{N}=119$ for Pb\texttt{+}Pb at $\sqrt{s_{NN}} = 5.02 $ TeV.  
However, after attempting $\mathcal{N}=119$ for Pb\texttt{+}Pb $\sqrt{s_{NN}} = 5.02 $ TeV and comparing to final multiplicity results, we found that a higher value of $\mathcal{N}$ was required to reproduce the data using our current parameters, both with and without \iccing{}. 
In fact, our current $\mathcal{N}=125$ fm  for both \TRENTo{}+\ccake{} and \TRENTo{}+\iccing{}+\ccake{}. We note that the inclusion of \iccing{} leads to an overall increase in our final multiplicity, for a fixed value of $\mathcal{N}$. 
This can be understood by considering the Gibbs relation between energy density $\varepsilon$ and entropy density $s$.  Using a compact vector notation,%
\footnote{Here we refer to the charge densities $\rho_B$, $\rho_S$, $\rho_Q$ collectively as the set $\vect{\rho}$ for brevity. Analogously, we can do the same procedure for the chemical potentials where $\mu_B$, $\mu_S$, $\mu_Q$ are referred to as $\vect{\mu}$.}
one can see that
\begin{align}\label{eq:gibbs}
	\varepsilon + p &= sT + \sum_{X \in (B,S,Q) } \rho_X \mu_X = sT + \vect{\rho}\cdot \vect{\mu},
\end{align}
where $p$ is the pressure and through this work we use the capital letters $X$, $Y$, \ldots, etc. to denote conserved charges. 
Then solving Eq.\ \eqref{eq:gibbs} for $s$ yields
\begin{equation}
    s=\frac{1}{T}\left[\varepsilon+p-\vect{\rho}\cdot \vect{\mu}\right].
\end{equation}
In \iccing{} the background energy is held fixed, whereas local fluctuations are sourced according to the term $\vect{\rho}\cdot \vect{\mu}$.   
When gluons split into $q\bar{q}$ pairs, both positive and negative charges appear and there is a non-trivial relationship with $\vec{\mu}$.
Thus, the term $\vec{\rho}\cdot \vec{\mu}$ is non-zero and changes the overall entropy of the system, even though energy is fixed.
This clarifies the slight increase in the total multiplicity of all charged particles when the effect of gluons splitting into quark anti-quark pairs is considered.

%%%%%%%%%%%%%%%%%%%%%%%%%%%%%%%%%%%%%%%%%%%%%%%%%
%%%%%%%%%%%%   PERTURBATIVE CUT-OFF   %%%%%%%%%%%%
%%%%%%%%%%%%%%%%%%%%%%%%%%%%%%%%%%%%%%%%%%%%%%%%%
\subsubsection{Perturbative cut-off}
Next, we discuss the perturbative cut-off that we employ to ensure that quark/anti-quark pairs are not produced at too low of energy densities.  The need for a perturbative cut-off was first discovered in \cite{Carzon:2023zfp} in the context of Greens functions that allowed one to propagate the charge distributions in time.  However, in this work we discovered a second reason the perturbative cut-off is needed. Essentially, if quark/anti-quark pairs are produced at energy densities which are too low, it is not possible to find a corresponding combination of densities $\left\{\varepsilon,\rho_B,\rho_S,\rho_Q\right\}$ in the lattice QCD EoS table, making it impossible to evolve the pair using hydrodynamics. Part of the problem may arise due to limitations in the lattice QCD EoS (see Sec.\ \ref{sec:EoS} for more details) or due to the breakdown of the perturbative assumption used in generating the quark/anti-quark pairs.  At the moment, it is not yet possible to tell without access to a more reliable EoS in the large density regime. 

We apply a perturbative cut-off such that fluid cells with energies below this value are not split into quark/anti-quark pairs. The perturbative cut-off is defined as
\begin{equation}
    \frac{E_q}{E_\text{Bg}} < P,
\end{equation}
where $P$ is the perturbative cut-off, $E_q$ is the energy of the quarks, and $E_\text{Bg}$ is the energy of the background.  Setting $P=1$ is equivalent to having no cut-off. Here we set $P=0.9$.  Fluid cells below this perturbative cut-off only contribute to the $\varepsilon$ distribution and not the BSQ charge distributions.
We find that applying this cut-off reduces the percentage of fluid cells which cannot be found in the EoS table, although we are not able to eliminate them entirely.  Thus, in Sec.\ \ref{sec:EoS} we develop alternative approaches to handle out-of-bounds fluid cells within \ccake{}.

\begin{figure*}[hbt]
    \centering
    \includegraphics[keepaspectratio, width=0.45\linewidth]{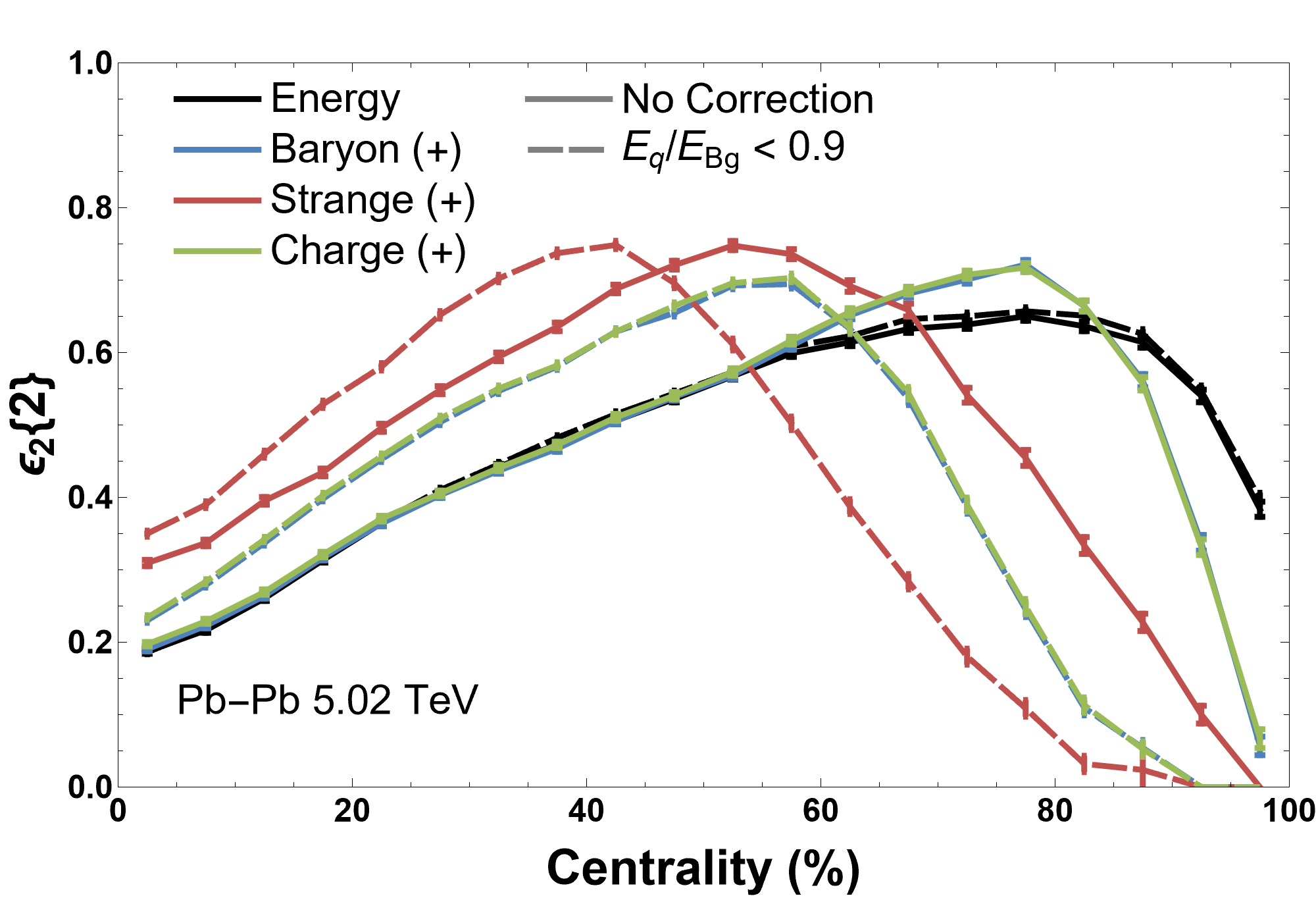}
    \includegraphics[keepaspectratio, width=0.45\linewidth]{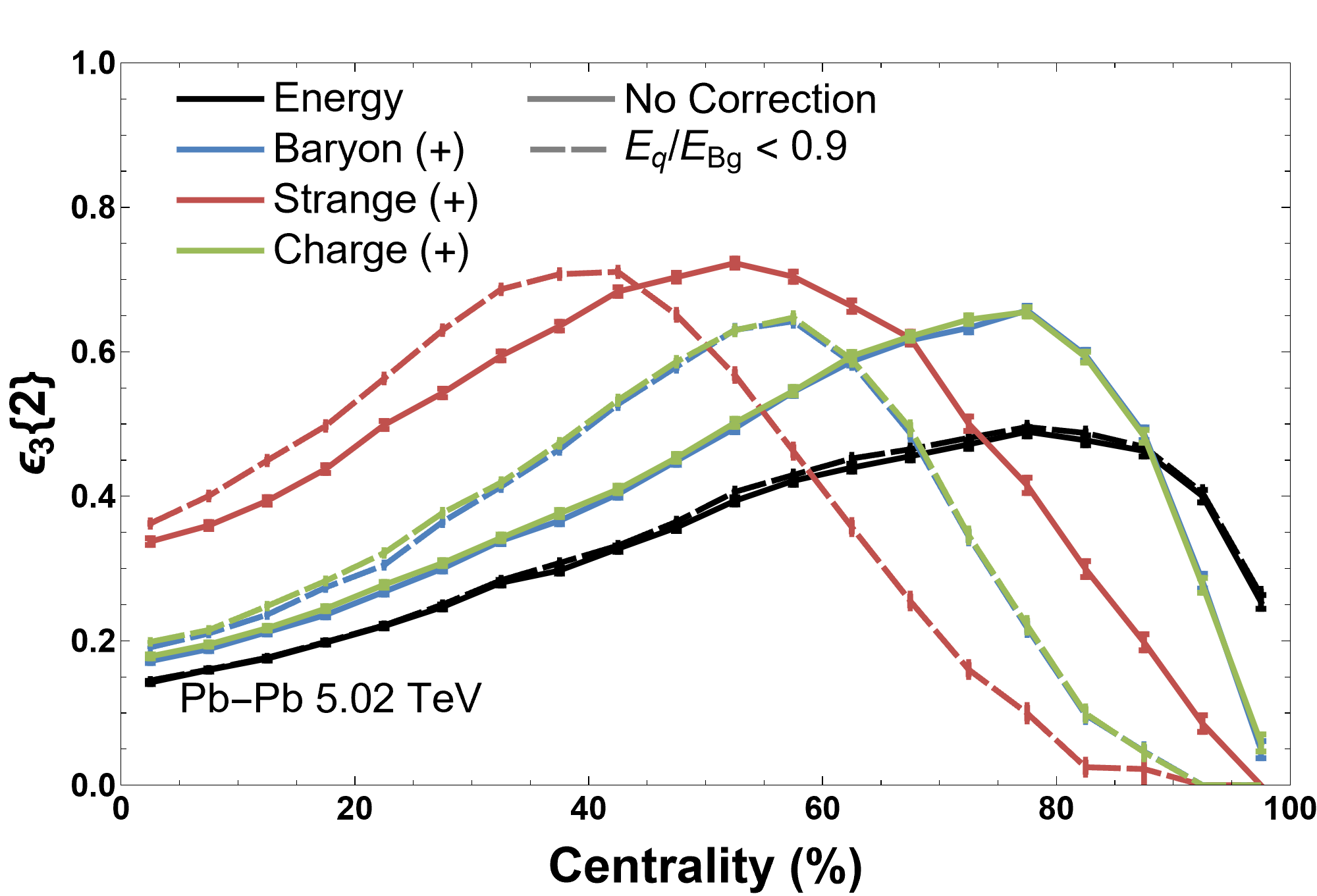}

    \caption{Eccentricities $\varepsilon_n\left\{2\right\}$ for $n=2$ (left) and $n=3$ (right) comparing either no perturbative cut-off vs a cut-off of such that the energy of the quark anti-quark pair must be less than $90\%$ of the background $E_q/E_{Bg}=0.9$. The eccentricities are shown for the energy densities and all 3 BSQ conserved charges.}
    \label{fig:PerpCUT}
\end{figure*}
We can check the impact of this perturbative cut-off on the final eccentricities\footnote{Note that eccentricities always come with a subscript $n$ and the number of particles correlated in the curly brackets, i.e., $\varepsilon_n \left\{2\right\}$  We point this out to avoid confusion with the energy density that is just $\varepsilon$.}, $\varepsilon_n$, to determine if it plays a significant role. In Fig.\ \ref{fig:PerpCUT} we compare the role of the perturbative cut-off on both the elliptical eccentricities (left) as well as the triangular eccentricities (right) for $\varepsilon_n \left\{2\right\}=\sqrt{\langle \varepsilon_n^2\rangle}$.  Note that the BSQ eccentricities cannot be defined in the same way as that for the energy density.  The energy density eccentricities are defined in the center of mass frame (see \cite{Carzon:2019qja} for further discussions) but this is not possible for a conserved charge since in \iccing{} we have no net charge density, i.e., $\langle \rho_B\rangle=\langle \rho_S\rangle=\langle \rho_Q\rangle=0$.  Thus, if one were to attempt to calculate an eccentricity in the center of charge frame, then one would have to normalize by zero.  To circumvent this issue, we calculate only the eccentricity of the quarks, not the anti-quarks for the time being.  There it is possible to define a center of positive charge and calculate the eccentricities in the usual fashion.   Additionally, future work is planned where new eccentricities are developed that can handle both charge and anti-charge.

With this understanding of the caveats of the eccentricities for BSQ conserved charges, we return to Fig.\ \ref{fig:PerpCUT} to better understand the influence of the perturbative cut-off.    We find that the cut-off has the effect of suppressing the eccentricities in peripheral collisions, which occurs because  quark production there is heavily suppressed (especially for strange quarks).  Note that unlike in \cite{Carzon:2023zfp}, we are not considering a time evolution from the Greens functions here, meaning that there are subtle differences in our figures with respect to those shown in that work.  Additionally, because we are not considering a time evolution, the initial times are different.  In this work we switch on hydrodynamics at $\tau_0=0.60$ fm$/c$ whereas in \cite{Carzon:2023zfp} the initial times where taken to be $\tau_0=0.1$ fm$/c$. 

%%%%%%%%%%%%%%%%%%%%%%%%%%%%%%%%%%%%%%%%%%%%%%%%%
%%%%%%%%%%%%   CHEMICAL POTENTIALS   %%%%%%%%%%%%
%%%%%%%%%%%%%%%%%%%%%%%%%%%%%%%%%%%%%%%%%%%%%%%%%
\subsubsection{Range in chemical potentials }
%%%%%%%%%%%%%%%%%%%%%%%%%%%%%%%%%%%%%%%%%%%%%%%%%
%
\begin{figure*}
    \centering
    \includegraphics[keepaspectratio, width=\linewidth]{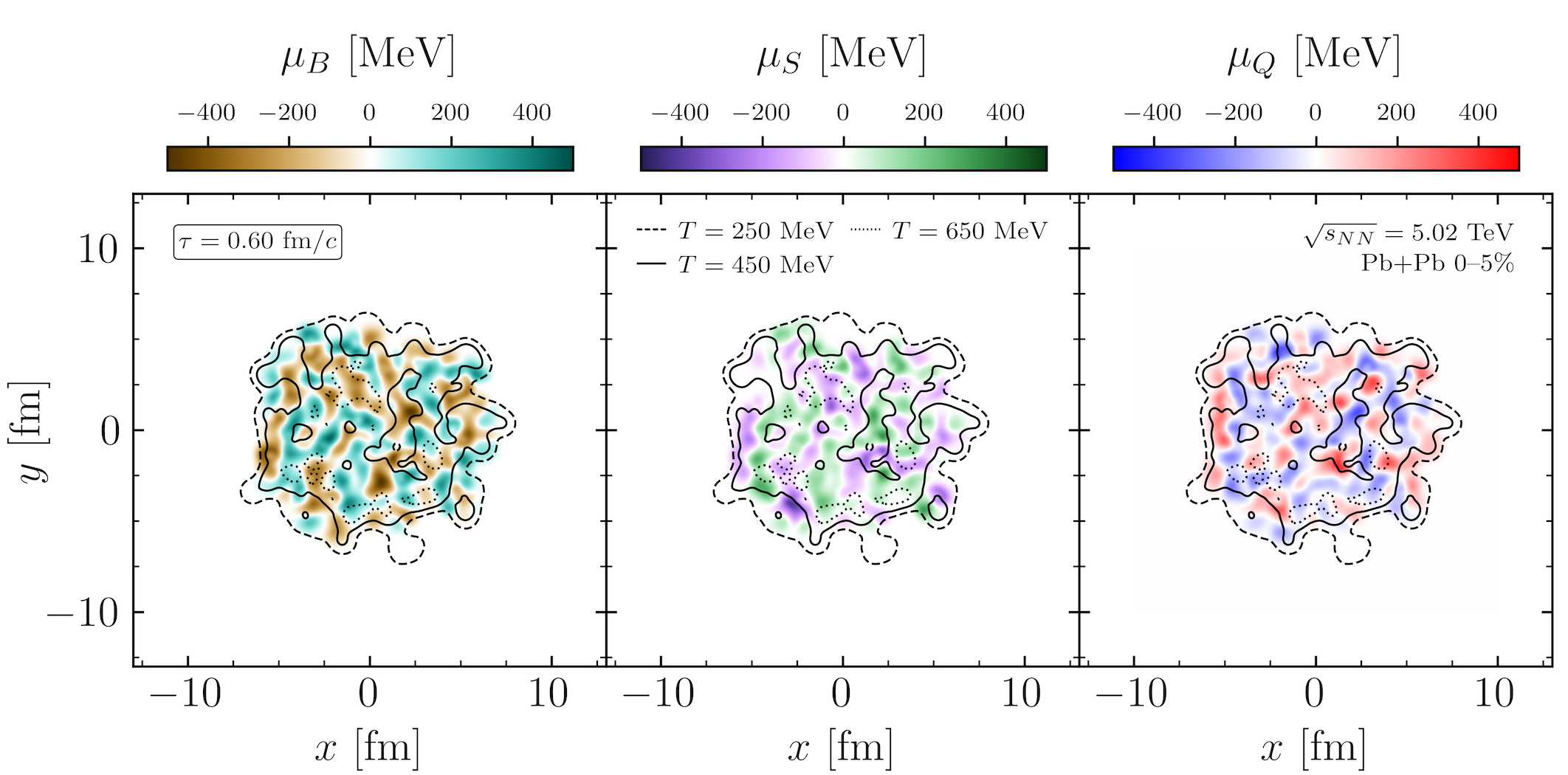}
    \caption{(color online) Mapping the $\{\rho_B, \rho_S, \rho_Q\}$ charge densities from \iccing{} into $\{\mu_B, \mu_S, \mu_Q\}$ chemical potentials through the lattice EoS at the initial time step $\tau_0=0.60$ fm/$c$ of the hydrodynamical evolution. Isothermal lines are shown for different initial temperatures. 
    }
    \label{fig:chemPOT}
\end{figure*}

Taking an example initial condition, we can now apply the lattice QCD EoS from \cite{Noronha-Hostler:2019ayj} to find the range of chemical potentials $\{\mu_B, \mu_S, \mu_Q\}$ that \iccing{} reaches.  The spatial distributions of $\{\mu_B, \mu_S, \mu_Q\}$ are shown in Fig.\ \ref{fig:chemPOT} wherein each chemical potential is plotted across the spatial coordinates for a given \iccing{} initial condition. 
Isothermal contours are shown as well to demonstrate hotter and colder regions within the initial state.
We find a surprisingly wide range of $\mu_B$ especially such that the initial state varies from $\mu_B\sim \pm 400$ MeV, which is approximately the range anticipated for the Beam Energy Scan.  However, we should point out that these $\mu_B$'s reached are at significantly higher temperatures than at the phase transition (ranging up to $T\approx 400-500$ MeV) such that we do not anticipate any influence from the critical point at freeze-out.
We remind the reader that typical hydrodynamic trajectories begin at large $T,\mu_B$ and in the limit with entropy is nearly conserved, $\mu_B$ decreases significantly with $T$ as freeze-out is approached (so called isentrope trajectories). 
Thus, we emphasize that a large $\mu_B$ at large $T$ does not necessarily imply large $\mu_B$ at freeze-out.
Additionally, these are only fluctuations in $\mu_B$ that occur around $\mu_B=0$, so after hydrodynamic expansion in \ccake{} we expect the range in $\mu_B$ to decrease significantly. 
Nevertheless, one can conclude that these conserved charge fluctuations may play a significant role in the hydrodynamic expansion, especially if there is a strong $\{\mu_B, \mu_S, \mu_Q\}$ dependence in the transport coefficients.
%%%%%%%%%%%%%%%%%%%%%%%%%%%%%%%%%%%%%%%%%%%%%%%%%
%%%%%%%%%%%%%%   ICCING PARAMETERS   %%%%%%%%%%%%
%%%%%%%%%%%%%%%%%%%%%%%%%%%%%%%%%%%%%%%%%%%%%%%%%
\subsubsection{Parameters and other details}
In this work, we use the parameter values from Table\ \ref{tab:ICCINGParameters} in \iccing{}.
\begin{table}
\centering
\begin{tabular}{ *2l }    \hline\hline
\emph{Initial State Parameter}     & \emph{Value}   \\
\hline
fluctuation (\TRENTo{})              & 1.6  \\ 
cross-section (\TRENTo{})            & 7.0 fm$^2$\\ 
nucleon-width (\TRENTo{})            & 0.3 fm \\
nucleon-min-dist (\TRENTo{})         & 0 fm \\
grid{\_}max (\TRENTo{}, \iccing{})  & 12 fm \\
grid{\_}step (\TRENTo{}, \iccing{}) & 0.03 fm \\
a{\_}TRENTo ($\mathcal{N}$)                   & 125 fm  \\
redistribution{\_}method              & "Default" \\
profile                            & "Kernel"  \\
chemistry                          & "GBW"     \\
charge{\_}type                        & "BSQ"     \\
perturbation{\_}cutoff                 & 0.9       \\
kappa                              & 1         \\
radius  ($r_g$)                           & 0.5 fm    \\
quark{\_}radius                       & 1.0 fm    \\
lambda                             & 1         \\
alpha{\_}s                         & 0.3        \\
alpha{\_}min                       & 0.01       \\
d{\_}max                           & 1 fm       \\
s{\_}chop                          & $10^{-20}$ fm$^{-3}$ \\
tau{\_}0                           & 0.6 fm/$c$    \\
e{\_}thresh ($E_\mathrm{thresh}$)                       & 0.25 GeV     \\
\hline\hline
\end{tabular}
\caption{List of all parameter values used in \iccing{} for this remainder of this analysis. The first four parameters are shared with \TRENTo{} (based on \cite{Bernhard:2016tnd}) and the value for $\mathcal{N}$ is specified.
}
\label{tab:ICCINGParameters}
\end{table}
The normalization constant is the same for both \TRENTo{} initial conditions and \TRENTo{}\texttt{+}\iccing{}.  
To obtain the correct normalization, we considered approximately 300 events in centrality collisions ($0$--$5\%$) and compared the produced spectra for all charged particles to $0$--$5\%$ centrality data from ALICE.  

%%%%%%%%%%%%%%%%%%%%%%%%%%%%%%%%%%%%%%%%%%%%%%%%%
%%%%%%%%%%%%%   2+1D HYDRODYNAMICS   %%%%%%%%%%%%
%%%%%%%%%%%%%%%%%%%%%%%%%%%%%%%%%%%%%%%%%%%%%%%%%
\subsection{2+1D viscous relativistic hydrodynamics}
\label{sec:Hydro}
%=============================================================================
\textit{Definitions}: In this work we employ the mostly
minus signature, i.e., $+,-,-,-$ in Minkowski space-time.  Natural units are used as well such that $\hbar=k_B=c=1$.
Throughout this paper we will be using hyperbolic coordinates $x^{\mu
}=(\tau, x, y,\eta )$ defined by 
\begin{eqnarray}
\tau &=&\sqrt{t^{2}-z^{2}},  \nonumber \\
\eta &=&\frac{1}{2}\ln \left( \frac{t+z}{t-z}\right) \,.
\end{eqnarray}%
Our 4-vectors are defined as:  $p^{\mu }$ is the 4-momentum,  $x^{\mu }$ is the
space-time coordinate, and $u^{\mu }$ is the flow field. In this paper we take our metric to be
\begin{equation}
    g_{\mu\nu}=\begin{pmatrix}
    1 & 0 & 0 & 0\\
    0 & -1 & 0 & 0\\
    0 & 0 &  -1& 0\\
    0 & 0 & 0 & -\tau^2\\
    \end{pmatrix},
\end{equation}
and $\sqrt{-g}=\tau$.
The boost-invariant flow is $u_{\mu }=\left( 
\sqrt{1+u_{x}^{2}+u_{y}^{2}},u_{x},u_{y},0\right) $. Thus, $u_\mu u^\mu=1$. 

Hydrodynamics is governed by conservation laws such that for an isolated system the energy-momentum must be conserved
\begin{equation}
    D_\mu T^{\mu\nu}=0,
\end{equation}
and all conserved charge currents (here we consider currents $X=B,S,Q$) such that
\begin{equation}
    D_\mu N_X^\mu=0,
\end{equation}
where $D_\mu$ is the covariant derivative for a general coordinate system
\begin{align}
D_\mu T^{\mu\nu} &= \frac{1}{\sqrt{g}}\partial_\mu (\sqrt{g} T^{\mu\nu}) + \Gamma^\nu_{\mu\alpha} T^{\mu\alpha} \,,\label{eq:T} \\
D_\mu N_X^{\mu} &= \frac{1}{\sqrt{g}}\partial_\mu (\sqrt{g} N^{\mu}_X) \,, \label{eq:N}
\end{align}
where the Christoffel symbols are defined by $\Gamma_{\mu\nu}^\alpha= \frac{1}{2} g^{\alpha\beta}(\partial_\nu g_{\mu\beta}+\partial_\mu g_{\beta\nu}-\partial_\beta g_{\mu\nu})$.

In this work, we assume the Landau frame such that
\begin{equation}\label{eq:LandauFrame}
    u_\nu T^{\mu\nu}=\varepsilon  u^\mu,
\end{equation}
where the energy density is $\varepsilon$ and the flow vector is $u^\mu$ that is normalized to $u^\mu u_\mu = 1$.  The Landau frame is most useful for high-energy collisions because it allows one to ignore contributions from heat flow. In contrast, most astrophysical simulations of relativistic fluids use the Eckart frame, which is more natural at large baryon densities.  Here, we specifically consider fluctuations around vanishing baryon densities such that we have large contributions both from positive \emph{and} negative densities so the Landau frame is a more natural choice.

Then, the energy-momentum tensor in the Landau frame where we also ignore contributions from vorticities is:
\begin{align}\label{eq:Energy-momentum_tensor_decomposition}
&T^{\mu\nu} = \varepsilon u^{\mu}u^{\nu}- (p+\Pi) \Delta^{\mu\nu} +\pi^{\mu\nu},
\end{align}
where the shear stress tensor is $\pi^{\mu\nu}$, the bulk pressure is $\Pi$,   the pressure is $p$, and the spatial projection operator is $\Delta^{\mu\nu}= g^{\mu\nu}-u^\mu u^\nu$.

We also consider multiple conserved charges where each current $N^{\mu}_X$ is defined as
\begin{align}\label{eq:Charge_current_decomposition}
N^{\mu}_X = \rho_X u^{\mu}+n^{\mu}_X\;,  \qquad   X\in\{B,S,Q\}
\end{align}
where the density of charge $X$ is defined as $\rho_X$, and the out-of-equilibrium contribution to the current is $n^{\mu}_X$.

By taking temporal and spatial projections of Eqs.~\eqref{eq:Energy-momentum_tensor_decomposition} and \eqref{eq:Charge_current_decomposition}, we find
\begin{align}
 D\varepsilon =& -(\varepsilon+P)\theta -\Pi \theta + \pi_{\mu\nu} \sigma^{\mu\nu}\,,\label{eq:e_conservation}\\
(\varepsilon+P)Du^\mu +\Pi Du^\mu =& \nabla^\mu(P{+}\Pi)- \Delta^{\mu\nu} \nabla^\lambda\pi_{\nu\lambda}\nonumber\\
&\; +\pi^{\mu\nu} Du_\nu\,,\label{eq:momentum_conservation}\\
D \rho_X =& -\rho_X \theta -\nabla_\mu n^\mu_X\;.
\label{eq:charge_conservation}
\end{align}
where the covariant time derivative is $D\equiv u_\mu D^\mu$,  the expansion scalar is $\theta\equiv D_\mu u^\mu$, the shear flow tensor is $\sigma^{\mu\nu} \equiv \nabla^{\langle\mu} u^{\nu\rangle}$,  and the spatial gradient in the local rest frame is $\nabla^\mu\equiv \Delta_{\mu\nu} \partial^\nu$.

At this point, to continue solving the equations of motion, one must make a choice of the definitions of the time dependence of the out-of-equilibrium corrections: $\pi^{\mu\nu}$, $\Pi$, and $n^\mu_X$.  Here we apply a minimal Israel-Stewart approach, although we plan to try other approaches in future work.  Additionally, we assume that the currents are ideal such that
\begin{equation}
    n^\mu_X\equiv 0,
\end{equation}
since significant complications occur in the SPH formalism once an out-of-equilibrium conserved current is considered. This assumption leaves us only with the time dependence of $\pi^{\mu\nu}$ and $\Pi$ to handle.

For the minimal set of Israel-Stewart equations of motion we obtain
\begin{align}
\Pi&=-\zeta\theta-\tau_{\Pi}\left(D\Pi+\Pi\theta\right) \label{eq:Bulk_equation_of_motion},\\
\pi^{\mu\nu} &= 2 \eta \sigma^{\mu\nu} - \tau_\pi  \bigg[ \Delta_{\alpha\beta}^{\mu\nu}  D \pi^{\alpha\beta}
    + \frac{4}{3} \pi^{\mu\nu} \theta \bigg]\,, \label{eq:Shear_equation_of_motion}
\end{align}
where
\begin{equation}\label{eq:Tensor_projector}
\Delta^{\mu\nu\alpha\beta}\equiv \frac{1}{2}\left[\Delta^{\mu\alpha}\Delta^{\nu\beta}+\Delta^{\mu\beta}\Delta^{\nu\alpha}-\frac{2}{3}\Delta^{\mu\nu}\Delta^{\alpha\beta}\right].
\end{equation}
These are the same equations of motion previously applied to \vUSPhydro{} in \cite{Noronha-Hostler:2013gga,Noronha-Hostler:2014dqa} and that were derived using the memory function approach in \cite{Koide:2006ef,Denicol:2009am,Denicol:2010tr}.

However, one difference from \vUSPhydro{} to \ccake{} is that time derivatives of thermodynamic variables have been added to the equations of motions (e.g., the ``$\dot{\beta}$" terms discussed in \cite{Dore:2020jye}).
However, in this work we only consider $\eta/s = \text{const.}$ at $\vect{\mu} = 0$, which generalizes to $T\eta/w = \text{const.}$ when $\vect{\mu} \neq 0$; in this case, the $\dot{\beta}_\pi$ term vanishes.
Conversely, a temperature dependent $[\eta/s](T, \vect{\mu} = 0)$ should include a time derivative of
\begin{equation}
    \beta_\pi=\frac{\tau_\pi}{2\eta T},
\end{equation}
where an extra term
\begin{equation}
    \frac{\tau_\pi\dot\beta_\pi}{2\beta_\pi}\pi^{\mu\nu}
\end{equation}
would appear in Eq.\ (\ref{eq:Shear_equation_of_motion}).
The new version of \ccake{} also includes the bulk viscosity term $\dot{\beta}_\Pi$ 
\begin{equation}
    \beta_\Pi=\frac{\tau_\Pi}{\zeta T},
\end{equation}
where for our specific choice of $\tau_\Pi$ leads to time derivatives of the speed of sound squared $c_s^2$ (that we will explain in more detail in the transport coefficient Sec.\ \ref{sec:TransportCoefficients}). Then, one would include a term in the $\Pi$ equation of
\begin{equation}
    \frac{\tau_\Pi}{2\beta_\Pi}\dot\beta_\Pi\Pi.
\end{equation}
In the code, the time derivative of $c_s^2$ is taken using a finite difference method.
However, for this work we rely on our previous parameterization from \vUSPhydro{} where the bulk viscosity is zero such that this term does not yet play a role. Thus, we leave a deeper study of this term for a future work.

%%%%%%%%%%%%%%%%%%%%%%%%%%%%%%%%%%%%%%%%%%%%%%%%%
%%%%%%%%   THE SPH LAGRANGIAN ALGORITHM   %%%%%%%
%%%%%%%%%%%%%%%%%%%%%%%%%%%%%%%%%%%%%%%%%%%%%%%%%
\subsection{The SPH Lagrangian algorithm}
\label{sec:SPHformalism}
%=============================================================================
The SPH Lagrangian algorithm proceeds by assigning to each fluid element a fictitious particle (known as an ``SPH particle") with a kernel function which locally smooths physical quantities over the volume of the system and provides a continuously differentiable representation of the state of the system at any given timestep.  
Each SPH particle then evolves independently according to its own equations of motion, in a way which depends on the locally smoothed densities in its vicinity.  
The kernel function, thus, acts to define the neighborhood over which the smoothed physical state of the system influences the subsequent evolution of a particle.

Here we use the same kernel function already defined in Eq.\ (\ref{eq:KernelQuarkDensity}) that was applied within \iccing{}.  However, one key difference is that the smoothing scale $h$ within \ccake{} does not necessarily have to correspond with $r_g/2$ described in Sec.\ \ref{sec:ICCING}.  In fact, in this work we will use $h=0.3$ fm for \ccake{} but $r_g/2=0.5$ fm for \iccing{}.
\begin{figure}
    \includegraphics[keepaspectratio, width=0.95\linewidth]{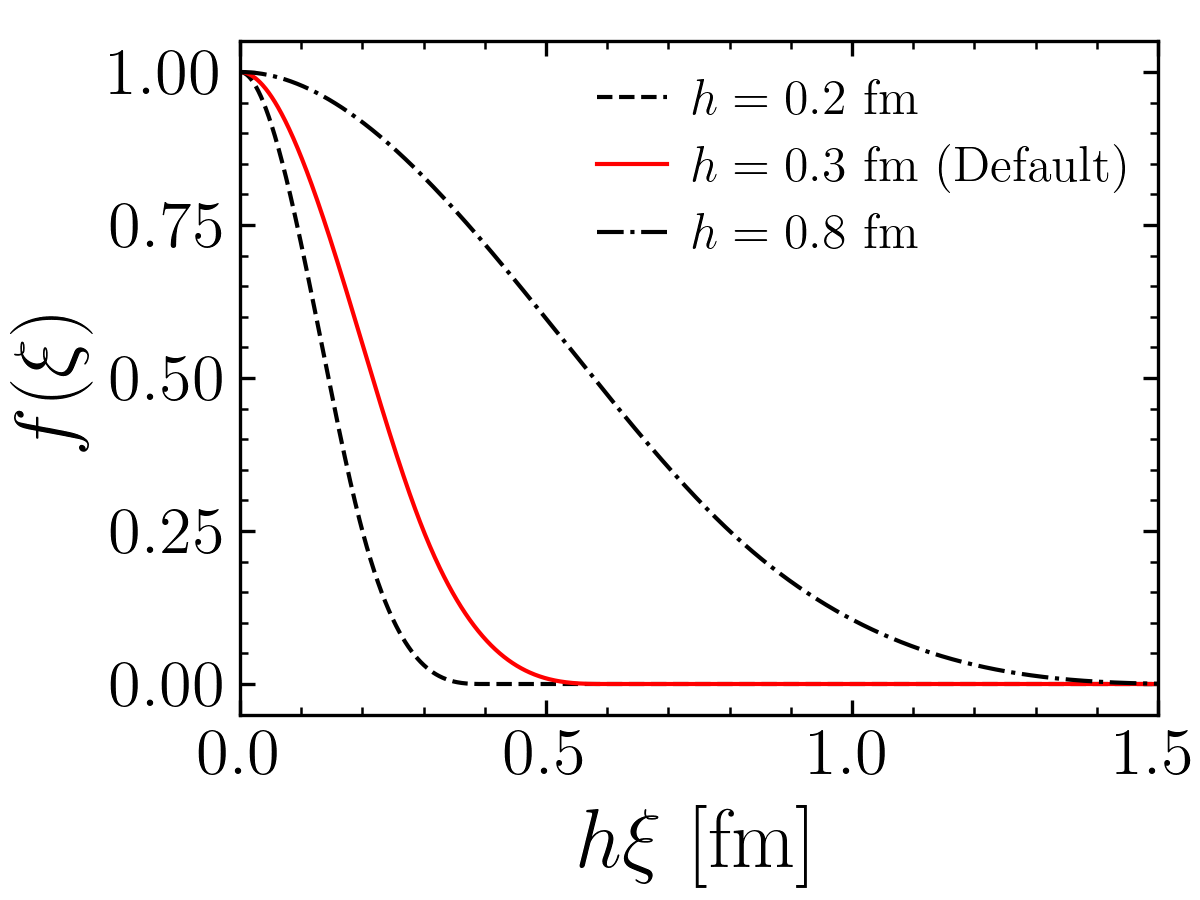}
    \caption{(color online) The smoothing kernel function described as a cubic spline, defined in Eqs.\ (\ref{eq:KernelQuarkDensity}) and (\ref{eq:CubicSpline}), implemented in \ccake{} for different values of the smoothing parameter $h$. In solid red, the smoothing kernel corresponding to $h=0.3$ fm, chosen by default in \ccake{}.
    }
    \label{fig:kernel_for_different_h}
\end{figure}
The kernel function has to obey certain symmetries, but most importantly for this work it is normalized according to
\begin{eqnarray}
\int W\left[\vect{r};h\right] d^2\vect{r} &=&1\,.
\end{eqnarray}
Here we demonstrate the kernel for different values of the smoothing scale $h$ in Fig.\ \ref{fig:kernel_for_different_h}. 
As one can see, a large value of $h$ incorporates SPH particles in a wider radius, thus, smoothing out the hydrodynamic gradients more.  
In contrast, a smaller value of $h$ allows for more short range fluctuations, see \cite{Noronha-Hostler:2015coa,Gardim:2017ruc}.

Within the SPH formalism, one must choose a conserved current over which the equations of motion are reconstructed using the SPH particles.  
In astrophysical codes \cite{Rosswog:2009sr,Rosswog:2020kwm}, this is typically done from the baryon current.  However, in heavy-ion collisions this is not possible because the baryon density may be positive or negative and on average $\langle \rho_B\rangle=0$ for \iccing{} initial conditions.  Thus, we choose a new current $J^\mu=\sigma u^\mu$ that is applicable in this regime known as the reference current \cite{Hama:2004rr,DerradideSouza:2015kpt}. 
Here $\sigma$ is the local density
of a fluid element in its rest frame and as the fluid flows in time, the fluid cell itself is deformed but the density always obeys:
\begin{equation}
    D\sigma +\sigma \theta=0,
\end{equation}
which in hyperbolic coordinates can be rewritten as
\begin{align}
\partial_\mu(\tau \sigma u^\mu)=0 .
\end{align}
One can then reconstruct the reference density in the lab frame using the SPH formalism
\begin{equation}
\tau \gamma \sigma \rightarrow \sigma^{*}\!\left(\vect{r},\tau\right)=\sum_{%
\alpha =1}^{N_\text{SPH}}\nu_{\alpha}\,W\left[\vect{r}-\vect{r}_\alpha(\tau);h%
\right],
\label{sigma}
\end{equation}
where $\nu_\alpha$ are constants attached to the Lagrangian coordinate $\vect{r}_\alpha(\tau)$ and $\sigma^{*}$ is a
sum of the distribution $\nu_\alpha W[\vect{r}-\vect{r}%
_\alpha(\tau);h]$ of the SPH particles. If one were to instead use the baryon current as the conserved quantity, then $\nu_\alpha$ would be the local baryon density of the fluid cell $\alpha$. However, in our case we use the initial grid size to set the $\nu_\alpha$ weight that remains fixed throughout the evolution of the code. Since our initial conditions are based on a fixed grid all SPH particles carry the same weight.
The vector current is then
\begin{gather}
\vect{j}^*\!\left( \vect{r},\tau\right) =\sum_{\alpha =1}^{N_\text{SPH}}\nu
_{\alpha }\frac{d\vect{r}_{\alpha }(\tau)}{d\tau}W[\vect{r}-\vect{r}%
_{\alpha }(\tau);h],  \label{current}
\end{gather}
such that the continuity equation is
\begin{align}
 \partial_\tau \sigma^{*}\!\left(\vect{r}%
,\tau\right)+\nabla_{\vect{r}} \cdot \vect{j}^*\!\left( \vect{r}%
,\tau\right)=0.
\end{align}
If we have some extensive quantity $A$ with an associated density is $a(\vect{r},\tau)$ for an individual SPH, the SPH description of this quantity is
\begin{equation}
a(\vect{r},\tau)=\sum_{\alpha =1}^{N_\text{SPH}}\nu_{\alpha}\,\frac{a(\vect{r}%
_\alpha(\tau))}{\sigma^{*}\!\left(\vect{r}_\alpha(\tau)\right)}W\left[%
\vect{r}-\vect{r}_\alpha(\tau);h\right]\,.
\end{equation}
For the zeroth component of the entropy current in the lab frame $s^*=s \gamma \tau$ we have
\begin{equation}
s^*(\vect{r},\tau)=\sum_{\alpha =1}^{N_\text{SPH}}\nu_{\alpha}\,\frac{s(\vect{%
r}_\alpha(\tau))}{\sigma\left(\vect{r}_\alpha(\tau)\right)}W\left[\vect{r%
}-\vect{r}_\alpha(\tau);h\right],
\end{equation}
and for the bulk term one finds
\begin{equation}
\Pi(\vect{r},\tau)=\sum_{\alpha=1}^{N_\text{SPH}}\nu _{\alpha }\frac{1}{\gamma
_{\alpha }\tau }\left( \frac{\Pi }{\sigma }\right) _{\alpha }W[\vect{r}-%
\vect{r}_{\alpha }(\tau);h]\,.
\end{equation}
The dynamical variables in the SPH method are then
\begin{equation}
\left\{ \vect{r}_{\alpha },\vect{u}_{\alpha },\left( \frac{s}{\sigma }%
\right) _{\alpha },\left( \frac{\Pi }{\sigma }\right)_{\alpha },\left( \frac{\pi^{\mu\nu} }{\sigma }\right)_{\alpha } ,\left( \frac{n^\mu_X }{\sigma }\right)_{\alpha }\right\},\label{dynamicalvariables}
\end{equation}
with $ \alpha=1,\ldots,N_{SPH}$, where each of these quantities is associated with the $\alpha $-th SPH particle, respectively. However, here we consider only ideal charge currents such that $\left( n^\mu_X /\sigma \right)_{\alpha }$ is dropped hereafter.  The fluid expansion rate for each SPH particle is
\begin{equation}
\theta_\alpha = (D_\mu u^\mu)_\alpha=\frac{d\gamma_\alpha}{d\tau}+\frac{%
\gamma_\alpha}{\tau}-\frac{\gamma_\alpha}{\sigma_\alpha^*}\frac{%
d\sigma_\alpha^*}{d\tau},
\end{equation}
where $\gamma_\alpha=1/\sqrt{1-v^2_\alpha}$.
We obtain the following equations associated with the momentum conservation Eq.\ (\ref{eq:momentum_conservation}) for each SPH particle
\begin{align}
\gamma \frac{d}{d\tau }\left[ \frac{\left( \varepsilon +p+\Pi \right) }{
\sigma }u^{\mu }\right] + \frac{1}{\tau \sigma} \partial_\mu (\tau \pi_{\mu i})&=\frac{1}{\sigma }\partial ^{\mu }\left( p+\Pi
\right)  \label{momentum_cons_sph}
\end{align}

The pressure gradients on the RHS of Eq.\ (\ref{eq:momentum_conservation}) are the SPH representation of the gradients of pressure and bulk viscosity and are calculated in the code using the following parametrization:
\begin{widetext}
\begin{align}
\sigma^{\ast }\frac{d}{d\tau }\left( \frac{\left( \varepsilon +p+\Pi \right)
_{\alpha }}{\sigma _{\alpha }}\,u_{i_\alpha }\right)+ \frac{1}{\tau \sigma} \partial_\mu (\tau \pi_{\mu i}) =
\tau \sum_{\beta=1}^{N_\text{SPH}}\nu _{\beta }\sigma _{\alpha }^{\ast }\left( \frac{p_{\beta}+\Pi _{\beta }}{\big( \sigma _{\beta }^{\ast }\big) ^{2}}+\frac{
p_{\alpha }+\Pi _{\alpha }}{\big( \sigma _{\alpha }^{\ast }\big) ^{2}}%
\right) \ \partial _{i}W[\vect{r}_{\alpha }-\vect{r}_{\beta }(\tau);h]\,.
\end{align}
\end{widetext}
where the $\frac{1}{\tau \sigma} \partial_\mu (\tau \pi_{\mu i})$ term is typically quite complex because one applies the product rule and then opens up the individual components of the $\pi^{\mu\nu}$.
The next equation of motion arises from the entropy production term 
\begin{align}
T D_\mu\left(s u^\mu\right)=-\Pi D_\mu u^\mu+\pi^{\mu \nu} D_\mu u_\nu
\end{align}

which can be written in SPH format as the following
\begin{equation}
\gamma_\alpha\frac{d}{d\tau}\left(\frac{s}{\sigma}\right)_\alpha+\left(\frac{%
\Pi}{\sigma}\right)_\alpha\left(\frac{\theta}{T}\right)_\alpha+\left(\frac{\pi^{\mu \nu}}{\sigma}\right) \left(\frac{D_\mu u_\nu}{T}\right)_\alpha =0\,,
\end{equation}
For the relaxation time equation of the bulk and shear viscosity, we then write for bulk,
\begin{equation}\label{eq:bulk_evolution_in_SPH}
\tau _{\Pi_{\alpha }}\gamma _{\alpha }\frac{d}{d\tau }\left( \frac{\Pi }{%
\sigma }\right) _{\alpha }+\left( \frac{\Pi }{\sigma }\right) _{\alpha
}+\left(\frac{\zeta}{\sigma}\right)_{\alpha }\theta _{\alpha }=0\,,
\end{equation}
Finally, the ideal expansion rate of the BSQ charges is expressed in the following format:
\begin{align}\label{eq:charge_evolution_in_SPH}
\gamma _{\alpha }\frac{d}{d\tau }\left( \frac{\rho_X }{%
\sigma }\right) _{\alpha }  +  \left(\frac{\rho_X}{\sigma}\right)_{\alpha }\theta _{\alpha } = 0\;.
\end{align}
\label{eq:charge_conservation_sph}
At this point in time, the shear stress tensor is not obtained using the SPH kernel but rather is simply re-constructed directly from the equations of motion, as described in Appendix \ref{sec:shear-rel-sph}.  
Previously in \cite{Noronha-Hostler:2014dqa} it was checked that the effect of this choice was negligible (except that the SPH version generally slowed down the hydrodynamic runs).  
However, in future work we plan to investigate this effect in more detail with the new additions to the code.

As a final note, in order to streamline the SPH process, \ccake{} uses linked lists to determine the nearest neighbors for each SPH particle in order to efficiently apply the kernel function.  At the beginning of each time step the linked list is formed for every SPH particle with all neighboring SPH particles that fall within a radius of $2h$ of the SPH particle itself. 

%%%%%%%%%%%%%%%%%%%%%%%%%%%%%%%%%%%%%%%%%%%%%%%%%
%%%%%%%%%%%   TRANSPORT COEFFICIENTS   %%%%%%%%%%
%%%%%%%%%%%%%%%%%%%%%%%%%%%%%%%%%%%%%%%%%%%%%%%%%
\subsection{Transport coefficients}
\label{sec:TransportCoefficients}
%=============================================================================
In this work, because we are only considering the minimal Israel-Stewart equations of motion, we only need to take into consideration the values of the shear viscosity to enthalpy ratio $\eta T/w$, bulk viscosity to enthalpy ratio ratio $\zeta T/w$, the shear relaxation time $\tau_\pi$, and the bulk relaxation time $\tau_\Pi$. Note that the enthalpy is $w=\varepsilon+p$.

A large number of Bayesian analyses have studied the temperature dependence of $\eta/s$ and $\zeta/s$ at $\mu_B=0$.  In this limit of $\vect{\mu}=0$, then the transport coefficients need only be normalized by the entropy:
\begin{eqnarray}
   \frac{\eta}{s}&=&\frac{\eta T}{w}\Big|_{\vect{\mu}=0}\\
   \frac{\zeta}{s}&=&\frac{\zeta T}{w}\Big|_{\vect{\mu}=0}.
\end{eqnarray}
From these studies, the posteriors for $\eta/s$ at $\vect{\mu}=0$ do not clearly indicate a dependence on the temperature \cite{Bernhard:2019bmu,JETSCAPE:2020mzn,Nijs:2020roc,Parkkila:2021tqq}.  Consequently, in this work we assume $\eta T/w=\text{const}$.
The corresponding relaxation time $\tau_\pi$ for shear viscosity is then taken from the usual form,
\begin{equation}
    \frac{\tau_\pi w}{\eta}=\text{const.},
\end{equation}
where we have taken that constant factor to be 5 in this particular paper.

Here we define our bulk viscosity to be of the form:
\begin{align}\label{eq:zetaTw}
    \l(\frac{\zeta T}{w}\r)(T)
    &=  f\l(\frac{T}{T_\text{transition}}, \frac{T_\text{transition}}{T_\text{scale}}\r) \nonumber\\
    &\times A\l(\frac{1}{3} - \mathrm{min}(c_s^2, 1)\r)^p,
\end{align}
where the prefactor
\begin{equation*}
    f(x,y) = \frac{1}{2}\l[\l(1 + x^p\r) + \l(1 - x^p\r) \tanh \l(y(x-1)\r) \r]
\end{equation*}
suppresses $\zeta T/w$ as $T \to 0$ according to
\begin{equation}
    \lim_{x \to 0} f(x,y) = \frac{1}{2} (1 - \tanh(y)).
\end{equation}
By default, we set $A = 8\pi/15$, $p=2$, and
\begin{equation*}
    T_\text{transition} = 150\text{ MeV},    \qquad
    T_\text{scale} = 10\text{ MeV}.
\end{equation*}
We note that $T_\text{transition}$ and $T_\text{scale}$ here are free parameters which need not coincide exactly with other quantities in the simulations (such as the freeze-out temperature $T_\mathrm{FO}$ at $\vect{\mu} = 0$), for instance.
Our bulk relaxation time is
\begin{equation}
    \frac{\tau_\Pi w}{\zeta} \left(\frac{1}{3}-c_s^2\right)^p=\text{const.}
\end{equation}
We have specifically chosen the form in Eq.\ (\ref{eq:zetaTw}) such that it can easily vary between $p=2$ for a weakly coupled system and $p=1$ for a strongly coupled system \cite{Czajka:2018bod}. While we do not use the bulk viscosity in our results section in this paper, a follow-up work is planned where we will vary the parameters $A$, $p$, $T_\text{transition}$, and $T_\text{scale}$.

%%%%%%%%%%%%
\begin{table}
\centering
\begin{tabular}{ *2l }    \hline\hline
\emph{Hydrodynamics Parameter}     & \emph{Value}   \\
\hline
ICtype:               & ICCING  \\
h:             & 0.3 fm \\ 
dt:            & 0.05 fm/$c$\\
t0:         & 0.6 fm/$c$ \\
e$\_$cutoff:  & 0.15 fm$^{-4}$ (0.02955 GeV/fm$^3$)\\
EoS$\_$Type: & table \\
EoS$\_$Path:      & Houston \\
freezeoutT:         & 150.0 MeV \\
etaMode:            & constant     \\
shearRelaxMode:        & default     \\
constant$\_$eta$\_$over$\_$s:       & 0.08       \\
zetaMode:                 & constant \\
constant$\_$zeta$\_$over$\_$s:     & 0.0  
 *  \\
bulkRelaxMode:            & default \\
cs2$\_$dependent$\_$zeta$\_$A:     & 1.67552 *\\
cs2$\_$dependent$\_$zeta$\_$p:     & 2 *\\
\hline\hline
\end{tabular}
\caption{a compact version of the input parameter file used in \ccake{}. A list of all parameter values used can be found in \ccake{}. In this work, we set $\zeta =0$. By default, the code sets the value of $\zeta = 0.005$ unless another constant is chosen by the user, and the parameters defined by * are therefore only relevant if zetaMode = constant and $\zeta =$ finite.
}
\label{tab:CCAKEParameters}
\end{table}
%%%%%%%%%%%%%%%

One advantage of our current format for $\zeta T/w$ is that it allows for a $\left\{T,\vect{\mu}\right\}$-dependent enhancement at the phase transition (see \cite{Dore:2020jye,Dore:2022qyz} for implications).  
However, as $\vect{\mu}$ increases, we anticipate that the phase transition bends towards lower temperatures (see previous lattice QCD studies \cite{Bellwied:2015rza,Borsanyi:2018grb,Borsanyi:2020fev,Bazavov:2020bjn,HotQCD:2018pds} on this matter for details).  
Thus, our parametrization of $\zeta T/w$ follows $c_s^2$ and will naturally capture its behavior.
Additionally, the bump will either sharpen or flatten depending on the sharpness of the dip in $c_s^2$ as one crosses the phase transition.  
In the  presence of a critical point, one could also add in critical scaling to the bulk viscosity as was done in \cite{Moore:2008ws,Monnai:2016kud,Dore:2020jye,Dore:2022qyz}.

In future work we plan to explore the possibility of allowing $\eta T/w$ to depend on $\vect{\mu}$ as well. 
Previous work has developed a phenomenological approach to study that behavior across both cross-over phase transitions and critical points \cite{McLaughlin:2021dph} with a first-order line. 
Within specific theories, it is also possible to calculate shear and bulk viscosity across the phase transition, as was shown in \cite{Soloveva:2019xph,Soloveva:2020hpr,Soloveva:2021quj,Grefa:2022sav}, in order to build intuition that can provide guidance for further phenomenological studies.  
Furthermore, there are now pQCD results available \cite{Danhoni:2022xmt} for finite $\vect{\mu}$ that would be quite interesting to use as guidance.   However, none of these studies have been performed yet for a 4D EoS so we leave that study for a follow-up paper.

%%%%%%%%%%%%%%%%%%%%%%%%%%%%%%%%%%%%%%%%%%%%%%%%%
%%%%%%%%%%%%%%   EQUATION OF STATE   %%%%%%%%%%%%
%%%%%%%%%%%%%%%%%%%%%%%%%%%%%%%%%%%%%%%%%%%%%%%%%
\subsection{Equation of State}
\label{sec:EoS}
%=============================================================================

The primary EoS employed in \ccake{} is the lattice QCD EoS based on a Taylor series expansion up to $\mathcal{O}(\mu_X^4)$ (where $X = B, S, Q$),  coupled to a Hadron Resonance Gas using the PDG2016+ \cite{Alba:2017mqu} list\footnote{Note that recently a new PDG2021+ was released \cite{SalinasSanMartin:2023idj} but it still has not yet been incorporated into an EoS so we leave that for a future work.} in the low temperature regime, and converging to the Stefan-Boltzmann limit at large temperatures \cite{Noronha-Hostler:2019ayj}.
However, in order to integrate this 4D EoS in the intensive variables $\left\{T,\mu_B,\mu_S,\mu_Q\right\}$ into BSQ hydrodynamic evolution, there are a number of specific challenges which first need to be overcome:
\begin{enumerate}
    \item Out-of-bounds fluid cells which are beyond the reach of the Taylor Series,
    \item 4D interpolation and root-finding,
    \item Accuracy and speed,
    \item Thermodynamic derivatives.
\end{enumerate}
We first briefly discuss each of these challenges.  Then, in the remainder of this section and Sec.\ \ref{sec:TimeDerivativesOfThermodynamicVariables}, we discuss how we address these challenges in greater depth.  Additional details are deferred to Apps. \ref{app:thermodynamic_derivatives}, $\ref{app:thermoRel}$, and \ref{app:Benchmark_checks}.

First we consider the challenge of {\bf out-of-bounds fluid cells}.
Ideally, the lattice EoS would perfectly represent the charge densities needed to evolve \iccing{} initial conditions with hydrodynamics.
In practice, however, we have inevitably found that a number of fluid cells either fall out of bounds of the EoS or are otherwise without a solution in the domain of the Taylor series EoS.
This can occur due to limitations in the reach of the Taylor series which cause it to break down at larger densities, or due to the inability of the low-temperature hadron resonance gas extension to yield an arbitrary combination of charge densities at a single $T$.

This breakdown is more pronounced in the case of \iccing{} than other initial state models for two reasons: (i), \iccing{} incorporates quark degrees of freedom whereas other initial state models incorporate only conserved charges at the hadronic level, and (ii), \iccing{} incorporates three conserved charges whereas most other codes only conserve baryon number (and occasionally electric charge) and then enforce strangeness neutrality exactly \cite{Monnai:2021kgu,Schafer:2021csj}.

Because \iccing{} incorporates quark degrees of freedom, it reaches different combinations of BSQ unreachable by hadronic models.
For example, if a large quantity of down quarks are clustered together, one could reach a region of the phase diagram with a positive baryon number but negative electric charge.
While this combination can occur in an HRG (with $\Delta^-$ baryons, for example), it is statistically suppressed in a hadronic model due to the much larger masses of the required resonance states (as compared to pions).
In contrast, in a quark phase $d$ quarks are abundant, and arbitrary charge configurations can be easily reproduced by distributing quarks appropriately.
Thus, in this work we require strategies to handle out-of-bounds fluid cells that do not lead to energy loss within our simulations.
Our main strategy, discussed below, will be to replace the existing Taylor series EoS with an alternative EoS.
However, a word of caution is that one cannot jump directly from the Taylor series to a different EoS because this will lead to energy loss within the code (such that energy is no longer conserved).
Thus, one must find a way of doing this `smoothly,' i.e., in a way which conserves energy within the simulations.

This challenge is dealt with by utilizing several `fallback' equations of state for those cases where the default EoS does not yield an acceptable solution.
The `fallback' EoSs are ordered in such a way as to minimize the violations of energy conservation which inevitably result when one discontinuously changes the EoS: in order, the `fallback' EoSs are the $\tanh$-conformal, the conformal, and conformal-diagonal EoS.
The functional forms of these `fallbacks' are described completely in the manual for \ccake{} \cite{ccakesite} and are also summarized below.
Their free parameters are chosen to provide optimal approximations to the default EoS, again serving to minimize violations of energy conservation.

The second challenge involves the use of {\bf 4D interpolation and root-finding} when numerically approximating the EoS during the hydrodynamic evolution.
For this purpose, we use the open-source GNU Standard Library (GSL) routines to perform numerical rootfinding.
This is done by first constructing multi-linear interpolants of $\varepsilon$, $s$ and $\vect{\rho}$ on a grid of uniformly spaced points in $T$ and $\vect{\mu}$.
The collection of multi-linear interpolants then provides a mapping between the $(T,\vect{\mu})$ space and the $(e,\vect{\rho})$ and $(s,\vect{\rho})$ spaces which can be passed to the GSL rootfinder.
The GSL functionality accepts a seed value for the estimated coordinates in $(T,\vect{\mu})$ space and returns either a converged solution or a flag which indicates the failure to find an acceptable solution.

The third challenge is to ensure that the EoS implementation is handled with efficiency {\bf efficiency and accuracy}. 
From convergence tests, Appendix \ref{app:Benchmark_checks} we have found that a grid size in the EoS with step sizes of $\Delta T = 5$ MeV and $\Delta \mu = 50$ MeV is sufficient to ensure accuracy of our simulations.
Changing these values to a finer ones with $\Delta \mu = 25 MeV$ affected the measurements of flow observables by less than $1\%$, see Fig. \ref{fig:EoSGrid}. 
 
However, a significant caveat for this is that we are considering an EoS with a only cross-over and hence no critical point.
From our experience with \cite{Dore:2022qyz} a critical point requires a significantly smaller grid spacing than the one used here.  We defer the consideration of these complications to future work.

An especially important way to improve efficiency is to avoid is taking thermodynamic derivatives during the run time of the code.
For example, in the SPH formalism, one requires a time derivative of the enthalpy $\frac{dw}{d\tau}$, which can both be costly in terms of time and can also lead to numerical errors.
To circumvent this, we apply the chain rule to rewrite thermodynamic time derivatives so that they contain a piece which is calculated directly from the EoS and another piece that is based upon quantities that are already calculated within hydrodynamics.
Thus,
\begin{equation}\label{eq:dwdt}
 \frac{dw}{d\tau}=\sum_{\varphi \in \Phi} \frac{\partial w}{\partial\varphi} \frac{d\varphi}{d\tau}
\end{equation}
where $\Phi$ is a set of natural hydrodynamic variables already computed within the simulation and each coefficient $dw/d\varphi$ is determined solely by the EoS.  For our purposes, a natural choice of $\Phi$ is the set of variables
\begin{equation}
 \Phi = \l\lbrace s, n_B, n_S, n_Q \r\rbrace
\end{equation}
We work out the corresponding derivatives of $w$ with respect to these variables in Sec.\ \ref{sec:TimeDerivativesOfThermodynamicVariables}.

The fourth challenge is to incorporate these new {\bf thermodynamic derivatives} into the EoS itself so that they can be quickly evaluated at runtime.
This requires an involved calculation to rewrite arbitrary thermodynamic derivatives in terms of susceptibilities and other derivatives of the pressure.
In this work, we have only incorporated this up to $2^{nd}$ order thermodynamic derivatives. 
Israel-Stewart has terms that require time derivatives of $c_s^2$ which would lead to third-order derivatives, although we have not yet included these terms in our code and we will therefore discuss them in a future work.

\begin{figure*}
    \centering
    \includegraphics[keepaspectratio, width=0.92\linewidth]{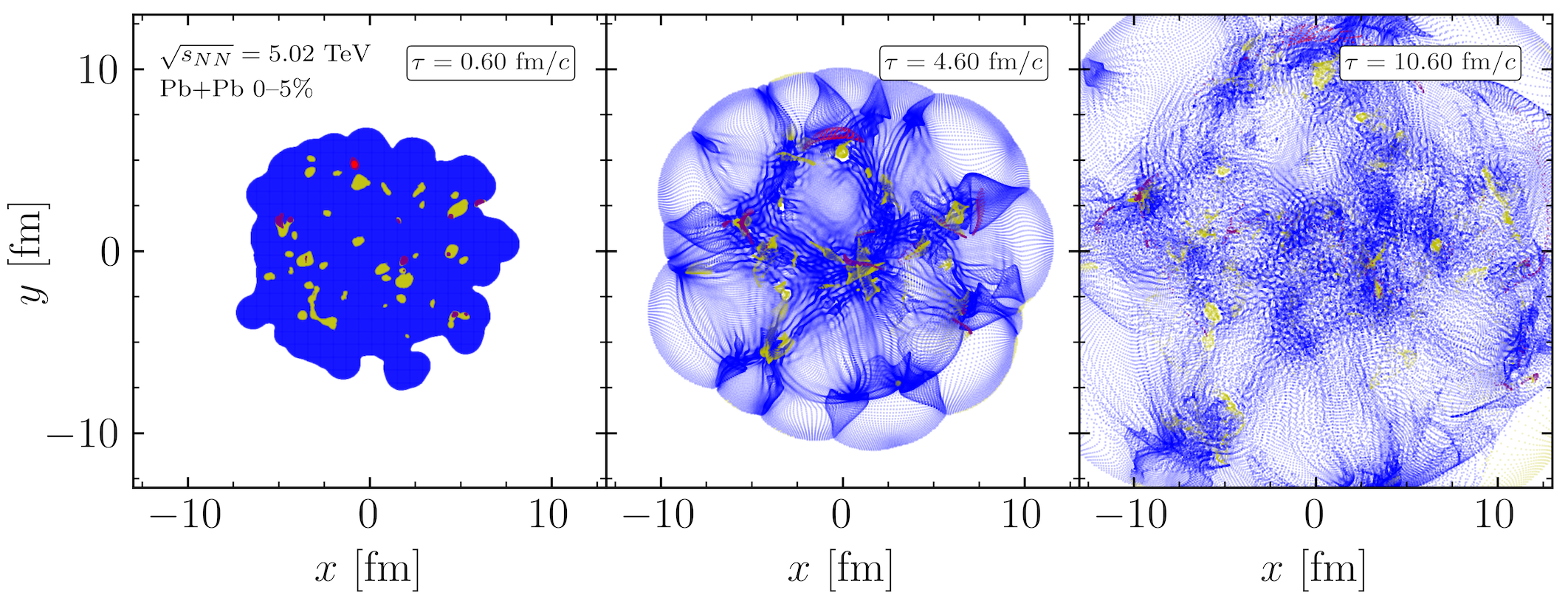}
    \caption{
    (color online) Equation of state distribution on the transverse plane at times $\tau=0.60$, $4.60$, and $10.60$ fm$/c$ in the evolution. In all cases, SPH particles are color coded according to the EoS used: blue $=$ lattice QCD, yellow $=$ $\tanh$-conformal, purple $=$ conformal, and red $=$ conformal-diagonal.
    }
    \label{fig:EoSexpand}
\end{figure*}

\subsubsection{Out-of-bound fluid cells}

We now discuss the first two of the four challenges mentioned above in greater depth.  The third challenge is addressed more thoroughly in Appendix \ref{app:Benchmark_checks}, while we deal with the fourth challenge in Sec.\ \ref{sec:TimeDerivativesOfThermodynamicVariables} and Appendix \ref{app:thermodynamic_derivatives}.
The primary role of the EoS is to take an input entropy density $s$, baryon density $\rho_B$, strangeness density $\rho_S$, and electric charge density $\rho_Q$ and to correlate it to the rest of the thermodynamic quantities such as pressure $p$, energy density $\varepsilon$, temperature $T$, and chemical potentials $\vect{\mu}=\left\{\mu_B,\mu_S,\mu_Q\right\}$.
Thus, our input from each fluid cell is
\begin{equation}
   \left\{s^0,\vect{\rho}^0\right\}= \left\{s^0,\rho_B^0,\rho_S^0,\rho_Q^0\right\},
\end{equation}
such that we must take a given EoS and convert that into
\begin{eqnarray}
T&=&f_T\left\{s^0,\vect{\rho}^0\right\},\\
\mu_X&=&f_{\mu_X}\left\{s^0,\vect{\rho}^0\right\},\quad X=B,S,Q\\
\varepsilon&=&f_{\varepsilon}\left\{s^0,\vect{\rho}^0\right\},\\
p&=&f_{p}\left\{s^0,\vect{\rho}^0\right\},\\
c_s^2&=&f_{c_s^2}\left\{s^0,\vect{\rho}^0\right\},\\
&\dots&
\end{eqnarray}
where the $\dots$ indicate higher-order derivatives as well. The functions $f_i$, where $i$ indicates the thermodynamic quantity that we are interested in, are dependent on which EoS we consider because the EoS dictates the mapping between all the thermodynamic quantities.

However, at this point, it is important to clarify that the natural free parameters of any given EoS model used in heavy-ion collisions are $\left\{T,\vect{\mu}\right\}$.
Thus, the EoS that comes from the Taylor series is given along a regular grid of values in $\left\{T,\vect{\mu}\right\}$, whereas the natural hydrodynamic variables are $\left\{s,\vect{\rho}\right\}$\footnote{Some hydrodynamic codes use $\varepsilon$ instead of $s$, but the problem still remains.}.
Thus, it is not just a problem of interpolating over a 4D grid to obtain the thermodynamic quantities; instead, one must make a choice between:
\begin{enumerate}
    \item Inverting the full 4D EoS into the natural hydrodynamic variables of $\left\{s,\vect{\rho}\right\}$ and then only interpolate directly over densities on-the-fly during simulations
    \item Using both a 4D root-finder and 4D interpolation over intensive variables on-the-fly within the hydrodynamic simulations
\end{enumerate}
Both of these methods have their advantages and disadvantages.
The first method is more direct but requires performing interpolation over an unstructured 4D grid.
One way to handle this is by constructing a Delaunay triangulation of the density coordinates in the EoS and then evaluating suitable interpolants defined on this triangulation.
This method is challenging to implement due to the complexity of constructing the triangulation and also locating the region in the the density space where a solution is likely to be found.
The second method proceeds by interpolating over a structured grid but relies on root-finding to convert the intensive variables to the densities, and thus tends to be the more robust method to changes in the EoS.
Both methods have comparable efficiency, yielding up to $\mathcal{O}(100)$ solutions per second.
If the purpose of the code is to vary the EoS often, it is significantly easier to include both the root-finding and interpolation within the code itself.
Furthermore, it is easier to build in the functionality of both root-finding and interpolation and then switch off the root-finding down the road if a pre-inverted EoS is used.
However, the inversion required by the second method is non-trivial: the mapping between $\rho_B$ and $\mu_B$ is highly non-linear and depends dramatically on the temperature $T$ (which is especially true in the presence of phase transitions).
This makes it difficult to obtain a `pre-inverted' EoS which also has a grid of constant step-size in densities.
In the version of \ccake{} presented here, we fix the root-finding approach as the default; subsequent code releases will include the option of Delaunay interpolation as well.

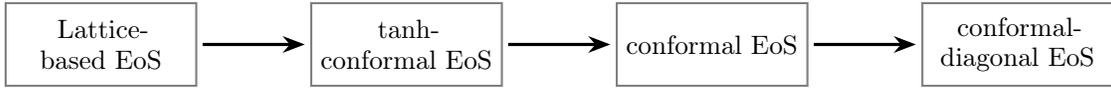
\begin{figure*}
    \centering
        \begin{tikzpicture}[auto,
                       > = Stealth,
           node distance = 1.5cm,
              box/.style = {draw=gray, thick,
                            minimum height=11mm, text width=23mm,
                            align=center},
       every edge/.style = {draw, ->, very thick, shorten >=2pt, shorten <=2pt},
every edge quotes/.style = {font=\footnotesize, align=center, inner sep=1pt}
                            ]
% from left to right
            \node (n1) [box]               {Lattice-based EoS};
            \node (n2) [box, right=of n1] {$\tanh$-conformal EoS};
            \node (n3) [box, right=of n2] {conformal EoS};
            \node (n4) [box, right=of n3] {conformal-diagonal EoS};
%Lines
            \draw (n1) edge (n2) (n2) edge (n3) (n3) edge (n4);
        \end{tikzpicture}
    \caption{The flowchart for iterating over equations of state in \code{rootfinder}. The default setting iterate over all the equations of state; however, the algorithm can iterate over only a subset based on the user's choice or purpose of the simulation. Other equations of state can also be added easily as well. 
}
    \label{fig:EoS_flowchart}
\end{figure*}

Our $f_i$ also depends on the underlying EoS used.
In Fig.\ \ref{fig:EoS_flowchart} we include a flow chart of our EoS module.  
We begin by assuming a lattice QCD EoS for all particles as the default.
If a fluid cell fails to yield a solution at any point, we then `fall back' to the remaining three EoSs, checking each systematically to see if it is possible to find our specific set of $\left\{s^0,\vect{\rho}^0\right\}$ within it. 
Occasionally, because of the limitations of the current lattice QCD EoS which is an expansion in terms of $\vect{\mu}/T$, certain regions of large $\vect{\mu}$ and small $T$ are especially susceptible to failure.  
We found that in the lattice QCD table there are certain regions of the QCD phase diagram that lead to either acausal $c_s^2>1$ or unstable $c_s^2<0$ points in the EoS. 
If we encounter such a point in the table then we automatically exclude that region from the interpolation function (to avoid compounding numerical error within neighboring points) and send any fluid cells that are consistent with this region of the phase diagram to a back-up EoS.

As indicated by the flowchart, we consider four different EoSs which provide systematically less robust approximations to the true QCD EoS and finite temperature and chemical potentials.
The four EoSs we use are: (i), the lattice QCD Taylor series EoS; (ii), an EoS designated ``$\tanh$-conformal'', for reasons we discuss below; (iii), a conformal EoS; and (iv), an EoS designated ``conformal-diagonal'', which we will also expound upon below.
The logic of this ordering is to begin with the optimal EoS, namely, lattice QCD; we therefore first attempt to find $\left\{s^0,\vect{\rho}^0\right\}$ in that framework.
If that is not possible we check the ``$\tanh$-conformal'' EoS which smoothly connects lattice QCD (at finite $T$ and $\vect{\mu}$) to a conformal EoS using a hyperbolic $\tanh$.
If this second EoS still does not yield a solution for the densities $\left\{s^0,\vect{\rho}^0\right\}$, we then check a conformal EoS for a solution.
Occasionally, however, even the conformal EoS does not yield an acceptable solution; in these cases, the rootfinder falls back to the fourth and final EoS, the ``conformal-diagonal'' EoS.
This final EoS is constructed in such a way as to guarantee that $\left\{s^0,\vect{\rho}^0\right\}$ possesses a solution. It is important to note that once a fluid cell has shifted to the right in our flow chart in Fig. \ref{fig:EoS_flowchart} then we never check an EoS to the left of it in subsequent time steps.
This implies that if a fluid cell is out-of-bounds for the lattice QCD EoS and moves to a $\tanh$-conformal EoS, in subsequent time steps we never check that fluid cell again within the lattice QCD EoS.
Instead we check first the $\tanh$-conformal EoS and if $\left\{s^0,\vect{\rho}^0\right\}$ is not found there then we next check the conformal EoS. Fluid cells are checked within each EoS and the final EoS, conformal-diagonal, always has a solution.

In the following, we discuss the four different EoS in greater detail and explain how we obtain the corresponding dimensionless pressure, $[p/T^4]\left(T,\vect{\mu}\right)$.  Once the pressure is obtained, it is straightforward to calculate all the other needed thermodynamic relations, which we have included in Appendix \ref{app:thermoRel} for completeness.

%
%%%%%%%%%%%%%%%%%%%%%%%%%%%%%%%%%%%%%%%%%%%%%%%%%
%%%%%%%%%%%%%    TAYLOR SERIES EoS   %%%%%%%%%%%%
%%%%%%%%%%%%%%%%%%%%%%%%%%%%%%%%%%%%%%%%%%%%%%%%%
\subsubsection{Taylor series EoS}
\label{sec:TaylorSeriesEoS}
%==============================================================================

Because of the fermionic sign problem \cite{Troyer:2004ge,Ratti:2018ksb} one cannot determine the finite density EoS directly from lattice QCD.  Instead one must obtain susceptibilities (derivatives of the pressure):
\begin{equation}
\chi_{ijk}^{BSQ}=\left.\frac{\partial^{i+j+k}(p/T^4)}{\partial(\mu_B/T)^i\partial(\mu_S/T)^j\partial(\mu_Q/T)^k}\right|_{\mu_B,\mu_S,
\mu_Q=0}
\end{equation}
that can then be used within a Taylor series to reconstruct the finite density EoS
\begin{equation}
\frac{p(T,\vect{\mu})}{T^4}=\sum_{i,j,k}\frac{1}{i!j!k!}\chi_{ijk}^{BSQ}
\left(\frac{\mu_B}{T}\right)^i\left(\frac{\mu_S}{T}\right)^j\left(\frac{\mu_Q}{T}\right)^k.
\end{equation}
where we expand the pressure in respect to $\mu_X/T$ where $X=B,S,Q$.

The Taylor series EoS is not valid throughout the entire QCD phase diagram, and so we must identify the range in $T$ and $\vect\mu$ where it is taken to be reliable.  For this, we require $T \in [T_\mathrm{min},T_\mathrm{max}]$ and $\mu \in [\mu_\mathrm{min},\mu_\mathrm{max}]$, where
$T_\mathrm{min} = 0$ MeV, $T_\mathrm{max} = 800$ MeV, and $\mu_\mathrm{max} = -\mu_\mathrm{min} = 450$ MeV.   The tabulated form of the Taylor series EoS used in the code has step-sizes of $\Delta T = 5$ MeV and $\Delta \mu = 50$ MeV for each conserved charge. 

We note that our $T_\mathrm{min}$ is lower than the choice of $T_\mathrm{min} = 30$ MeV made in \cite{Noronha-Hostler:2019ayj}.
It is important to note here that SPH requires a thermodynamically consistent EoS down to $T=0$ in order to preserve energy conservation.  
Previous studies have resolved this issue by simply attaching a pion gas EoS at nearly vanishing temperatures (see e.g. \cite{Alba:2017hhe}). 
However, the pion gas EoS only works at vanishing $\mu_B$ and $\mu_S$ as an adequate solution.  Thus, instead we use a different approach to develop a well-behaved EoS at low-$T$ that can be found in Appendix \ref{sec:lowT} where we smoothly extrapolated the parametrization of \cite{Noronha-Hostler:2019ayj} to zero temperature.
It is important to point out that the details of this low-$T$ EoS are irrelevant to the simulations themselves since these temperatures are well below the freeze-out temperature. However, it is crucial for the stability of the code to preserve energy conservation.

%%%%%%%%%%%%%%%%%%%%%%%%%%%%%%%%%%%%%%%%%%%%%%%%%
%%%%%%%%%%%%    TANH-CONFORMAL EoS   %%%%%%%%%%%%
%%%%%%%%%%%%%%%%%%%%%%%%%%%%%%%%%%%%%%%%%%%%%%%%%
\subsubsection{\texorpdfstring{$\tanh$}{tanh}-conformal EoS}
\label{sec:tanhConformalEoS}
%==============================================================================
The ``$\tanh$-conformal'' EoS is similar to a conformal EoS but includes a modulating hyperbolic tangent factor which depends on the temperature $T$ and more closely approximates the lattice-based EoS than a simple conformal EoS.
Although it ensures that an approximate conformal limit is still retained at high temperatures, the modulating factor obviously removes the conformality, meaning that the ``$\tanh$-conformal'' EoS is not a true conformal EoS.
The precise functional form is given by
    \begin{align}
     p_\mathrm{tc}(T,\vect{\mu})
            &= \frac{1}{2} A_0 T_0^4 \l(1 + \tanh\left(\frac{T-T_c}{T_s}\right) \r) \nonumber\\
           & \times \left[ \left( \frac{T}{T_0} \right)^2
             +\sum_{X=B,S,Q} \left( \frac{\mu_X}{\mu_{X,0}} \right)^2 \right]^2
    \end{align}
where $A_0$ and $(T_0,\vect{\mu}_{0})$ are free parameters fixed by the following constraints:
\begin{align}
    A_0 &\equiv p_{T,0}/T_{\mathrm{scale}}^4 \nonumber \\
    T_0 &\equiv 1 \text{ fm}^{-1} \label{tanhConformalConstraints} \\
    \mu_{X,0} &\equiv \frac{A_0^{1/4} T_0 \mu_{X,{\max}}}{\sqrt{\sqrt{p_{X,\max}} - \sqrt{p_{T,0}}}}, \nonumber
\end{align}
where $X = B,S,Q$ and
\begin{align*}
    p_{T,0} &\equiv p_\mathrm{table}(T_{\mathrm{scale}}, \vect{0}) \\
    p_{B,\max} &\equiv p_\mathrm{table}(T_{\mathrm{scale}}, \mu_{B,{\max}}, 0, 0) \\
    p_{S,\max} &\equiv p_\mathrm{table}(T_{\mathrm{scale}}, 0, \mu_{S,{\max}}, 0) \label{EoS_constraints_definitions} \\
    p_{Q,\max} &\equiv p_\mathrm{table}(T_{\mathrm{scale}}, 0, 0, \mu_{Q,{\max}})
\end{align*}
Here, $T_{\mathrm{scale}} = 1.1 \, T_{\mathrm{FO}}$, and $T_c = 220$ MeV and $T_s = 120$ MeV are additional free parameters which control the location and strength of the modulating factor.  In this study the values of $T_c$ and $T_s$ have been fixed by hand.
The effects of varying them will explored in future work.

%
%%%%%%%%%%%%%%%%%%%%%%%%%%%%%%%%%%%%%%%%%%%%%%%%%
%%%%%%%%%%%%%%%    CONFORMAL EoS   %%%%%%%%%%%%%%
%%%%%%%%%%%%%%%%%%%%%%%%%%%%%%%%%%%%%%%%%%%%%%%%%
\subsubsection{conformal EoS}
\label{sec:ConformalEoS}
%==============================================================================
The next EoS in Fig.\ \ref{fig:EoS_flowchart} is a truly conformal EoS.  The pressure is defined by
    \begin{equation}
        p_\mathrm{c}(T,\vect{\mu})
        = A_0 T_0^4 \left[ \left( \frac{T}{T_0} \right)^2
             +\sum_{X=B,S,Q} \left( \frac{\mu_X}{\mu_{X,0}} \right)^2 \right]^2,
    \end{equation}
When the conformal EoS is being used as a fallback EoS, the free parameters $A_0$ and $(T_0,\vect\mu_{0})$ are fixed by Eqs.\ \eqref{tanhConformalConstraints}.
When the conformal EoS is used as the default EoS (e.g., for the Gubser benchmarks), we set $A_0 \equiv \frac{\pi^2}{90}(2(N_c^2-1)+\frac{7}{2} N_c N_f)$, $T_0 = \vect\mu_{0} = 1 \text{ fm}^{-1}$, and $N_c = 3$ and $N_f = 2.5$.

Note that the conformal EoS, like the the tanh-conformal EoS \eqref{tanhConformalConstraints}, may have cross-terms between the set of variables $\l\{T,\vect{\mu}\r\}$ which are quadratic in any two of them (e.g., terms proportional to $T^2 \mu_Q^2$).
%

%%%%%%%%%%%%%%%%%%%%%%%%%%%%%%%%%%%%%%%%%%%%%%%%%
%%%%%%%%%%    CONFORMAL-DIAGONAL EoS   %%%%%%%%%%
%%%%%%%%%%%%%%%%%%%%%%%%%%%%%%%%%%%%%%%%%%%%%%%%%
\subsubsection{conformal-diagonal EoS}
\label{sec:ConformalDiagonalEoS}
%==============================================================================
The final EoS in Fig.\ \ref{fig:EoS_flowchart} is the conformal-diagonal EoS.  This EoS implements a simpler version of a conformal EoS which eliminates cross-terms between the temperature and chemical potentials in the system and permits a simple, analytical solution for a given combination of energy/entropy and charge densities.
The pressure is defined by
    \begin{equation}\label{eqn:pressureConDiag}
        p_\mathrm{cd}(T, \vect{\mu} )
        = A_0 T_0^4 \left[ \left( \frac{T}{T_0} \right)^4
             +\sum_{X=B,S,Q} \left( \frac{\mu_X}{\mu_{X,0}} \right)^4\right] ,
    \end{equation}
where
\begin{align}
    A_0 &\equiv p_{T,0}/T_{\mathrm{scale}}^4 \nonumber \\
    T_0 &\equiv 1 \text{ fm}^{-1} \\
    \mu_{X,0} &\equiv T_0 \mu_{X,{\max}}\l(\frac{A_0}{p_{X,\max} - p_{T,0}}\r)^{1/4} \nonumber
\end{align}
and we adopt the same notation here as that used in Sec.\ \ref{sec:tanhConformalEoS} above.

In Appendix \ref{sec:emin}, we show that a necessary and sufficient condition for the existence of a solution is the following criterion which the energy and charge densities must respect:
    \begin{align}
      \varepsilon &\geq \varepsilon_\mathrm{min}\left( \vect{\rho} \right)
      \equiv \frac{3}{4\cdot 2^{2/3} (A_0 T_0^4)^{1/3}} \sum_{X=B,S,Q}\left( \mu_{X,0} \left| \rho_X \right| \right)^{4/3}.
      \label{emin_definition}
    \end{align}
The existence of an exact, closed-form solution for the phase diagram coordinates which provide a given combination of input densities also allows this class to be used to make an \textit{educated guess} for the seed value which can be supplied to the RootFinder.

\begin{figure}
    \centering
    \includegraphics[keepaspectratio, width=0.98\linewidth]{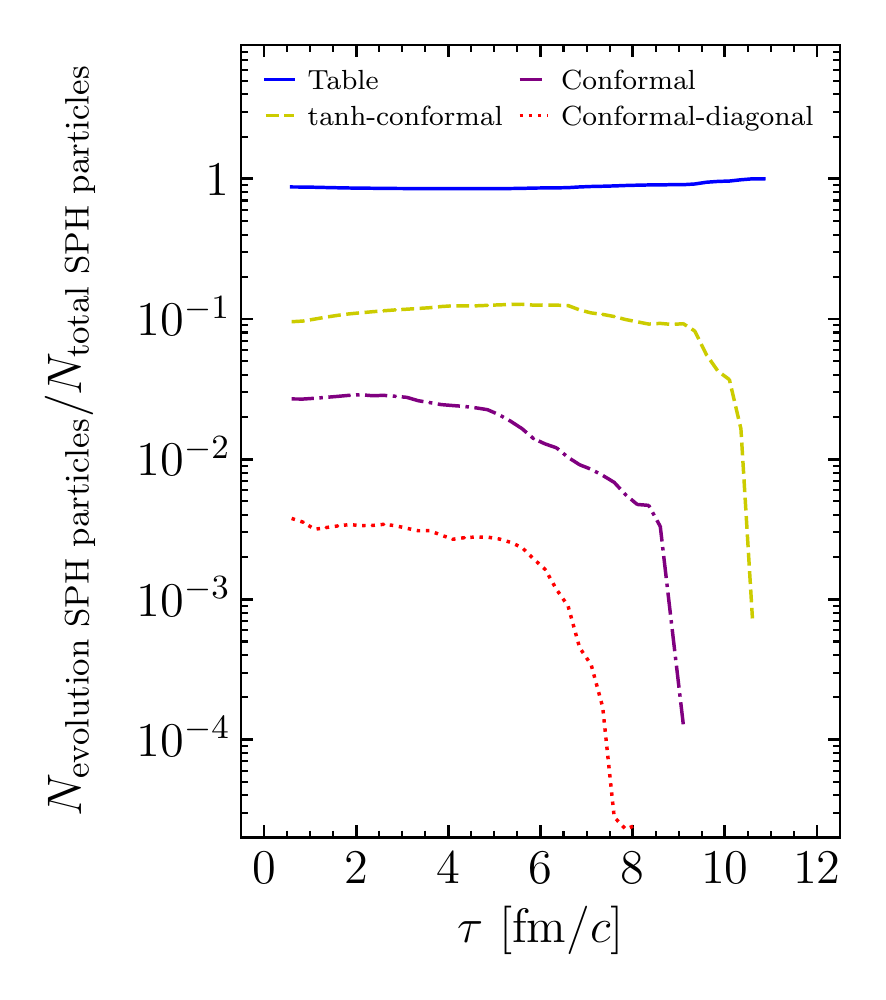}
    \caption{
    (color online) Fraction of SPH fluid cells using a given EoS (Table, tanh-conformal, Conformal, Conformal-diagonal) as functions of proper time $\tau$. Here only fluid cells that that have not yet frozen out, i.e., $\varepsilon>\varepsilon_\mathrm{FO}$, are considered.  Note that the relative ordering of the different fractions can change significantly from one event to the next, especially at late times when only a small number of particles remain above freeze out.
    }
    \label{fig:EoSfraction}
\end{figure}
%

%%%%%%%%%%%%%%%%%%%%%%%%%%%%%%%%%%%%%%%%%%%%%%%%%
%%%%%%%%%%%%%%    ADDITIONAL EoS   %%%%%%%%%%%%%%
%%%%%%%%%%%%%%%%%%%%%%%%%%%%%%%%%%%%%%%%%%%%%%%%%
\subsubsection{Influence of the additional (non-lattice QCD) EoS}
%%%%%%%%%%%%%%%%%%%%%%%%%%%%%%%%%%%%%%%%%%%%%%%%%
Due to the limitations of lattice QCD at finite densities, we have a number of out-of-bounds fluid cells in our simulations that end up using our alternative EoS shown in Fig.\ \ref{fig:EoS_flowchart}.  Here we quantify both the location of these fluids cells as well as the fraction of fluid cells that utilize each respective EoS. 

In Fig.\ \ref{fig:EoSexpand} an example event from Pb\texttt{+}Pb $\sqrt{5.02}$ TeV in central collisions of $0$--$5\%$ at three different time steps: $\tau=\tau_0=0.60$ fm$/c$ (left), $\tau=4.60$ fm$/c$ (middle), and $\tau=10.60$ fm$/c$ (right).
The lattice QCD fluids cells are shown in blue and are clearly the dominant EoS.
Next we have $\tanh$-conformal in yellow, followed by a handful of fluid cells from the conformal EoS in purple, and an extremely small number from the diagonal conformal in red. Since \ccake{} is a Lagrangian code, the individual fluid cells are free to move about.
Thus, we can see that while they are on a uniform grid at $\tau_0$ (although there is a specific shape to the event because fluids cells with very low energy densities/zero densities are not included in simulations), over time they spread apart and begin to exhibit structure.

While the out-of-bounds fluid cells are scattered across the event, more are found at the edges than in the center.
We can also see that red/purple cells are rarely found in isolation but rather are surrounded by the $\tanh$-conformal EoS.
This implies that these fluid cells are likely close to the lattice QCD EoS but are slightly off.
Further extensions to the lattice QCD EoS would likely ensure that these fluid cells are no longer out-of-bounds.
We do not see a significant increase in out-of-bounds fluid cells over time, which indicates that testing the EoS at the initial state is likely sufficient to quantify the effect of out-of-bounds fluid cells.

In Fig.\ \ref{fig:EoSfraction} we plot the fraction of fluid cells in a given event that make up a given EoS at a certain point in time, $\tau$.
In this figure we consider only fluids cells that are above freeze-out such that over time the total number of fluid cells considered decreases.
The fluid cells labeled as Table are the lattice QCD EoS and they are the dominant EoS throughout the entire evolution.
Next are the $\tanh$-conformal EoS that has over an order of magnitude fewer fluid cells, followed by the Conformal, and Conformal-diagonal.  Our results are consistent with our findings in Fig.\ \ref{fig:EoSexpand}.
One new thing that we learn from Fig.\ \ref{fig:EoSfraction} is that the out-of-bounds fluid cells clearly must be near freeze-out because of how quickly the fraction of those fluid cells drop over time.
This effect is most obvious for the Conformal-diagonal EoS where the fraction of its fluid cells drops by orders of magnitude over time and eventually ends approximately 2 fm/$c$ before the hydrodynamic evolution is entirely frozen out.
Similarly, the Conformal EoS fluid cells and the $\tanh$-conformal EoS fluid cells also drop over time (although not quite as quickly).
Thus, this implies that across the entire evolution fluid cells that are not yet frozen out (i.e., the high temperature fluid cells) are predominately still coming from the lattice QCD table. 
Note that this need not be true in every event at sufficiently late times: since only particles above freeze out are shown in Fig.~\ref{fig:EoSfraction}, some events may exhibit a different relative ordering once the vast majority of particles has frozen out and only a few particles remain above freeze out.
In this case, however, the ordering is no longer statistically significant and it remains true that the Table EoS overwhelmingly dominates the system's dynamical evolution.

These findings are consistent with what one would expect from the error of a Taylor series expanded in $\vect{\mu}/T$.  Essentially, our out-of-bounds fluid cells are more likely to occur at lower temperatures and larger $\vect{\mu}$, which is where our series expansion is most likely to break down. 
\begin{figure}
    \centering
    \includegraphics[width=\linewidth]{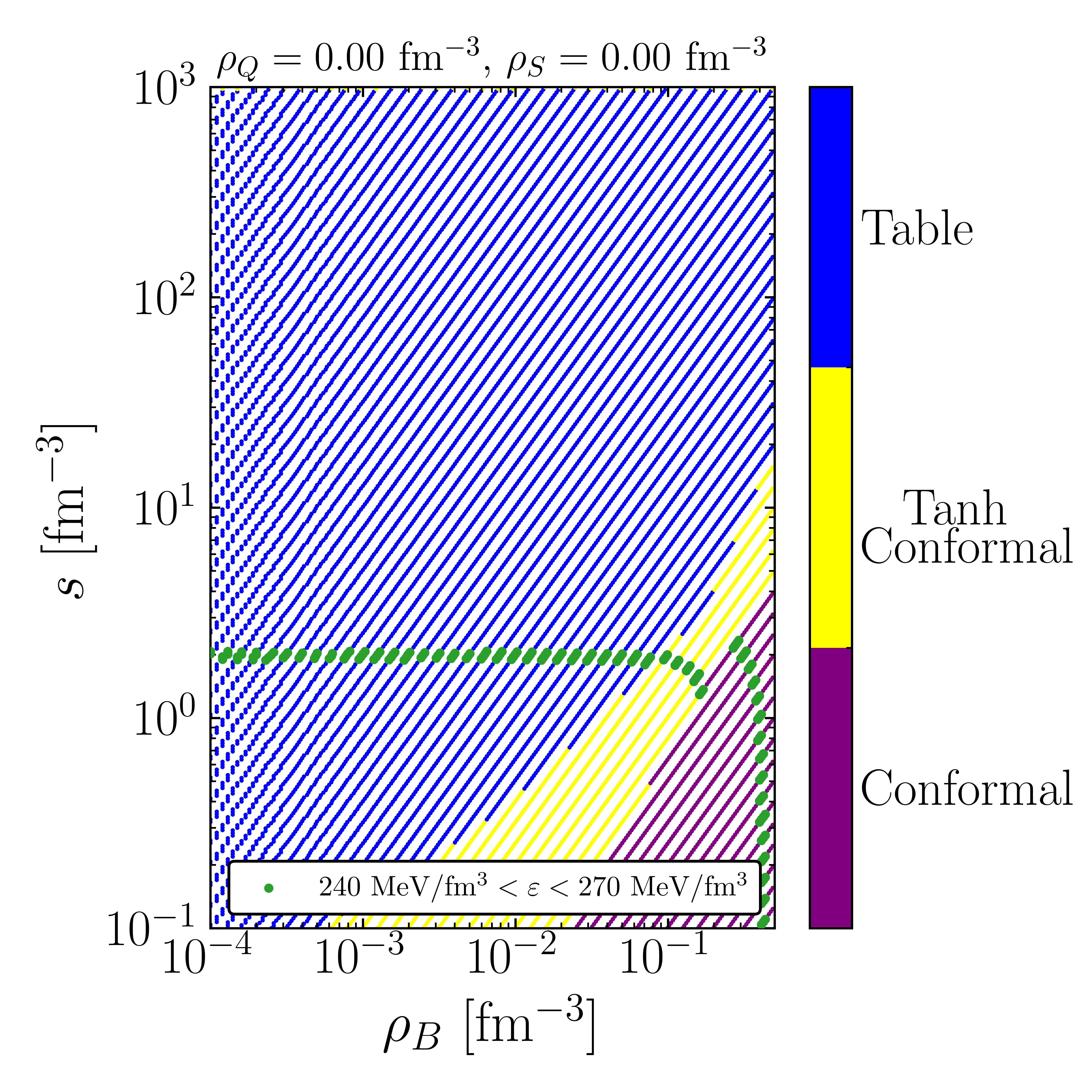}\\
    \includegraphics[width=\linewidth]{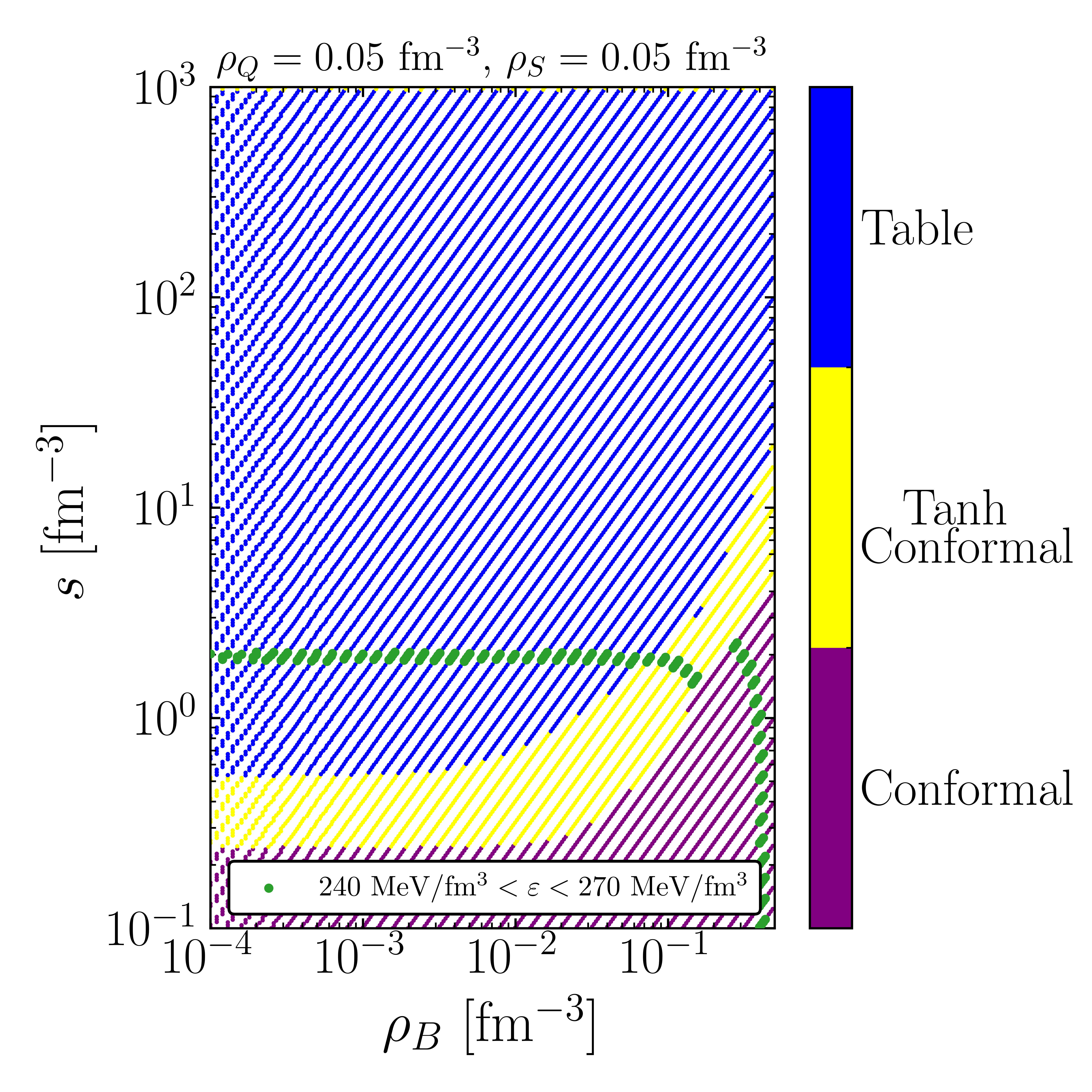}
    \caption{(color online) Plot of the EoS used for particular entropy vs baryon densities for zero electric charge and strangeness (top) and finite strangeness and electric charge densities (bottom).  The green line demonstrates our freeze-out criteria.}
    \label{fig:Isentropes_EoS}
\end{figure}
Since the expansion for the lattice QCD EoS is in terms of $\mu/T$ and there is a non-linear mapping between densities and $\vect{\mu}$, then it is also of interest to understand how the use of these different EoS map into densities.
In Fig.\ \ref{fig:Isentropes_EoS} we plot the type of EoS used for individual isentropes both for the scenario where there is only one conserved charge (just baryon density, shown in the top panel) as well as the scenario where all 3 conserved charges are finite (bottom panel).
We also compare these different types of EoS along isentropes to our freeze-out criteria, which is shown in green.
We find then in Fig. \ref{fig:Isentropes_EoS} that the vast majority of fluid cells, before they freeze-out, use the lattice QCD EoS.
However, there is some small subset that uses the tanh-conformal or the conformal EoS.
After freeze-out there are a significant number of fluid cells at low entropies that use either the tanh-conformal or conformal EoS and almost none that use the conformal-diagonal EoS.  

One further consideration is that fluid cells carrying multiple conserve charges more quickly have troubles matching to the lattice QCD EoS, especially for low entropy densities. 
Note that here we show results only for relatively small $\rho_B$ (saturation is approximately $\rho_{sat}\sim 0.16$ fm$^{-3}$ and falls at the right edge of the figure). 
Later on we will show that the remaining baryon densities are relatively small at freeze-out such that Fig.\ \ref{fig:Isentropes_EoS} is a reasonable comparison for BSQ fluctuations at the LHC.
However, for future studies at the beam energy scan where large $\rho_B$'s are reached at freeze-out it may be better to explore new methods for the low entropy regime of the EoS to ensure that most fluid cells have matches within the correct EoS. 

%%%%%%%%%%%%%%%%%%%%%%%%%%%%%%%%%%%%%%%%%%%%%%%%%
%%%%%%%%%%%%%    TIME DERIVATIVES   %%%%%%%%%%%%%
%%%%%%%%%%%%%%%%%%%%%%%%%%%%%%%%%%%%%%%%%%%%%%%%%
\subsection{Time derivatives of thermodynamic variables}
\label{sec:TimeDerivativesOfThermodynamicVariables}
%==============================================================================

In Sec.\ \ref{sec:EoS}, we have discussed four separate challenges to implementing the thermodynamics of QCD matter in hydrodynamic simulations.
We now discuss the fourth of these challenges -- namely, the evaluation of general thermodynamic derivatives in terms of derivatives with respect to the pressure.
In the absence of conserved charges, the EoS can generally be parametrized in terms of a single quantity, such as the temperature $T$.
For multiple conserved charges, however, one must supplement the temperature with the corresponding chemical potentials.
For BSQ conservation, one thus has $\left\{T,\mu_B,\mu_S,\mu_Q\right\}$, resulting in a 4D EoS.

For most EoS models, it is most natural to calculate thermodynamic observables along grids of $\left\{T,\mu_B,\mu_S,\mu_Q\right\}$.  This makes it significantly easier to take derivatives along trajectories in $\left\{T,\mu_B,\mu_S,\mu_Q\right\}$-space, rather than along lines of constant entropy density, $s$, for instance. Relativistic hydrodynamics, however, formulates the system evolution in terms of \textit{densities}, such as the energy density $\varepsilon$ or entropy density $s$, along with the 3 BSQ densities: $\left\{\rho_B,\rho_S,\rho_Q\right\}$. 
In \ccake{}, the latter set of densities is used in the evolution.  One must therefore have a way to translate to the more natural intensive variables of the EoS $\left\{T,\mu_B,\mu_S,\mu_Q\right\}$ into the more natural extensive variables required by hydrodynamics $\left\{s,\rho_B,\rho_S,\rho_Q\right\}$.

This translation is accomplished by first separating time derivatives of the extensive densities $\left\{s,\rho_B,\rho_S,\rho_Q\right\}$ from thermodynamic derivatives, as in Eq.~\eqref{eq:dwdt}, and then formulating the latter entirely in terms of the intensive variables $\left\{T,\mu_B,\mu_S,\mu_Q\right\}$.  In the remainder of this subsection, we outline this procedure as it is implemented in \ccake{}. 

In the momentum conservation equation Eq.~\eqref{eq:momentum_conservation}, one deals with the time derivative of the enthalpy density ($w=\varepsilon+p$). The enthalpy density is related to the other thermodynamic quantities by the Gibbs relation Eq.~\eqref{eq:gibbs}.
Applying this relation to the time derivative of $w$ we find:
\begin{align} \label{eq:ep1}
	\frac{d}{d\tau} \Big( \varepsilon + p \Big) &=
	\left( \frac{ds}{d\tau} \right) \, T +
	s \, \left( \frac{dT}{d\tau} \right) \nonumber\\
	&+
	 \left( \frac{d\vect{\rho}}{d\tau} \right)  \cdot\vect{\mu} +
	\vect{\rho} \cdot \left( \frac{d\vect{\mu}}{d\tau} \right).
\end{align}
Here all thermodynamic quantities are effectively functions of a single variable $s = s(\tau), \rho_X = \rho_X (\tau)$, etc., where $\tau$ parametrizes a particular trajectory through the phase diagram by means of a complete set of thermodynamic variables, such as $\{ s(\tau), \vect{\rho}(\tau) \}$ or $\{ T(\tau), \vect{\mu} (\tau) \}$.

In order to translate between intensive and extensive quantities, we treat $\{ T, \vect{\mu} \}$ as dependent functions of our chosen independent variables $\{ s, \vect{\rho} \}$.  We have
\begin{align}
	T &= T(\tau) = T\big( s(\tau) , \vect{\rho} (\tau) \big),
	\notag \\ \notag \\
	\frac{dT}{d\tau} &=
	\left(\frac{\partial T}{\partial s}\right)_{\vect{\rho}} \, \frac{d s}{d \tau}
	+ \sum_X
	\left(\frac{\partial T}{\partial \rho_X}\right)_{s \, , \, \rho_X \neq \rho_Y} \, \frac{d \rho_X}{d \tau}
\end{align}
and similarly for $\frac{d\mu_X}{d\tau}$.    Thus $\left(\frac{\partial T}{\partial s}\right)_{\vect{\rho}}$ is a derivative with all of $\{ \rho_B , \rho_S , \rho_Q \}$ held constant and $\left(\frac{\partial T}{\partial \rho_X}\right)_{s \, , \, \rho_X \neq \rho_Y}$ is a derivative with respect to only the one density $\rho_X$, with the others being held fixed.

Inserting this back into \eqref{eq:ep1} allows us to eliminate $\frac{dT}{d\tau}$ and $\frac{d\mu_X}{d\tau}$, yielding:
\begin{widetext}
\begin{align}  \label{eq:ep2}
	\frac{dw}{d\tau} &=
	\left[
	T +
	s \left(\frac{\partial s}{\partial T}\right)_{\vect{\rho}}^{-1}
	+ \sum_X \rho_X \left(\frac{\partial s}{\partial \mu_X}\right)_{\vect{\rho}}^{-1}
	\right] \frac{ds}{d\tau}+
	\sum_Y
	\left[
	\mu_Y
	+
	s \left(\frac{\partial \rho_Y}{\partial T}\right)_{s \, , \, \rho_X \neq \rho_Y}^{-1}
	+ \sum_X \rho_X \left(\frac{\partial \rho_Y}{\partial \mu_X}\right)_{s \, , \, \rho_X \neq \rho_Y}^{-1}
	\right] \frac{d\rho_Y}{d\tau}	,
\end{align}
\end{widetext}
where we have replaced the first-order derivatives with their inverses, which presumably works as long as the functions are nonzero.  The time evolution of the enthalpy density $w$ therefore involves two types of derivatives: (i) \textit{time} derivatives of extensive quantities $\{ s(\tau), \vect{\rho}(\tau) \}$ which are propagated dynamically by the hydrodynamic equations of motions; and (ii), thermodynamic derivatives with respect to intensive variables $\{ T, \vect{\mu} \}$ (appearing in the square brackets in Eq.~\eqref{eq:ep2}) which depend only on the EoS and therefore need only be evaluated once in a given simulation.  The evaluation of dynamical derivatives has already been outlined in Sec.~\ref{sec:SPHformalism}; here, we focus on the thermodynamic derivatives.

The derivatives in Eq.\ \eqref{eq:ep2} can be divided into the following four subcategories:
\[\left(\dfrac{\partial s}{\partial T}\right)_{\vect{\rho}}, \left(\dfrac{\partial s}{\partial \mu_X}\right)_{\vect{\rho}}, \left(\dfrac{\partial \rho_X}{\partial T}\right)_{s,\, \rho_X \neq \rho_Y}, \left(\dfrac{\partial \rho_Y}{\partial \mu_X}\right)_{s,\, \rho_X \neq \rho_Y}.
\]
It is convenient to group these terms in this way because then the derivatives have the same form for different charges $X,Y \in (B,S,Q)$.

In the remainder of this subsection, we illustrate how to evaluate the first of these four types of derivative in the simple case of a single conserved charge.  Extending this treatment to all four subcategories in the presence of three conserved charges ($B$, $S$, $Q$) is straightforward but tedious; the details are deferred to Appendix \ref{app:thermodynamic_derivatives}.

To evaluate $\left(\partial s / \partial T \right)_{\rho_B}$, we begin by treating $\{T,\mu_B\}$ as independent variables.  Using the notation introduced in Appendix \ref{app:thermoRel}, these equations take the differential form
\begin{align}\label{eq:ds}
    ds &= \left(\dfrac{\partial s}{\partial T} \right)_{\mu_B}dT + \left(\dfrac{\partial s}{\partial \mu_B} \right)_{T}d\mu_B \notag\\
    &= (\partial^2_T p)\, dT + (\partial^2_{\mu_B,T}p)\, d\mu_B \notag\\
    &= \chi_{TT}\, dT + \chi_{TB}\, d\mu_B,
\end{align}
where $\chi_{ab} = \partial^2 p/\partial a \partial b$ are second-order susceptibilities of the pressure,
and similarly,
\begin{align}\label{e:dnb}
    d \rho_B &= \chi_{TB}\, dT + \chi_{BB}\, d\mu_B.
\end{align}
We impose baryon number conservation by setting $d\rho_B = 0$.  Using Eq. \eqref{e:dnb}, this implies
\begin{align}
    \label{eq:dtdmuB}
    \left(\dfrac{dT}{d\mu_B}\right)_{\rho_B} &= - \dfrac{\chi_{BB}}{\chi_{TB}}.
\end{align}
Finally, we employ Eqs.~\eqref{eq:ds} and \eqref{eq:dtdmuB} to evaluate the derivative we are interested in:
\begin{align}
    \left(\dfrac{\partial s}{\partial T} \right)_{\rho_B
    } & = \dfrac{\chi_{TT}\, dT + \chi_{TB}\, d\mu_B}{dT}\Bigg|_{\rho_B = \text{const.}} \notag\\
    %& = \dfrac{\chi_{TT} (dT/d\mu_B)_{\rho_B} + \chi_{TB} }{(dT/d\mu_B)_{\rho_B}}, \notag\\
    & = \chi_{TT} - \dfrac{\chi_{TB}^2 }{\chi_{BB}} .
\end{align}
As the procedure above illustrates, the thermodynamic derivatives needed for the time derivative of the enthalpy density $w$ (see Eq.~\eqref{eq:ep2}) can be evaluated entirely in terms of second-order derivatives of the pressure $p$ with respect to the temperature and chemical potentials.  This provides us with the connection we need between the system's thermodynamic properties and its hydrodynamic evolution, thereby allowing us to close the equations the motion.

%%%%%%%%%%%%%%%%%%%%%%%%%%%%%%%%%%%%%%%%%%%%%%%%%
%%%%%%%%%%%    DYNAMICAL EVOLUTION   %%%%%%%%%%%%
%%%%%%%%%%%%%%%%%%%%%%%%%%%%%%%%%%%%%%%%%%%%%%%%%
\subsection{The dynamical evolution and numerical tests}
\label{sec:BenchmarkTest}
%===========================================================================%
The previous version of the code (i.e., \vUSPhydro{}) has been benchmarked against the Gubser solution, TECHQM, energy conservation, convergence tests of the smoothing scale $h$ and grid size, and efficiency checks.
However, with the new version of \ccake{}, significant changes have been made such that one must rerun these benchmark checks, specifically with conserved charges in mind.
Hence, the Gubser check was performed both for the case of shear viscosity as well as for an ideal fluid with three conserved charges.
In Appendix \ref{app:Benchmark_checks} we have also studied the energy loss, efficiency, and convergence checks relate to the performance of \ccake{}.
Additionally, we have developed a new method in Appendix \ref{app:SPHbuffer} that includes ``buffer'' particles to ensure numerical stability in certain extreme cases (although the default within the code is to turn this option off).

%%%%%%%%%%%%%%%%%%%%%%%%%%%%%%%%%%%%%%%%%%%%%%%%%
%%%%%%%%%%%%%%%    GUBSER CHECK   %%%%%%%%%%%%%%%
%%%%%%%%%%%%%%%%%%%%%%%%%%%%%%%%%%%%%%%%%%%%%%%%%
\subsubsection{Gubser check}
\label{sec:GubserCheck}

Analytic and semi-analytic solutions to relativistic hydrodynamic equations of motion are only available when certain symmetry constraints are met and they provide an invaluable tool for testing and quantifying the accuracy of numerical implementations. For example, the well-known Hwa-Bjorken solution relies on translation and rotational invariance in the transverse plane and boost-invariance along the beamline \cite{Bjorken:1982qr}; despite the usefulness of this solution, it neglects the description of transverse expansion. A non-trivial radial flow dependence can be obtained for conformal fluids by demanding azimuthal symmetry around the transverse plane instead of translation invariance. This solution is known as Gubser flow and was originally obtained by Gubser \cite{Gubser:2010ze}, further developed by Gubser and Yarom \cite{Gubser:2010ui} for ideal systems, expanded to shear viscous systems in \cite{Marrochio:2013wla}, and then extended to one conserved charge (baryon density) in \cite{Denicol:2018wdp} but for ideal equations of motion. 

The Gubser flow solution assumes a system that possesses boost-invariant longitudinal evolution and a non-trivial transverse flow; additionally, it assumes a conformal equation of state (see Sec.~\ref{sec:ConformalEoS}). The conformal symmetry is a crucial assumption and restricts the types of fluids that can obey the solution. For instance, only shear viscosity can be considered and the shear viscosity to entropy density ratio must be constant. In practice, this solution results in a rigorous test for nuclear collision simulations.

We have checked that \ccake{} reproduces the Gubser solution in two different scenarios: (i) evolution with finite conserved charge densities and zero shear viscosity; and (ii), evolution with finite shear viscosity and no conserved charge densities.

For the {\bf ideal case}, here we have derived the Gubser test for multiple conserved charges (BSQ), extending the work done in \cite{Denicol:2018wdp}. We found the resulting analytic expressions for the energy density
\begin{equation}
     \varepsilon(r,\tau) =
        \frac{\varepsilon_0}{\tau^4}
        \frac{(2 \kappa \tau)^{8/3}}
            {\l[ 1 + 2 \kappa^2 (\tau^2 + r^2) + \kappa^4 (\tau^2 - r^2)^2 \r]^{4/3}},
\end{equation}
flow velocity,
\begin{equation}\label{eq:GubserRadialFlow}
     u^r(r, \tau) =
        \frac{2 \kappa^2 r \tau }
            { \sqrt{ 1 + 2 \kappa^2 (\tau^2 + r^2) + \kappa^4 (\tau^2 - r^2)^2 }},
\end{equation}
and charge densities,
\begin{equation}
    \rho_X(r, \tau) =
        \frac{\rho_{X,0}}{\tau^3}
        \frac{4 \kappa^2 \tau^2}
            {1 + 2 \kappa^2 (\tau^2 + r^2) + \kappa^4 (\tau^2 - r^2)^2},
    \label{eq:GubserConservedChargeDensities}
\end{equation}
where $X = B,S,Q$. In the above equations, $r=\sqrt{x^2+y^2}$ is the transverse radius, $\kappa$ is a free parameter in the system that is inversely related to the system size (i.e., a small $\kappa$ implies a larger system) where we take $\kappa=1$ fm$^{-1}$ to be consistent with previous works; the initial conditions at $r=0$ fm for the initial energy density $\varepsilon_0$ and initial charge densities $\rho_{X,0}$ are free parameters chosen as $\varepsilon_0=1$ fm$^{-4}$, $\rho_{B,0}=\rho_{S,0}=\rho_{Q,0}=0.5$ fm$^{-3}$.

\begin{figure}
    \centering
    \includegraphics[keepaspectratio, width=0.92\linewidth]{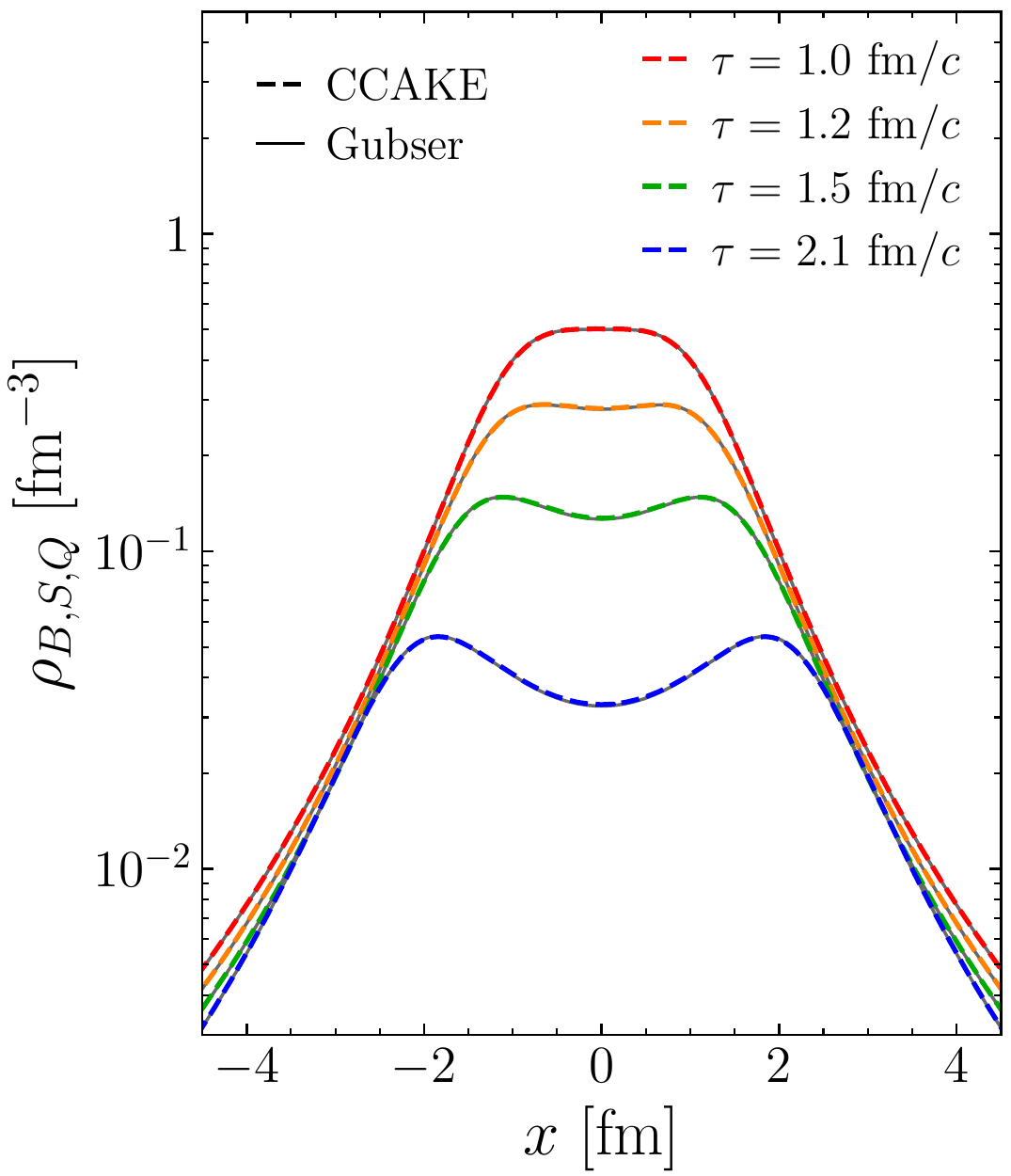}
    \caption{(color online) Comparison between the analytic results from Eq.~\eqref{eq:GubserConservedChargeDensities} (solid) and \ccake{} (dashed) at times $\tau=1.0$, $1.2$, $1.5$, and $2.1$ fm/$c$ along the $y=0$ fm axis.
    }
    \label{fig:GubserConservedChargeDensities}
\end{figure}

Figure\ \ref{fig:GubserConservedChargeDensities} shows the comparison between the Gubser analytic and the \ccake{} numerical solutions for the spatial profile of multiple charge densities along the $y=0$ fm axis at different times of the hydrodynamic evolution. For all conserved charge densities, the numerical solution is in full agreement with the analytic solution.

\begin{figure*}
    \centering
    \includegraphics[keepaspectratio, width=0.8\textwidth]{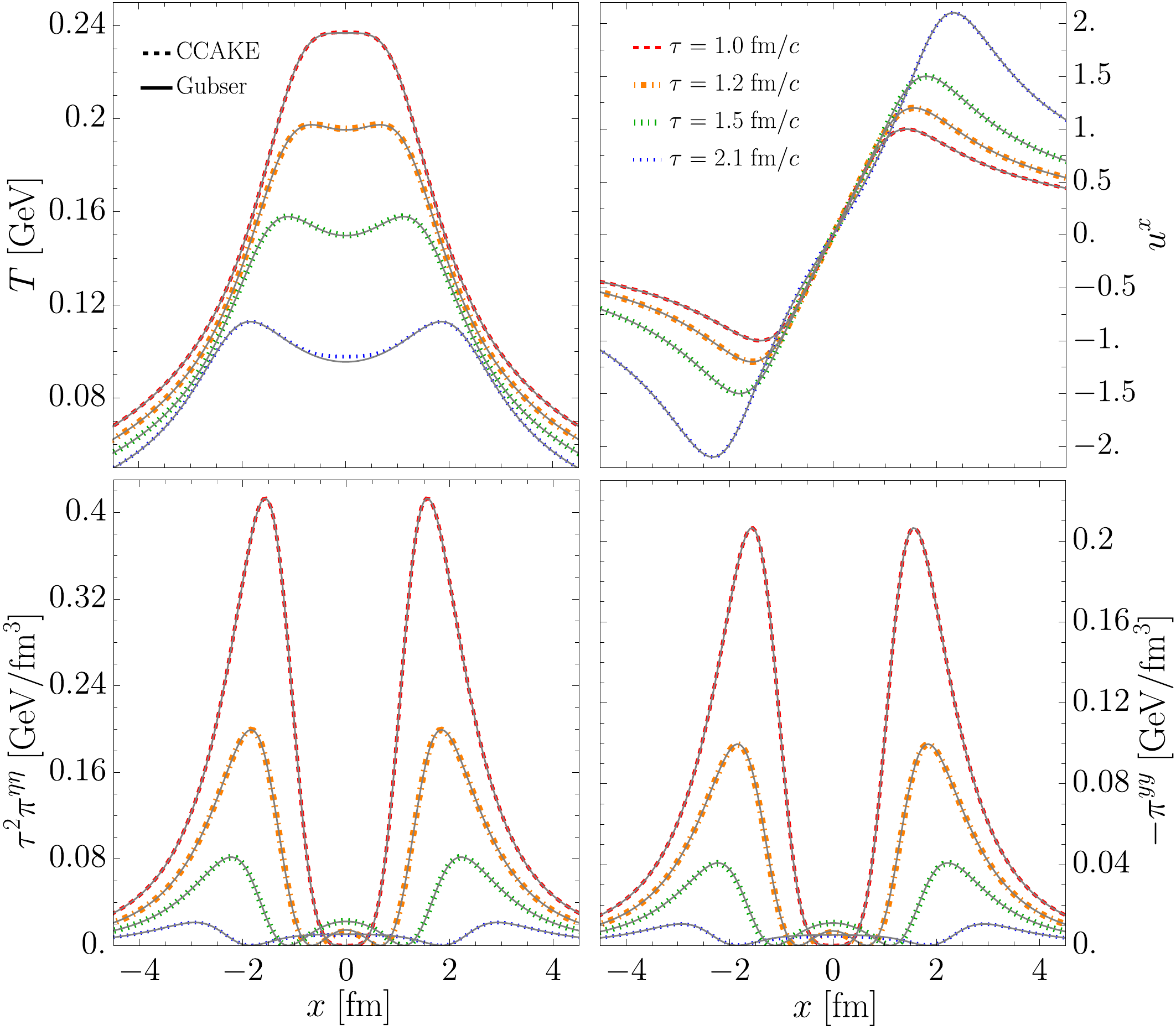}
    \caption{(color online) Comparison between the semi-analytic results from Ref.~\cite{Marrochio:2013wla} (solid) and \ccake{} (dashed) at times $\tau=1.0$, $1.2$, $1.5$, and $2.1$ fm$/c$ along the $y=0$ axis.}
    \label{fig:GubserGrid}
\end{figure*}

\begin{figure*}
    \centering
    \includegraphics[keepaspectratio, width=0.42\textwidth]{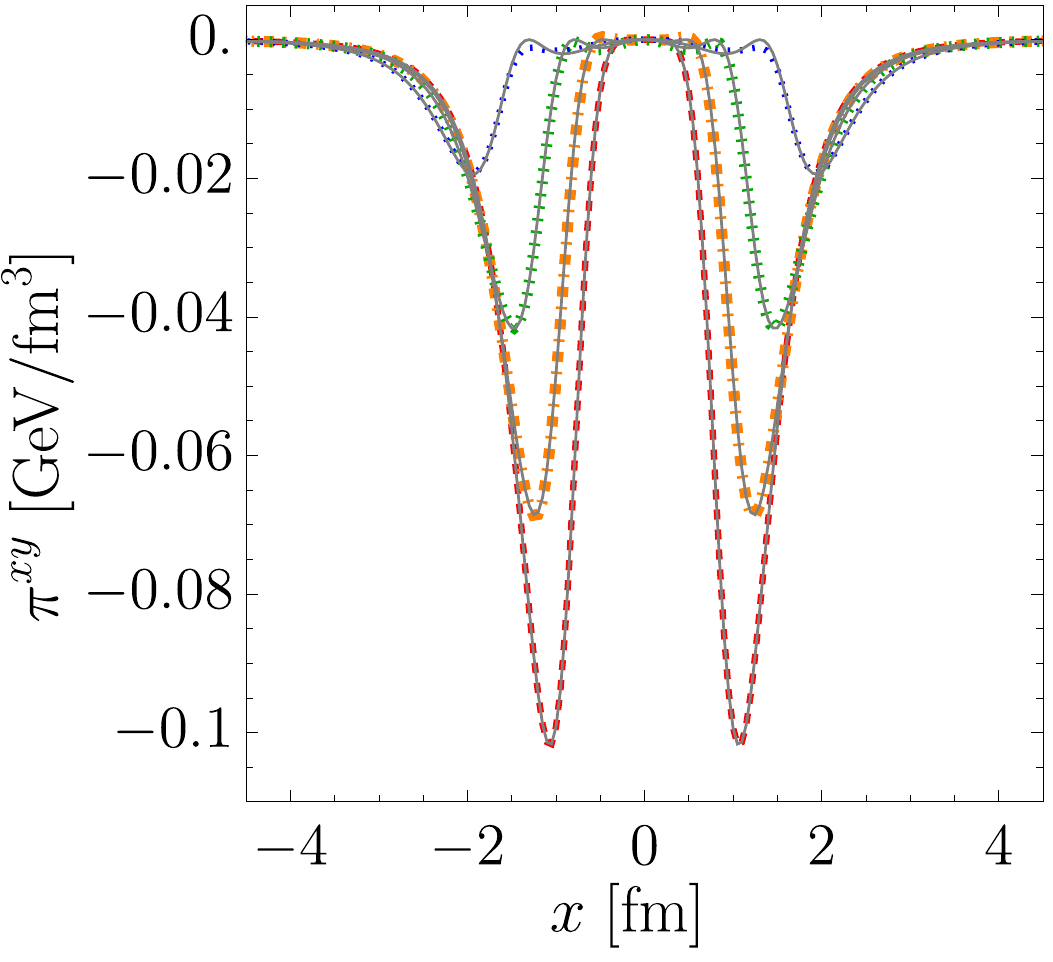}
    \caption{(color online) Comparison between the semi-analytic results from Ref.~\cite{Marrochio:2013wla} (solid) and \ccake{} (dashed) at times $\tau=1.0$, $1.2$, $1.5$, and $2.1$ fm$/c$ along the $y=x$ axis.}
    \label{fig:GubserPixy}
\end{figure*}

For the {\bf shear viscous case} we can also reproduce the Gubser test. The high degree of symmetry of the system completely dictates the flow, which is the same as in Eq.~\eqref{eq:GubserRadialFlow}. However, both the temperature and the shear-stress tensor follow the relaxation-type equations given by \cite{Marrochio:2013wla,Denicol:2018pak}
\begin{align}\label{eq:GubserIsraelStewart}
    \frac{1}{T}\frac{d\hat{T}}{d\rho} + \frac{2}{3}\tanh\rho - \frac{1}{3}\bar{\pi}_\eta^\eta(\rho)\tanh\rho&= 0, \nonumber \\
    \frac{c}{\hat{T}}\frac{\eta}{s}\l[ \frac{d\bar{\pi}_\eta^\eta}{d\rho} + \frac{4}{3}\l(\bar{\pi}_\eta^\eta\r)^2\tanh\rho\r] + \bar{\pi}_\eta^\eta &= \frac{4}{3}\frac{\eta}{s\hat{T}}\tanh\rho,
\end{align}
where $\bar{\pi}_\eta^\eta = \hat{\pi}_\eta^\eta/(\hat{T}\hat{s})$. These ordinary differential equations are obtained by using a coordinate transformation on Eqs.~\eqref{eq:e_conservation} and \eqref{eq:Shear_equation_of_motion} from hyperbolic space to the so-called Gubser coordinates $\hat{x}^\mu = (\rho,\theta,\phi,\eta)$ defined by
\begin{align}
    \sinh\rho &= - \frac{1 - \kappa^2\tau^2 + \kappa^2 r^2}{2\kappa\tau}, \\
    \tan\theta &= \frac{2\kappa r}{1 + \kappa^2\tau^2 - \kappa^2 r^2}.
\end{align}
In these conformal coordinates, Eqs.~\eqref{eq:GubserIsraelStewart} can be solved numerically, resulting in a semi-analytical solution. In hyperbolic coordinates, the solutions for the temperature and the components of the shear-stress tensor take the form

\begin{equation}
    T(\tau,x,y) = \frac{\hat{T}}{\tau} ,
\end{equation}
and
\begin{align}
    \pi^{xx}(\tau,x,y) &= -\frac{1}{2\tau^4}\l[ 1 + \l(\frac{x}{r}\r)^2\sinh^2\xi \r] \hat{\pi}^\eta_\eta, \\
    \pi^{yy}(\tau,x,y) &= -\frac{1}{2\tau^4}\l[ 1 + \l(\frac{y}{r}\r)^2\sinh^2\xi \r] \hat{\pi}^\eta_\eta, \\
    \pi^{xy}(\tau,x,y) &= -\frac{1}{2\tau^4}\l[ \l(\frac{xy}{r^2}\r)\sinh^2\xi \r] \hat{\pi}^\eta_\eta, \\
    \tau^2 \pi^{\eta\eta}(\tau,x,y) &= \frac{1}{\tau^4} \hat{\pi}^\eta_\eta,
\end{align}
where $\hat{T}$ and $\hat{\pi}^{\eta}_{\eta}$ are the numerical solutions of Eqs.~(12)-(13) in \cite{Marrochio:2013wla} and
\begin{equation}
    \sinh\xi = \mathrm{arctanh}\l( \frac{2\kappa^2\tau r}{1 + \kappa^2\tau^2 + \kappa^2 r^2} \r).
\end{equation}
The semi-analytical solution profile at $\tau=\tau_0=1$ fm/$c$ serves as an initial condition for \ccake{}. In the numerical solution we have used a grid of maximum size $x_\text{max} = y_\text{max} = 5$ fm and spacing of $\delta x = \delta y = 0.05$ fm, as well as a smoothing scale parameter of $h=0.1$ fm, $\eta/s=0.2$, $c=5$, and $\delta \tau = 0.01$ fm/$c$ as time step. As seen in Figs. \ref{fig:GubserGrid} and \ref{fig:GubserPixy}, our numerical solution reproduces well the semi-analytical solutions for the temperature $T$, radial flow, and shear stress tensor until at least 2.0 fm$/c$.


%%%%%%%%%%%%%%%%%%%%%%%%%%%%%%%%%%%%%%%%%%%%%%%%%
%%%%%%%%%%%%%%%%    TIME-CHECKS   %%%%%%%%%%%%%%%
%%%%%%%%%%%%%%%%%%%%%%%%%%%%%%%%%%%%%%%%%%%%%%%%%
\subsubsection{Time checks}
\label{sec:TimeCheck}

We document here the typical runtimes associated to the most significant parts of the \ccake{} evolution.  For typical events, evolved using a second-order Runge-Kutta method, we find that a single time-step takes $\mathcal{O}(30\text{ s})$ to complete with maximum optimizations enabled.  This translates to an overall runtime $\mathcal{O}(1-2\text{h})$ for the largest central events generated by \iccing{}.

The largest contributions to the runtime of a single time-step include the smoothing of particle fields, gradients over their nearest neighbors, and the evaluation of the particle thermodynamics, which includes the root-finding procedure described in Sec.~\ref{sec:EoS}.  
Obtaining the thermodynamic quantities of a given SPH particles has a linear complexity in the number of particles, i.e., it is $\mathcal{O}(N_\text{SPH})$.  The smoothing of particle fields and gradients depends on both the number of particles and the average number of neighbors per particle, which is expected to scale quadratically with the smoothing scale $h$, and thus has complexity $\mathcal{O}(N_\text{SPH} h^2)$.  Under maximum optimizations, we find that the smoothing procedure can process up to $\mathcal{O}\l(10^3\r)$ particles per second, while the root-finding procedure can achieve up to $\mathcal{O}\l(10^4\r)$ solutions per second.  
% .

%%%%%%%%%%%%%%%%%%%%%%%%%%%%%%%%%%%%%%%%%%%%%%%%%
%%%%%%%%%%%%%%%%    FREEZE-OUT   %%%%%%%%%%%%%%%%
%%%%%%%%%%%%%%%%%%%%%%%%%%%%%%%%%%%%%%%%%%%%%%%%%
\subsection{Freeze-out}
\label{sec:Freeze-out}
%%%%%%%%%%%%%%%%%%%%%%%%%%%%%%%%%%%%%%%%%%%%%%%%%
In the previous version of \vUSPhydro{}, freeze-out was developed assuming a constant freeze-out temperature, $T_\mathrm{FO}$.
While this is a reasonable assumption when $\vect{\mu}=0$, at finite densities it is known that both the hadronization \cite{Bellwied:2015rza,Borsanyi:2018grb,Borsanyi:2020fev,Bazavov:2020bjn,HotQCD:2018pds} and chemical freeze-out temperatures \cite{Cleymans:2005xv,Alba:2014eba,Borsanyi:2014ewa,Alba:2015iva,Bellwied:2018tkc,Alba:2020jir} lower with increasing $\vect{\mu}$'s such that the assumption of $T_\mathrm{FO}=\text{const.}$ breaks down.
Thus, for \ccake{} we have rewritten our freeze-out criterion such that it is instead implemented at a constant energy density.
Such an approach naturally captures the behavior of the freeze-out line bending down to lower $T_\mathrm{FO}$ as one increases $\vect{\mu}$.
Additionally, it follows well the same curvature as the chiral phase transition \cite{Bellwied:2015rza,Borsanyi:2020fev,HotQCD:2018pds} as shown from lattice QCD calculations.

In order to switch from a $T_\mathrm{FO}=\text{const.}$ to a $\varepsilon_\mathrm{FO}=\text{const.}$ freeze-out hypersurface, there were a number of changes that had to be implemented within the code.  The general algorithm as well as the changes (highlighted in bold) are as follows:
\begin{itemize}
    \item $T_\mathrm{FO}$ at $\vect{\mu}=0$ is chosen
    \item {\bf Change:} Using the EoS, one determines $\varepsilon_\mathrm{FO}(T_\mathrm{FO})$ at $\vect{\mu}=0$
    \item Every fluid cell is tracked in time and is only frozen out if it remains below $\varepsilon_\mathrm{FO}$ after two subsequent time steps (to avoid fluid cells that return immediately back to the fluid).
    \item Tiny fluctuations around $\varepsilon_\mathrm{FO}$ are allowed which occur between $d\tau$ time steps.   
    \item {\bf Change:}  The normal vectors are calculated assuming $\varepsilon_\mathrm{FO} \approx \text{const.}$, (see Sec.\ \ref{sec:constant-efo})).
    \item The hypersurface is reconstructed using the SPH formalism that allows one to switch integrals into summations (see Sec.\ \ref{sec:CF_SPH}).
    \item {\bf Change:}  Using this hypersurface Cooper-Frye freeze-out is performed in such a way as to include finite $\vect{\mu}$ and also to contain corrections to the distribution function from shear and bulk viscosity
    \item The primordial spectra of all particles are calculated from the particle data group (PDG) 2016 list \cite{Alba:2017mqu} that includes all * to **** star states.  This list is known as the PDG2016+. 
    \item Direct decays from the PDG2016+ using an adapted version of the direct decay code from \cite{Sollfrank:1990qz}
    %,Kolb:2000sd,Kolb:2002ve,Kolb:2003dz
    are performed in order to obtain the final spectra of stable particles.
    \item The adaptations from the original direct decay code in \cite{Sollfrank:1990qz} include: new particle resonances, improved input/output, full range of $\phi$ angle, and {\bf change:} finite $\vect{\mu}$.
    \item Using these spectra all final state observables are calculated.
\end{itemize}

Much of this procedure has been standard in the \vUSPhydro{} framework for years. The main changes occur because of the switch from $T_\mathrm{FO}\rightarrow \varepsilon_\mathrm{FO}$ and the addition of $\vect{\mu}$ within Cooper-Frye. In the following sections, we discuss those specific changes. 

%%%%%%%%%%%%%%%%%%%%%%%%%%%%%%%%%%%%%%%%%%%%%%%%%
%%%%%%%%%%%%%%    NORMAL VECTORS   %%%%%%%%%%%%%%
%%%%%%%%%%%%%%%%%%%%%%%%%%%%%%%%%%%%%%%%%%%%%%%%%
\subsection{Normal vectors for \texorpdfstring{$\varepsilon_\mathrm{FO}=\text{const.}$}{εᶠᵒ=const.}}
\label{sec:constant-efo}

%\jaki{Dekrayat, please add}
The normal vector to the constant energy hypersurface in the 2+1D case is
\begin{align}
    n_\mu = \frac{(\partial_\tau \varepsilon, \partial_x \varepsilon, \partial_y \varepsilon)}{\sqrt{(\partial_\tau \varepsilon)^2-(\partial_x \varepsilon)^2-(\partial_y \varepsilon)^2}}\,.
\end{align}
To calculate the hypersurface normal we use the general relation 
\begin{equation}
D \varepsilon=\gamma \,\partial_\tau \varepsilon +\left(\vect{u}\cdot\vect{\nabla}\,\varepsilon\right),
\end{equation}
and, using the energy density conservation, i.e., Eq.~\eqref{eq:e_conservation}, we obtain
 \begin{equation}
 \partial_\tau \varepsilon = \frac{1}{\gamma}\left(- (\varepsilon+P)\theta - \Pi \theta + \pi_{\mu\nu} \sigma^{\mu\nu} - \vect{u}\cdot\vect{\nabla}\,\varepsilon \right).
 \end{equation}

 For the spatial gradients, we follow the SPH parametrization: 
% \left(\partial_i p\right)_\beta=\sum_\alpha \nu_\alpha s_\beta^*\left(\frac{p_\alpha}{\left(s_\alpha^*\right)^2}+\frac{p_\beta}{\left(s_\beta^*\right)^2}\right) \nabla_\beta W_{\alpha \beta},
 %
 \begin{align}
 \left(\nabla_i \varepsilon \right)_\beta=\sum_\alpha \nu_\alpha s_\beta^*\l(\frac{\varepsilon_\alpha}{\big(s_\alpha^*\big)^2}+\frac{\varepsilon_\beta}{\big(s_\beta^*\big)^2}\r) \nabla_\beta W_{\alpha \beta},
\end{align}
\begin{figure}
    \centering
    \includegraphics[keepaspectratio, width=\linewidth]{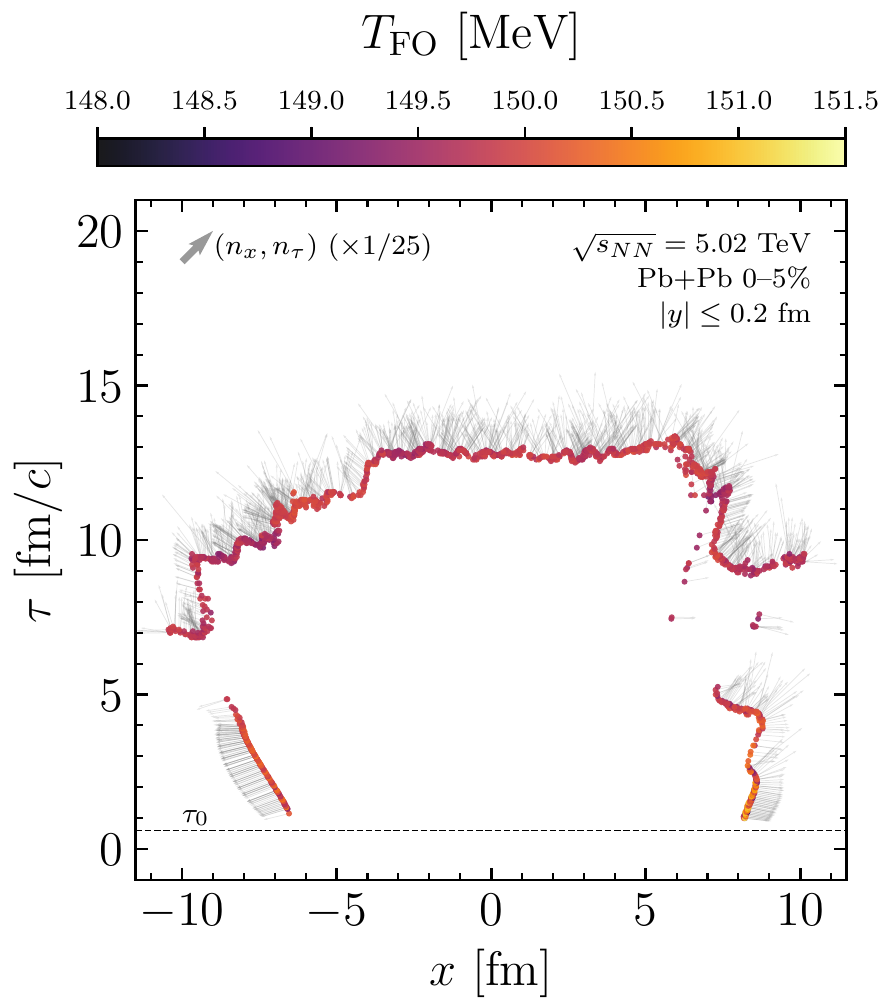}
    \caption{
    (color online) Constant energy density hypersurface of a central Pb+Pb at $\sqrt{s_{NN}}=5.02$ TeV event from \TRENTo{}\texttt{+}\iccing{}\texttt{+}\ccake{} at $\lvert y\lvert\; \leq 0.2$ fm, see \ref{sec:constant-efo}. The color and the grey arrows of every SPH particle indicate the temperature and the normal vectors at freeze-out, respectively.
    }
    \label{fig:hypersurface}
\end{figure}

An example hypersurface for the constant energy density approach is shown in Fig.\ \ref{fig:hypersurface}.  The hypersurface vectors are color coded to indicate the corresponding freeze-out temperature that $\varepsilon_\mathrm{FO}$ relates to. Clearly, the lower freeze-out temperatures occur for fluid cells that freeze-out at large chemical potentials.  From the figure we find that most fluid cells freeze-out in temperatures ranging from $T_\mathrm{FO}\sim 140-150$ MeV where a significantly smaller amount fall at the lower end of this range as compared with the upper end.

%%%%%%%%%%%%%%%%%%%%%%%%%%%%%%%%%%%%%%%%%%%%%%%%%
%%%%%%%%%%%%%%%    COOPER-FRYE   %%%%%%%%%%%%%%%%
%%%%%%%%%%%%%%%%%%%%%%%%%%%%%%%%%%%%%%%%%%%%%%%%%
\subsection{Cooper-Frye in SPH at finite \texorpdfstring{$\vect{\mu}$}{chemical potential}}\label{sec:CF_SPH}

Once the hypersurface $\Sigma$ is determined, we can calculate the total number of particles of a given species $j$ using
\begin{equation}
    N_j= \int \rho^{\mu}_j d \Sigma_{\mu},
\end{equation}
where $\rho^{\mu}_j$ is the particle number density. Using the Cooper-Frye formula \cite{Cooper:1974mv}, one determines the particle momentum spectra at freeze-out
\begin{equation}
   E\frac{d^3N_j}{dp^3}=\frac{g_j}{(2\pi)^3}\int_\sigma f_j({p}) p^\mu d\sigma_\mu,
\end{equation}
where $p^\mu$ is the momentum, $g_j$ is the degeneracy, $f_j$ is the distribution function, and $d\sigma_{\mu}$ is an element of the hypersurface. 

The distribution function itself has both equilibrium contributions, $f^{(i)}_{eq,j}$, as well as out-of-equilibrium contributions from shear $\delta f^{(i)Shear}_{j}$, bulk $\delta f^{(i)Bulk}_{j}$, and diffusion $\delta f^{(i)diff}_{j}$ such that
\begin{equation}
f^{(i)}_{j}=f^{(i)}_{eq,j}\left[1+\left(1-a_jf^{(i)}_{eq,j}\right)\delta f^{(i)}_{j}\right],
\end{equation}
where $j$ is again a specific particle species, $i$ is the $i^{\text{th}}$ SPH particle. Note that here we have redefined the $\delta f^{(i)}_{j}$ term slightly from previous works wherein we have pulled out the $f^{(i)}_{eq,j}\left(1-a_jf^{(i)}_{eq,j}\right)$ contribution from the individual dissipative $\delta f$ contributions for simplicity's sake. Then, 
\begin{equation}
    \delta f_j^{(i)}=\delta f^{(i)Bulk}_{j}+\delta f^{(i)Shear}_{j}+\delta f^{(i)diff}_{j},
\end{equation}
is the out-of-equilibrium correction to the distribution function.
 The ideal distribution, $f^{(i)}_{eq,j}$, is defined as
\begin{equation}\label{eqn:f_eq}
f^{(i)}_{eq,j}=\frac{1}{e^{( p\cdot u- \vect{\mu}_i \cdot \vect{X}_i)/T_i}+a_j},
\end{equation}
where $a_j=1$ for fermions, $a_j=-1$ for bosons, and $a_j=0$ for Boltzmann.
We have also defined the vector $\vect{X}_i=\left\{B_i,S_i,Q_i\right\}$ and adopted the notation:
\begin{equation}
    p \cdot u \equiv  p^{\mu,i} u_{\mu,i}
\end{equation}
where it is understood that we substitute the values of momentum and flow for the $i^{th}$ SPH particle.
The correction term for the shear viscosity is
\begin{equation}
\delta f^{(i)Shear}_{j}=\frac{\pi^{\mu\nu}p_{\mu}p_{\nu}}{2\left(\varepsilon_i+p_i\right)T^2},
\end{equation}
the correction term for the bulk is
\begin{equation}
\delta f^{(i)Bulk}_{j}=\Pi\left[B_0^{(i)}+D_0^{(i)}\left( p\cdot u\right)+E_0^{(i)}\left( p\cdot u\right)^2\right],
\end{equation}
where the coefficients $B_0$, $D_0$, and $E_0$ are taken from the 14 moments approximation,
and the correction term for the diffusion \emph{for a single component} is
\begin{eqnarray}\label{eq:Tensor_projectorf_diff}
  \delta f^{(i)diff}_j &=&  
 \sum_{X=B,S,Q}\frac{n_X^\mu p_{\langle\mu\rangle}}{\beta_X}\left(\frac{\rho_X}{\varepsilon_i {+} p_i} - \frac{X_i}{ p\cdot u}\right),
\end{eqnarray}
where $\beta_X=\kappa_X/\tau_X$ is the ratio of the diffusion to relaxation time and $p_{\langle\mu\rangle}\equiv \Delta_{\mu\nu}p^\nu$. There are, of course, cross-correlations when it comes to BSQ charges \cite{Greif:2017byw,Fotakis:2019nbq,Fotakis:2022usk} but the $\delta f$ corrections have not yet been derived for the entire diffusion matrix.  Rather, Eq.\ (\ref{eq:Tensor_projectorf_diff}) is equivalent to the diagonal terms in this diffusion matrix.
In this work we assume ideal conserved charge currents (i.e., $n_X^{\mu}\rightarrow 0$) such that we obtain $\delta f^{(i)diff}_j= 0$.

Now that the equations for Cooper Frye freeze-out are established, we must next make a choice about:
\begin{enumerate}
    \item The coordinate system: here we choose hyperbolic coordinates.
    \item The numerical hydrodynamic algorithm: SPH significantly simplifies the hypersurface calculation.
\end{enumerate}

First, the coordinate system dictates the form of our four-vectors such that in Cartesian coordinates:
\begin{equation}
p^{\mu}\cdot n_{\mu}=E n_t+p^x n_x+p^y n_y +p^z n_z,
\end{equation}
where $n_\mu$ are the normal vectors of the constant energy density hypersurface defined in Sec.\ \ref{sec:constant-efo}.
In hyperbolic coordinates 
\begin{eqnarray}
p^{\mu}&=& \left( \begin{matrix} m_{\perp} \text{cosh}(\eta-y),&p^x,&p^y,& m_{\perp} \text{sinh}(\eta- y) \end{matrix}\right)\nonumber\\
p^{\mu}\cdot u_{\mu}&=&  m_{\perp} u_{\tau} \text{cosh}\left(\eta-y\right) - \vect{p}_{\perp}\cdot \vect{u}_{\perp}  \\
p^{\mu}\cdot n_{\mu}&=&  m_{\perp} n_{\tau} \text{cosh}\left(\eta-y\right) - \vect{p}_{\perp}\cdot \vect{n}_{\perp} \\
u^{\mu} \cdot n_{\mu}&=& u_{\tau}n_{\tau}+u^{x}n_x+u^y n_y\\
\pi^{\mu\nu}p_{\mu}p_{\nu}&=& m_{\perp}^2 \left[\pi^{00} \text{cosh}^2(\eta- y) +\tau^4\pi^{33}  \text{sinh}^2(\eta- y) \right]\nonumber\\
&+&p^2_x\pi^{11}+p_y^2\pi^{22}+2p_x p_y \pi^{12}
\end{eqnarray}
where we considered all possible contributions needed for Cooper-Frye with both shear and bulk viscosity (diffusion contributions would require other terms).

Next, we use the features of the SPH formalism to simplify the integral over the isothermal hypersurface \cite{Osada:2001hw,Hama:2004rr} such that it is written in terms
of a sum of SPH particles as 
\begin{equation}
\frac{d^3N}{dydp_T^2} = \frac{g}{(2\pi)^3}\,\sum_{i=1}^{N_\text{SPH}}%
\int_{-\infty}^{\infty}d\eta_i\frac{(p\cdot n)_i}{ (n\cdot u)_i }\frac{\nu_i}{\sigma_i}
\,f^{(i)}_{j},
\end{equation}
where $N_{SPH}$ is the total
number of SPH particles, $(n_\mu)_i$ is the normal vector of the
isothermal hypersurface reconstructed using the $i$-th SPH particle, $%
(u_\mu)_i$ is the 4-velocity of the SPH particle,  $\Pi_i$ is
the bulk viscosity of the SPH particle, d$\eta_i$ is the integral over the rapidity of the $i$th particle, and $\pi^{\mu\nu}_i$ is
the shear stress tensor of the SPH particle. Then, in hyperbolic coordinates the ideal distribution function is
\begin{equation}
f^{(i)}_{eq,j}=e^{ \vect{p}_{\perp}\cdot \vect{u}_{\perp}/T_0}e^{- m_{\perp} u_{\tau}/T_0 \text{cosh}\left(\eta-y\right)},
\end{equation}
and the integral over the isothermal hypersurface becomes
\begin{widetext}
\begin{eqnarray}
\frac{d^3N}{dydp_T^2}& =& \frac{g}{(2\pi)^3}\,\sum_{i=1}^{N_\text{SPH}}\frac{1}{ (n\cdot u)_i } \frac{\nu_i}{\sigma_i} \left\{  m_{\perp} n_{\tau} 
\int_{-\infty}^{\infty}d\eta_i \, \text{cosh}\left(\eta-y\right)
f^{(i)}_{j}  + \left[p^xn_x +p^y n_y\right]
\int_{-\infty}^{\infty}d\eta_i 
\,f^{(i)}_{j}\right\},
\end{eqnarray}
\end{widetext}
where it should be clear that $f^{(i)}_{j}$ is a function of the following hydrodynamic variables at freeze-out: $T$, $u^\mu$, $p^\mu$, $\pi^{\mu\nu}$, $\Pi$, $n^{X,\mu}$.  For further details on the numerical implementation with the freeze-out code, see Appendix B from \cite{Noronha-Hostler:2014dqa} for details on shear viscosity and Appendix B from \cite{Noronha-Hostler:2013gga} for details on ideal and  bulk viscosity. 
The out-of-equilibrium corrections for diffusion will be explored in a future work.

One challenge that arises when including finite $\vect{\mu}$ is that occasionally fluid cells freeze-out at such large chemical potentials that the $\vect{\mu}$ term in Eq.\ (\ref{eqn:f_eq}) is larger than the energy term, i.e., 
\begin{equation}
    p\cdot u < \vect{\mu}_i\cdot \vect{X}_i,
\end{equation}
such that it can lead to diverging contributions to the multiplicity.  These contributions are especially problematic at low temperatures. In order to prevent these contributions to the multiplicity, we insert a check  such that:
\begin{equation}
    f_{eq,j}^{(i)} = 
    \begin{cases}
    \left[ e^{( p\cdot u- \vect{\mu}_i \cdot \vect{X}_i)/T_i}+a_j \right]^{-1} & \text{if}\quad p\cdot u \geq \vect{\mu}_i\cdot \vect{X}_i \\
    0 & \text{otherwise}.
    \end{cases}
\end{equation}
This essentially ensures that a single fluid cell has enough energy to actually produce a certain particle, given its local $\vect{\mu}$.  

%%%%%%%%%%%%%%%%%%%%%%%%%%%%%%%%%%%%%%%%%%%%%%%%%
%%%%%%%%%%%%%%%%%    RESULTS   %%%%%%%%%%%%%%%%%%
%%%%%%%%%%%%%%%%%%%%%%%%%%%%%%%%%%%%%%%%%%%%%%%%%
\section{Results: Fluctuations in BSQ at freeze-out}
\label{sec:Results}

Having presented the main elements of the BSQ framework (consisting of \TRENTo{}\texttt{+}\iccing{}\texttt{+}\ccake{}) in Sec.~\ref{sec:BSQframework}, we now apply this framework to the description of fluctuations of BSQ conserved charges in Pb\texttt{+}Pb collisions at the LHC $\sqrt{s_{NN}}=5.02$ TeV collision energy.
We first consider in Sec.\ \ref{sec:BSQdyn} how the $\vect{\rho}(\tau)$ evolve during the hydrodynamic expansion of a central event (in the $0$--$5\%$ centrality class).
Next, we show in Sec.\ \ref{sec:Trajectory} how the evolution of conserved charges is reflected in the dynamical trajectories of individual fluid cells which pass through the QCD phase diagram as they expand and cool over time.
The resulting trajectories span a wide range in chemical potentials as a consequence of event-by-event fluctuations in the initial state.  In Sec.\ \ref{sec:BSQDensityProfiles}, we study the extent to which these fluctuations in $\vect{\mu}$ survive until the system freezes out.
Then in Sec.\ \ref{sec:densityT} we discuss how it is possible that such large BSQ density fluctuations in the initial state can lead to relatively small chemical potentials in the final state. 
Finally, we discuss in Sec.\ \ref{sec:h0} how BSQ charges affect standard experimental observables (including spectra and flow) at the level of a single event.

%
%%%%%%%%%%%%%%%%%%%%%%%%%%%%%%%%%%%%%%%%%%%%%%%%%
%%%%%    BSQ CHARGE DYNAMICAL EVOLUTION   %%%%%%%
%%%%%%%%%%%%%%%%%%%%%%%%%%%%%%%%%%%%%%%%%%%%%%%%%
\subsection{Dynamical evolution of BSQ charge densities}\label{sec:BSQdyn}
%==============================================================================
\begin{figure}
    \centering
    \includegraphics[keepaspectratio, width=0.99\linewidth]{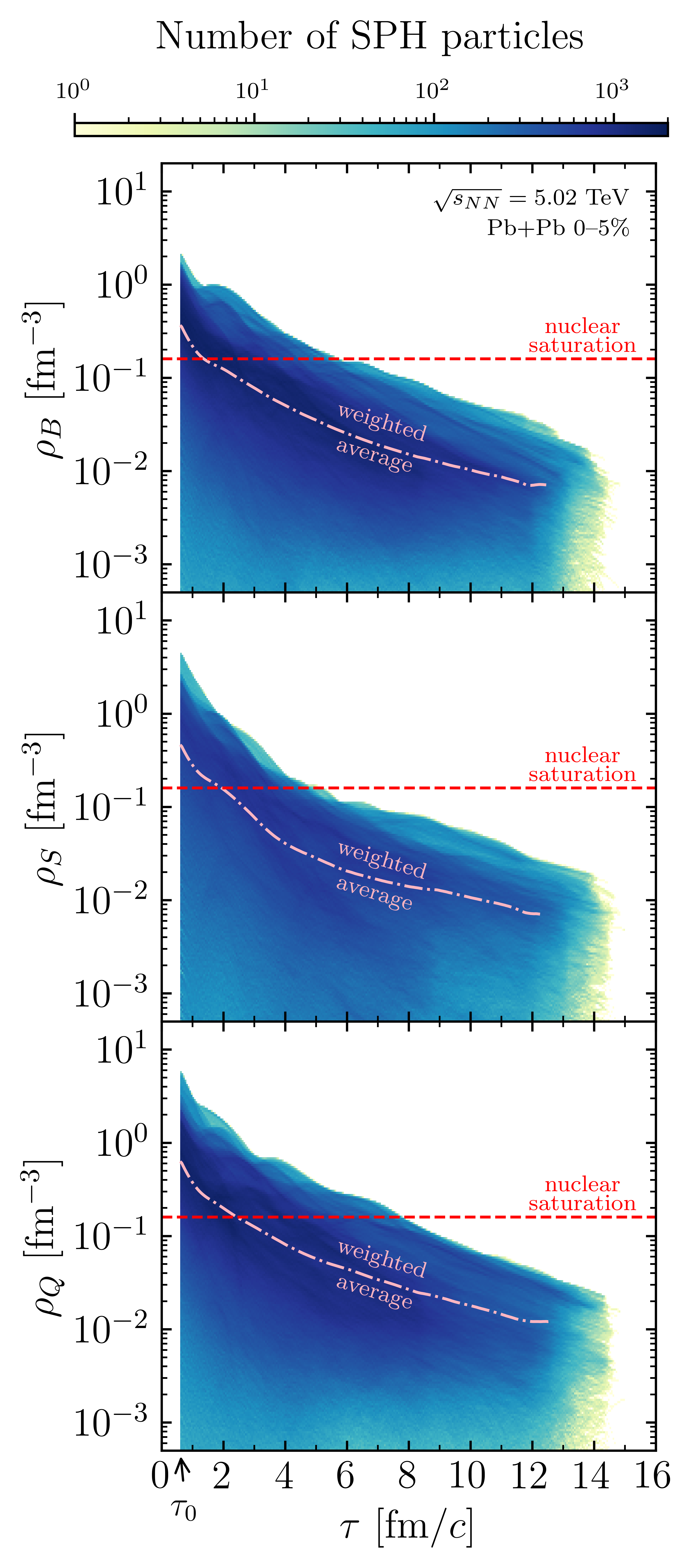}
    % \includegraphics[keepaspectratio, width=0.99\linewidth]{figures/density_evolution/e_vs_tau}
    % \\
    % \includegraphics[keepaspectratio, width=0.99\linewidth]{figures/density_evolution/B_vs_tau}
    % \\
    %  \includegraphics[keepaspectratio, width=0.99\linewidth]{figures/density_evolution/S_vs_tau}
    %  \\
    %   \includegraphics[keepaspectratio, width=0.99\linewidth]{figures/density_evolution/Q_vs_tau}
    \caption{(color online) Dynamical evolution of conserved densities $\{\rho_B,\rho_S,\rho_Q\}$ (top, middle, bottom) in the hydrodynamic evolution starting from their initial time until the final hydrodynamic step.  Here we only show the SPH particles in \ccake{} which have positive charged densities and energy density above freeze-out.
     }
    \label{fig:density-profile}
\end{figure}

Because of the strong expansion of the system, one expects the charge densities $\rho_B(\tau)$, $\rho_S(\tau)$, $\rho_Q(\tau)$ to decrease over time.  However, due to the finite lifetime of the system and complications arising from out-of-equilibrium effects, the rate of decrease is far from obvious.  This leaves it an open question whether the effects of finite charge density will still be relevant at freeze out.

To explore the time evolution of $\rho_B(\tau)$, $\rho_S(\tau)$, $\rho_Q(\tau)$, we plot the distribution in the density of SPH particles at each point in time, $\tau$ in Fig.\ \ref{fig:density-profile}.
We show only distributions of the positive charge densities ($\vect{\rho} > 0$) as functions of $\tau$, and use the color scales to indicate the number of SPH fluid cells at a specific density.
The top panel shows $\rho_B(\tau)$, the middle panel shows $\rho_S(\tau)$, and the bottom panel shows $\rho_Q(\tau)$.
On average, all three densities exhibit an approximate $1/\tau$ dependence on time which is characteristic of Bjorken expansion (the scaling is not exact here, since the system also contains fluctuations and strong transverse expansion, which are absence from the Bjorken model).
The densities also have a similar order of magnitude, although close inspection reveals relatively minor quantitative differences in their evolution.
For instance, $\rho_Q$ remains roughly twice as large as $\rho_B$ (and approximately $50\%$ larger than $\rho_S$) for much of the evolution.
This is reasonable in light of the fact that the scale of the average positive densities (dash-dotted curves) is set by the typical scale of the quantum numbers on the $u$, $d$, and $s$ quarks sampled in the \iccing{} model.
In particular, the up (anti-)quarks carry a quantum numbers of $\l| Q \r| = +\text{\sfrac{2}{3}}$ and $\l| B \r| = +\text{\sfrac{1}{3}}$ and are produced in larger quantities than down and strange quarks due to their smaller masses, thereby resulting in a larger scale for $\rho_Q$ than $\rho_B$.
We emphasize that the charge densities shown in Fig.~\ref{fig:density-profile} arise only from \emph{fluctuations} in the initial state and do not reflect the \emph{average} relationships between densities mentioned earlier, specifically $\langle \rho_Q\rangle = (Z/A) \langle \rho_B\rangle$.
To account for this additional constraint, one would need to include a \emph{net} BSQ density when constructing the initial state.  This feature has not yet been implemented in the current framework.

In addition to the qualitative behaviors noted above, we also observe several non-trivial features in Fig.~\ref{fig:density-profile}, such as small increases and decreases in densities at a various point during the evolution.
These behaviors are likely due to interactions between the background fluctuations as well as the 4D EoS.
If diffusion was also included in our model, we would anticipate even further complex behavior.
The panels of Fig.~\ref{fig:density-profile} also show that the charge densities initially reach values 2-4 times in excess of the nuclear saturation density $\rho_0 \approx 0.16$ fm$^{-3}$.
However, after approximate $\tau\sim 2.5$ fm/$c$ we find that all densities have dropped below $\rho_0$, with their final values an order of magnitude smaller than $\rho_0$.
%

%%%%%%%%%%%%%%%%%%%%%%%%%%%%%%%%%%%%%%%%%%%%%%%%%
%%%%    PASSAGE THROUGH QCD PHASE DIAGRAM   %%%%%
%%%%%%%%%%%%%%%%%%%%%%%%%%%%%%%%%%%%%%%%%%%%%%%%%
\subsection{Passage through the QCD phase diagram at the LHC}
\label{sec:Trajectory}
%==============================================================================

%
\begin{figure*}
    \centering
    \includegraphics[keepaspectratio, width=0.9\linewidth]{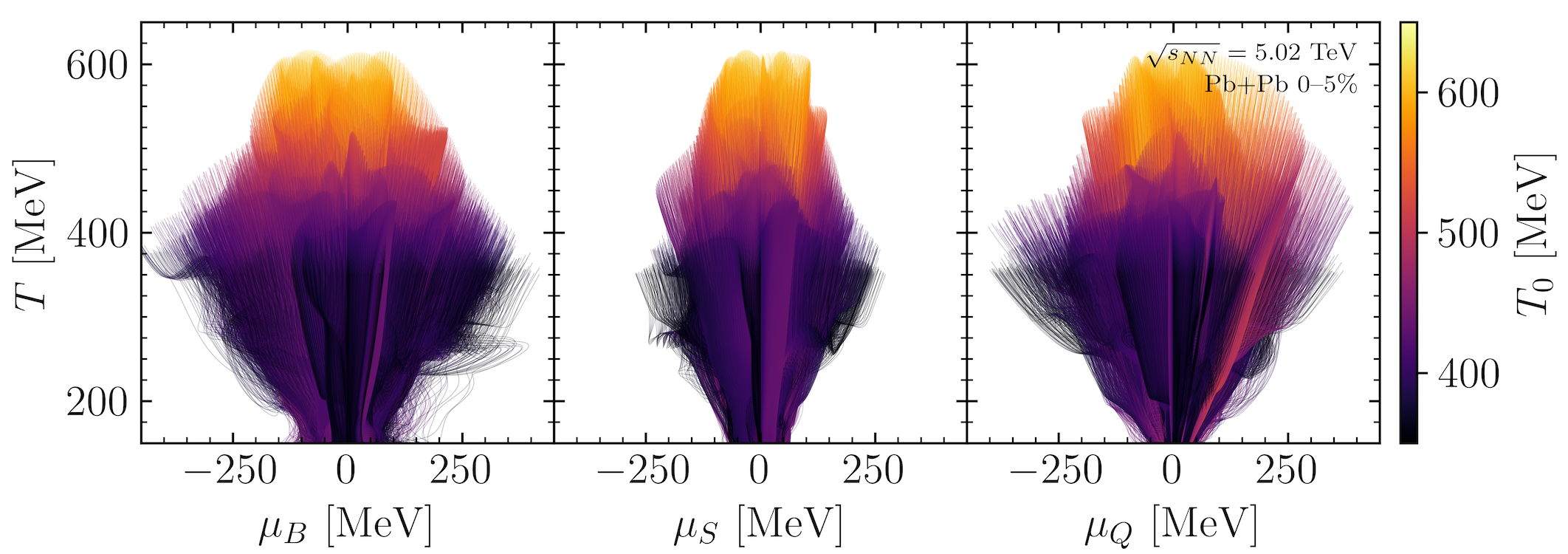}
    \caption{(color online) Trajectories of fluid cells that have initial temperatures above $T_0 \geq 350$ MeV across the phase diagram in $\left\{ T, \mu_B\right\}$ (left), $\left\{ T, \mu_S\right\}$ (middle), $\left\{ T, \mu_Q\right\}$ (right). The trajectories are color coded by their initial temperature, $T_0$. The BSQ local charge fluctuations survive during the hydrodynamical response and result in finite BSQ chemical potentials at the freeze-out hypersurface.
    }
    \label{fig:Tmu_trajs}
\end{figure*}

In ideal fluid dynamic systems, one can determine isentropic trajectories through the QCD phase diagram wherein the total entropy over baryon number is held constant, $S/N_B=\text{const}$.  Isentropes have been applied in many different cases in the field of heavy-ion collisions (see e.g. \cite{Ejiri:2005uv,Schmid:2008sy,Bellwied:2016cpq,Noronha-Hostler:2019ayj,Motornenko:2019arp}).  However, it was recently  pointed out in \cite{Dore:2020jye} with other follow-up papers on the work in \cite{Du:2020zqg,Chattopadhyay:2022sxk} that large deviations from isentropic trajectories should be anticipated for viscous fluids due to entropy production (this also changes the passage over the first-order phase transition \cite{Feng:2018anl}). In our case, our initial conditions are still ``ideal" in the sense that the initial $T^{\mu\nu}_0$ provided by \TRENTo{}\texttt{+}\iccing{} do not yet contain out-of-equilibrium components (although initial work in that direction can be found in \cite{Carzon:2023zfp} and the \KOMPOST{} group \cite{Kurkela:2018wud,Kurkela:2018vqr} will likely have this capability in the near future). However, because we incorporate shear  viscosity in our hydrodynamic simulations, entropy production over time still leads to deviations from the isentropic trajectories. Here we will study these out-of-equilibrium trajectories and the regions of the QCD phase diagram that are covered within our model.

In Fig.\ \ref{fig:Tmu_trajs} we plot the trajectories across $\left\{ T, \mu_B\right\}$ (left), $\left\{ T, \mu_S\right\}$ (middle), $\left\{ T, \mu_Q\right\}$ (right).  Each line represents the trajectory of a single fluid cell with \ccake{} and we select only fluid cells that start at high temperatures $T_0>300$ MeV, in order to make the qualitative behavior of the trajectories more easily visible.  The trajectories are color coded by their initial temperatures. 

Looking first at $\left\{ T, \mu_B\right\}$ trajectories in Fig.\ \ref{fig:Tmu_trajs} we find a very wide initial spread in $\mu_B$ for high temperatures in the range of $\Delta\mu_B^0\sim\pm 400$ MeV.  As the fluid expands and cools, we find that the fluctuations in $\mu_B$ are significantly suppressed but do not disappear entirely. Around freeze-out, i.e., $T_\mathrm{FO}\approx 150$ MeV, we can see that the range is still quite broad, with $\Delta\mu_B^{FO}\approx \pm 100$ MeV.  Thus, we anticipate non-trivial effects in both our spectra and flow observables.  Note that, since we have only selected fluid cells with initial temperatures above $T_0>300$ MeV, the actual spread in $\mu_B$ at freeze-out is even wider, arising from fluid cells that begin at lower temperatures.

The $\left\{ T, \mu_S\right\}$ trajectories are shown in the center panel of Fig.\ \ref{fig:Tmu_trajs}.  We find a somewhat narrower initial distribution in the fluctuations of $\mu_S$ than in the case of $\mu_B$, with $\Delta\mu_S^0\sim\pm 250$ MeV.  The distribution at freeze-out is also narrower, but we nevertheless observe fluctuations at a scale of approximately $\Delta\mu_S^{FO}\sim \pm 50$ MeV.  Since the fluctuations in $\mu_S$ are non-vanishing even at freeze-out, we anticipate that strange observables may still be affected by them.

Finally, we consider the $\left\{ T, \mu_Q\right\}$ trajectories in Fig.\ \ref{fig:Tmu_trajs} and find a similar initial range of electric charge chemical potentials to that of baryons, in the range of $\Delta\mu_Q^0\sim\pm 350$ MeV.
By freeze-out, however, the systems narrowed to a similar range of the strangeness chemical potentials, such that the scale of fluctuations is about $\Delta\mu_Q^{FO}\sim \pm 50$ MeV.

Thus, after comparing the ranges of the various chemical potentials at freeze-out, we anticipate that baryon fluctuations are likely to be more important than fluctuations of strangeness or electric charge, although one must check this expectation systematically against a number of experimental observables.
The picture is also complicated by the fact a larger spread in chemical potential need not always correspond to a larger spread in density, as evidence by the preceding discussion of Fig.~\ref{fig:density-profile}.
Additionally, the fluid cells with lower $T_0$ may enhance the fluctuations of some charges over others at freeze-out.

\begin{figure}
    \centering
    \includegraphics[keepaspectratio, width=0.99\linewidth]{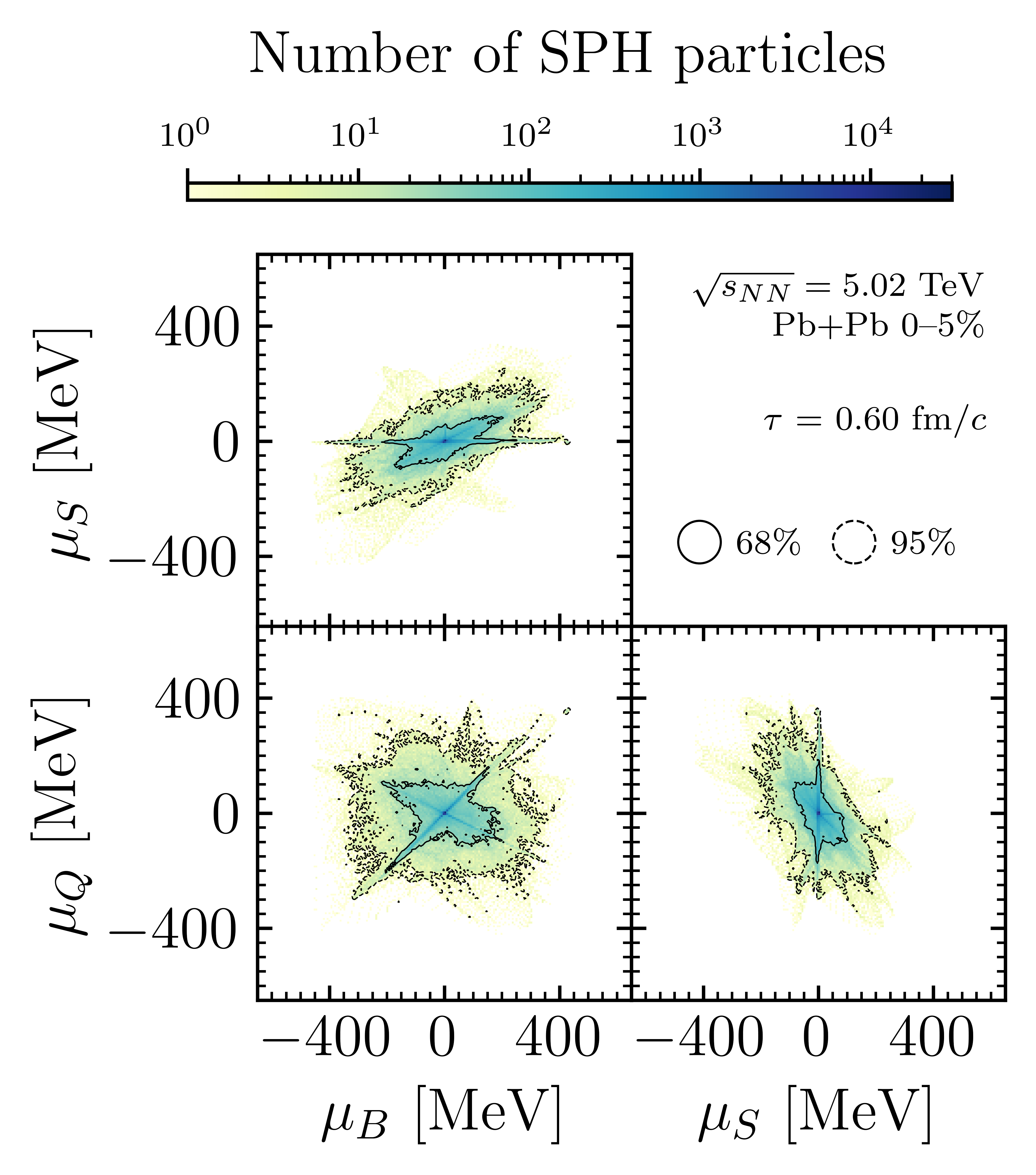}
    \caption{
    (color online) Two-dimensional histogram of the correlation of $\{\mu_S,\mu_B\}$ (top-left), $\{\mu_Q,\mu_B\}$ (bottom-left), and $\{\mu_Q,\mu_S\}$ (bottom-right) chemical potentials at $\tau=0.60$ fm/$c$ for a Pb\texttt{+}Pb system at $\sqrt{s_{NN}}=5.02$ TeV. 
    In all cases, a solid contour enclosing 68.27\% and a dashed contour enclosing 95.45\% of the SPH particles is displayed.
    }
    \label{fig:vecmu}
\end{figure}

Up until this point, we have considered only slices of the QCD phase diagram which plot temperature against individual chemical potentials.
However, all chemical potentials are also related to each other through the EoS and by the charge correlations which arise from the quarks initially seeded by \iccing{}.
We therefore have non-trivial correlations between all three charges (both densities and chemical potentials) in our approach.
To explore these correlations, we next plot different chemical potentials against one another, in order to see clearly how the charge correlations are mapped to the full 4D phase diagram of QCD.

In Fig.\ \ref{fig:vecmu} we show the  correlations between different pairs of chemical potentials in the initial state obtained from \iccing{}.
In the upper left panel, we see that $\mu_B$ exhibits a noticeably positive correlation with $\mu_S$.
Likewise, in the lower left panel we observe little correlation between $\mu_B$ and $\mu_Q$, whereas $\mu_S$ and $\mu_Q$ exhibit a significant anti-correlation.
We stress that these correlations need \emph{not} imply similar correlations between the corresponding densities.
This is because the presence of multiple conserved charges on a single quark automatically generates highly non-trivial correlations between the charge densities which need not be reflected in the correlations between the chemical potentials.

As an illustration of this last point, consider a scenario where $T = 800$ MeV, $\mu_B = 450$ MeV, and $\mu_S = \mu_Q = 0$.
In this case, by straightforwardly evaluating the pressure and charge densities of the table EoS described in Sec.~\ref{sec:TaylorSeriesEoS}, we find approximately that
\begin{align}
    \frac{p}{T^4} = 4.10,\, \frac{\rho_B}{T^3} = 0.177,\,  \frac{\rho_S}{T^3} = -0.179, \text{ and } \frac{\rho_Q}{T^3} = 0 \label{charge_correlations_example}
\end{align}
We observe that $\rho_B/\rho_S \approx -1$ and $\rho_Q \approx 0$.
We can understand this result by first recalling that the $u$, $d$, and $s$ quarks each carry the quantum numbers
\begin{alignat}{4}
&    \quad && B   \quad && S \quad && Q \nonumber\\
& u: \quad && 1/3 \quad && 0 \quad && 2/3\nonumber\\
& d: \quad && 1/3 \quad && 0 \quad && -1/3\nonumber\\
& s: \quad && 1/3 \quad && -1 \quad && -1/3\nonumber
\end{alignat}
Setting $\mu_B \neq 0$, $\mu_S = \mu_Q = 0$ thus creates a system which has equal parts of each quark flavor with a total baryon density proportional to $B_u + B_d + B_s = \text{\sfrac{1}{3}} + \text{\sfrac{1}{3}} + \text{\sfrac{1}{3}} = +1$, total strangeness density proportional to $S_u + S_d + S_s = 0 + 0 + (-1) = -1$, and total electric charge density proportional to $Q_u + Q_d + Q_s = \text{\sfrac{2}{3}} + (-\text{\sfrac{1}{3}}) + (-\text{\sfrac{1}{3}}) = 0$.
These results are consistent with the patterns observed in \eqref{charge_correlations_example}.

Similar considerations apply in the case where \textit{two} chemical potentials are non-zero.
For example, one finds by a similar calculation for $T = 800$ MeV, $\mu_B = 3\mu_S = 450$ MeV, and $\mu_Q = 0$ that approximately
\begin{align}\label{charge_correlations_example2}
    \frac{p}{T^4} = 4.09,\, \frac{\rho_B}{T^3} = 0.117,\,  \frac{\rho_S}{T^3} = 0, \text{ and } \frac{\rho_Q}{T^3} = 0.060. 
\end{align}
This charge configuration corresponds to a roughly equal number of $u$ and $d$ quarks and few or no $s$ quarks, thus agreeing well with the situation found in \iccing{} \cite{Carzon:2019qja}.
Moreover, if one fixes $\mu_B = 3\mu_S$ and decreases both chemical potentials to 0 (thereby mimicking the positive correlations visible in the upper left panel of Fig.\ \ref{fig:vecmu}), the charge densities in Eq.\ \eqref{charge_correlations_example2} decrease accordingly, while still maintaining the approximate ratio $\rho_B/\rho_Q \approx 2$ and $\rho_S \approx 0$.
We conclude that a correlation between two chemical potentials (in this case, $\mu_B$ and $\mu_S$) \textit{cannot} be naively interpreted as a correlation between the corresponding charge densities (i.e., $\rho_B$ and $\rho_S$).
Again, this is ultimately a consequence of the fact that different quark flavors carry different charge combinations.

In addition to being influenced by the non-trivial charge combinations carried by the $u$, $d$, and $s$ quarks, correlations between chemical potentials are further modified by the correlations implicit in the hadronic contributions to the Taylor EoS.
For example, for thermal models in global equilibrium, in general the following is approximately true \cite{Monnai:2021kgu}:
\begin{equation}
    \mu_Q \sim -0.1 \mu_B
\end{equation}
This is also influenced by the constraint that electric charge is conserved in the collision and most heavy-ion collisions have an initial condition of $Z/A \sim 0.4$.

We see, however, from Fig.\ \ref{fig:vecmu} (lower left panel) that once one considers the effect of $q\bar{q}$ fluctuations that we no longer have a clear hierarchy between $\mu_B$ and $\mu_Q$, such as one would expect in global equilibrium.  Rather, there are a number of fluid cells that find $\mu_B$ and $\mu_Q$ to have similar orders of magnitude.   

Likewise, in Fig.\ \ref{fig:vecmu} (upper left), the correlation between $\mu_S$ and $\mu_B$ agrees well with the scaling typically found in thermal models:
\begin{equation}
    \mu_S\sim \mu_B/3.
\end{equation}

The anti-correlation between $\mu_S$ and $\mu_Q$ visible in the lower right panel of Fig.\ \ref{fig:vecmu} can be understood in a similar way.  Like the above chemical potential correlations, it represents a complex interplay of charge correlations at both the partonic and hadronic levels.  Nuclear collisions therefore probe the QCD phase diagram in rich and highly non-trivial ways.

%
%
%==============================================================================
\subsection{Fluctuations of BSQ chemical potentials at freeze-out}
\label{sec:BSQDensityProfiles}
%==============================================================================

So far we have explored the spread in $T$ and $\vect{\mu}$ covered by BSQ fluctuations at the LHC during the evolution itself.
Given that Fig.\ \ref{fig:density-profile} exhibits a steady decrease in $\vect{\rho}(\tau)$ over time, however, one might worry that the BSQ fluctuations would be negligible by the time the system freezes out, so that their effects would be washed out in the hadronization process.
In order to assess this possibility, we now consider how large the fluctuations and correlations in chemical potential are anticipated to be at freeze-out.

\begin{figure}
    \centering
    \includegraphics[keepaspectratio, width=0.99\linewidth]{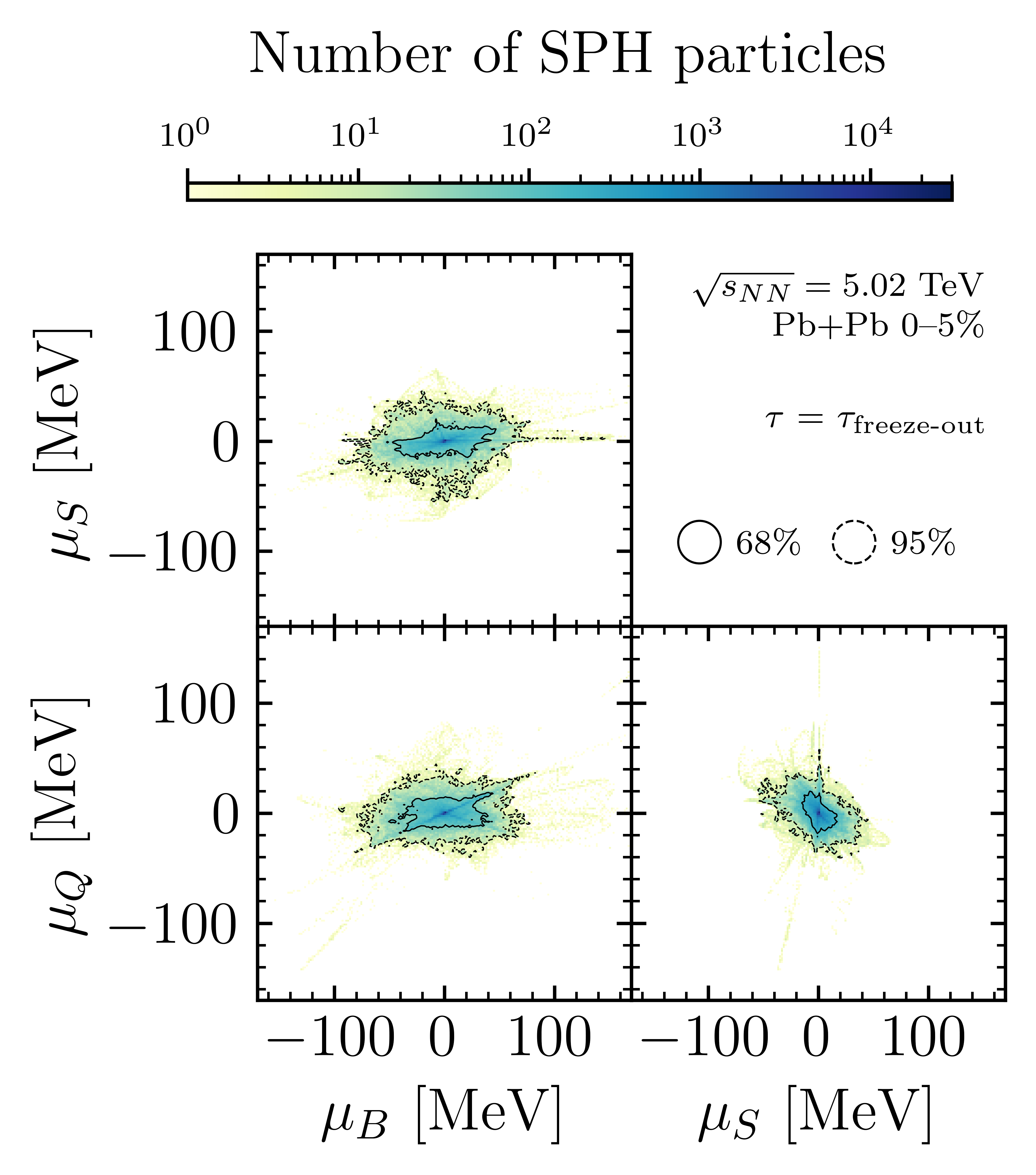}
    \caption{ 
    (color online) Two-dimensional histogram of the correlation of $\{\mu_S,\mu_B\}$ (top-left), $\{\mu_Q,\mu_B\}$ (bottom-left), and $\{\mu_Q,\mu_S\}$ (bottom-right) chemical potentials at freeze-out after hydrodynamical evolution using \ccake{} for a Pb\texttt{+}Pb system at $\sqrt{s_{NN}}=5.02$ TeV. 
    In all cases, a solid contour enclosing 68.27\% and a dashed contour enclosing 95.45\% of the SPH particles is displayed.
    }
    \label{fig:fo-vecmu}
\end{figure}

In Fig.\ \ref{fig:fo-vecmu} we show scatter plots of the correlations between all combinations of $\vect{\mu}$ at the point of freeze-out.  We emphasize that each SPH particle's chemical potentials are plotted at the proper time at which that particle freeze out.  We denote this in the Figure by writing $\tau = \tau_\mathrm{freeze-out}$, where $\tau_\mathrm{freeze-out}$ thus assumes different values for different particles.

We make two key observations.  First, note that the correlations present in the initial state (and observed above in Fig.\ \ref{fig:vecmu}) appear to be qualitatively preserved at freeze-out.  
For instance, a visual comparison of Figs.\ \ref{fig:vecmu} and \ref{fig:fo-vecmu} suggests a weakly positive correlation between $\mu_S$ and $\mu_B$ (upper left), a noticeably negative correlation between $\mu_Q$ and $\mu_S$, and no discernible correlation between $\mu_Q$ and $\mu_B$.  
The correlations born in the initial state are thus to some extent preserved in the final state.
However, the correlations appear to be significantly dampened by the time the final state, which leads up to believe that hydrodynamics has a smearing effect on the initial state correlations between $\vec{\mu}$ that are formed from an \iccing{} initial condition. 

Second, the range of fluctuations in Fig.\ \ref{fig:fo-vecmu} is considerably reduced with respect to that in Fig.\ \ref{fig:vecmu}.  
Using the 95\% contours to guide the eye, we estimate that that $|\mu_B| \lesssim 75$ MeV,  $|\mu_S| \lesssim 50$ MeV, and  $|\mu_Q| \lesssim 25$ MeV, in rough agreement with the estimates made using Fig.\ \ref{fig:Tmu_trajs}.  
The fluctuations in chemical potentials are thus reduced during the evolution by an approximate factor of 5-10, but are still non-vanishing. 
We therefore expect the final state to be a qualitatively accurate (but quantitatively suppressed) reflection of the chemical potential correlations in the initial state.  

%
%%%%%%%%%%%%%%%%%%%%%%%%%%%%%%%%%%%%%%%%%%%%%%%%%%%%%%%%%%%
\subsection{Densities at finite T}\label{sec:densityT}
%%%%%%%%%%%%%%%%%%%%%%%%%%%%%%%%%%%%%%%%%%%%%%%%%%%%%%%%%%
%
\begin{figure}
    \centering
    \includegraphics[width=\linewidth]{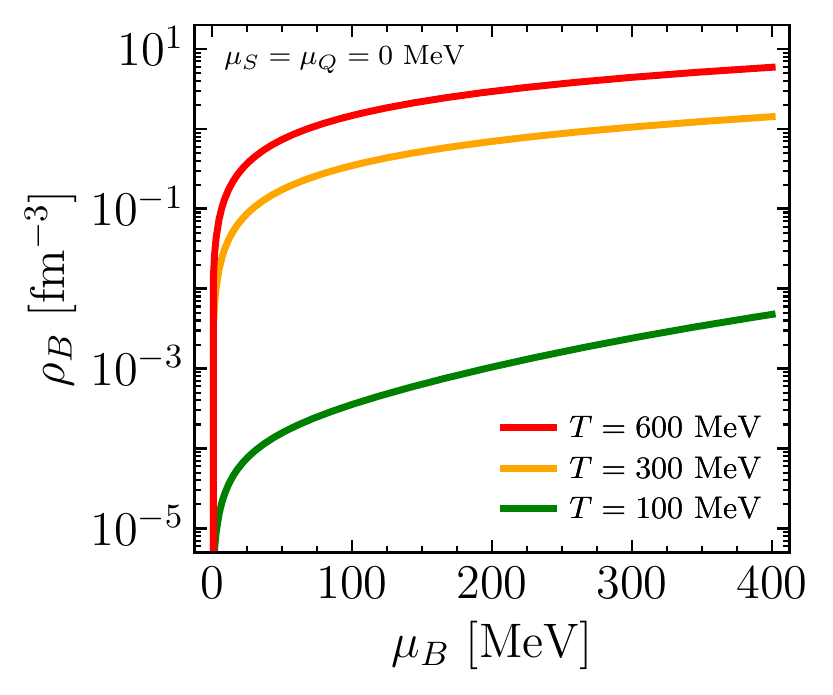}
    \caption{Baryon density $\rho_B$ vs baryon chemical potential along fixed slices of temperature. For a fixed $\mu_B$ the $\rho_B$ grows significantly with $T$.
    }
    \label{fig:nB_finiteT}
\end{figure}
A reader may wonder how it it possible that an \iccing{} event leads to such large $\vec{\rho}$ fluctuations in the initial state that is many times larger than $\rho_{sat}$.
However, let us consider the local density contribution of a single strange quark that carries $B=1/3$, $S=-1$, $Q=-1/3$. 
Let us assume this quark is produced with a radius of $r_s=0.5$ fm, then the density is
\begin{equation}
    \rho_X=\frac{X}{4/3\pi r^3}
\end{equation}

for $\rho_S=-1.9$ fm$^{-3}$ and $\rho_B=\rho_Q=0.6$ fm$^{-3}$, which is quite significant! 
In fact, from the production of just a single quark the local fluctuations end up being many times the nuclear saturation density. 
If multiple gluon splittings occur near each other then the local charge densities can be quite large, which is evident already in Fig.\ \ref{fig:GausKern} where fluctuations can reach up to $\rho_X=\pm 4$ fm$^{-3}$ $=25\,\rho_{sat}$. 
For strange quarks, it's easiest to see that the maximum densities are reached when just 2-3 $s\bar{s}$ pairs are produced close enough to each other that the densities overlap.

Another common questions is if these large local fluctuations of $\vec{\rho}$ are realistic in terms of the EoS? 
Aren't they many times denser than even the core of a neutron star (around $\sim 6\,\rho_{sat}$ vs $25\,\rho_{sat}$ that we find here)?
Indeed, it is true that the densities that we obtain in \iccing{} appear to be very large at first glance, however, one must realize that this is a feature of the EoS and the very non-trivial relationship between $\rho$ and $\mu$ at finite $T$. 

In Fig.\ \ref{fig:nB_finiteT} we plot $\rho_B$ vs $\mu_B$ along different $T$ slices. 
We find that for a fixed $\mu_B$ that increasing $T$ dramatically increases $\rho_B$. 
At $\mu_B=400$ MeV, for instance, $\rho_B$ ends up increases approximately 3 orders of magnitude from $T=100$ MeV to $T=600$ MeV.
Thus, these large densities do not translate to extremely large chemical potentials.
In fact, as we showed in Sec.\ \ref{sec:BSQDensityProfiles} the initial chemical potentials decrease significantly over time (similar to what is found from isentropic trajectories).
Thus, what may at first appear as extremely large densities produced from \iccing{} end up leading to very reasonable and relatively small final state fluctuations in chemical potentials.
%%%%%%%%%%%%%%%%%%%%%%%%%%%%%%%%%%%%%%%%%%%%%%%%%%%%%%%%%%%%%%%%%%%%%%%%%%%%%%%%%%%%%%%%%%%%%%%%%%%%%%%%%%%
\subsection{Effect of BSQ fluctuations on a single event: multiplicity, mean transverse momentum, and flow}\label{sec:h0}
%%%%%%%%%%%%%%%%%%%%%%%%%%%%%%%%%%%%%%%%%%%%%%%%%%%%%%%%%%%%%%%%%%%%%%%%%%%%%%%%%%%%%%%%%%%%%%%%%%%%%%%%%%%%

%
\begin{figure}
    \centering
    \includegraphics[keepaspectratio, width=\linewidth]{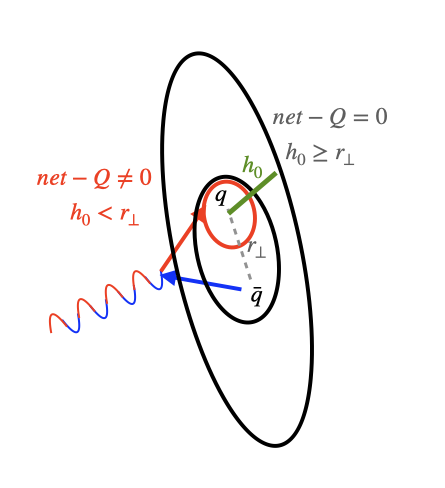}
        \caption{(color online) Cartoon of a gluon splitting into a quark anti-quark pair ($q\bar{q}$).  The separation distance between the $q\bar{q}$ pair is $r_\perp$.  The smoothing scale or rather the scale at which structure can be resolved is $h_0$ for the initial state. If the smoothing scale is larger than the separation scale ($h_0\geq r_\perp$), the charge fluctuations cannot be resolved and the initial state appears to have no charge fluctuations. If the smoothing scale is less than the separation scale ($h_0<r_\perp$), the charge fluctuations can be resolved and the initial state appears to have BSQ charge fluctuations.
        }
    \label{fig:scaleQQ}
\end{figure}
\begin{figure}
    \centering
    \includegraphics[keepaspectratio, width=\linewidth]{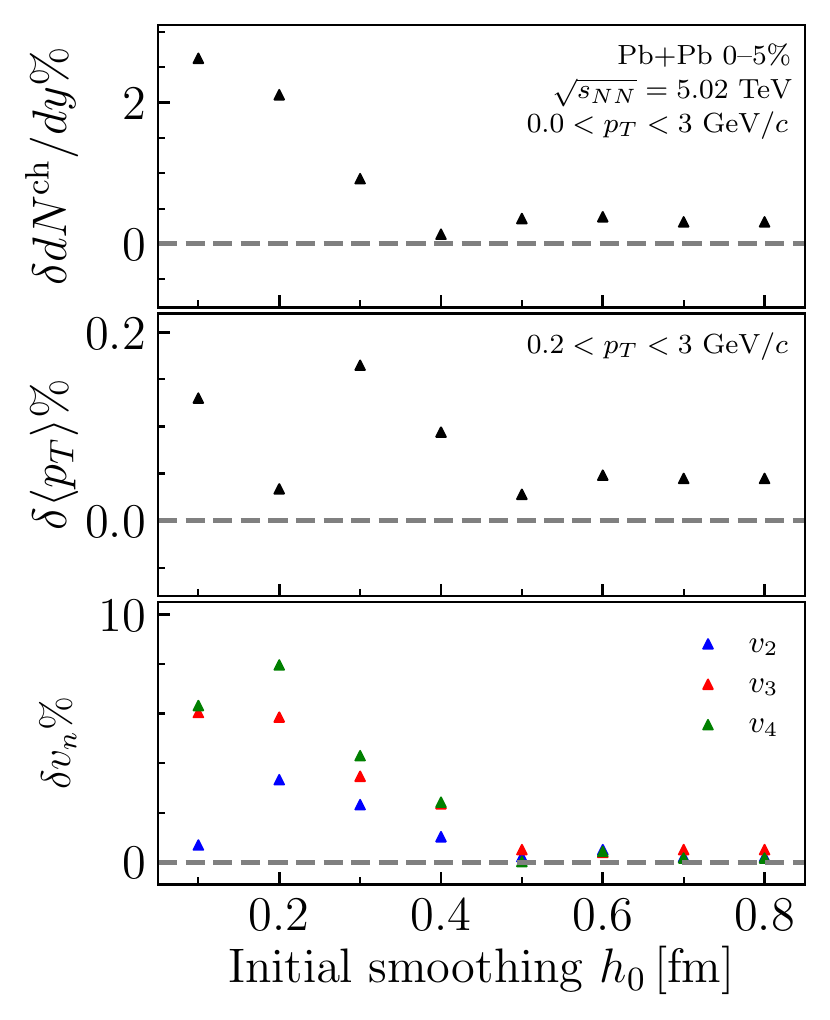}
        \caption{(color online) Here, $\delta \mathcal{O}\% \equiv 100 \cdot |\mathcal{O}_{\rm ICCING}-\mathcal{O}_{\rm TRENTo}|/\mathcal{O}_{\rm TRENTo}$. Initial smoothing of the \iccing{} events with different values of the smoothing parameter $h_0$ results in washing out the BSQ charge contribution to the flow. Both \iccing{} and \TRENTo{} events converge when smoothed by around $h_0 = 0.5$ fm. }
    \label{fig:initial_smoothing}
\end{figure}

The effect of the BSQ charge fluctuations is an inherently short-range correlation.  
The BSQ fluctuations arise from gluons splitting into quark anti-quark pairs and the distance between the splitting, $r_\perp$, determines the importance of the BSQ charge fluctuations as a function of length scale throughout the dynamical evolution. 
If $q\bar{q}$ pairs are produced at distances below the resolution scale of hydrodynamics, then their charges will cancel and the initial state will appear to have no charge fluctuations. 
However, if hydrodynamics can resolve structure below $r_\perp$ then the BSQ charge fluctuations will be relevant to the dynamics of a heavy-ion collision. 

Using the method of SPH we can test precisely this effect of the resolution of hydrodynamics on BSQ charge fluctuations.  
The SPH kernel function described in Sec.\ \ref{sec:SPHformalism} can be applied to just the initial state (as was previously done in \cite{Noronha-Hostler:2015coa,Gardim:2017ruc}), allowing us to define the initial smoothness scale as $h_0$ (which should not be confused with the usual smoothing scale $h=0.3$ fm used during the subsequent hydrodynamic evolution).  
Thus, we can then test our previous mentioned assumption that if $h_0< r_\perp$ we anticipate effects from BSQ charge fluctuations to play a role on experimental observables. 
However, if $h_0\geq  r_\perp$ then we expect that BSQ charge fluctuations will be washed out and there will be no effect on experimental observables. 
We summarize this discussion with an illustration in Fig.\ \ref{fig:scaleQQ} that demonstrates the interplay of $h_0$ and $r_\perp$ on a single $q\bar{q}$. 

The value of $r_\perp$ is sampled from the $q\bar{q}$ splitting probability which is a complicated function that depends on the underlying color glass condensate model, the strong coupling constant, and momentum transferred (see Fig.\ 2 in \cite{Carzon:2019qja}). 
Given our \iccing{} parameters used in this work, most $r_\perp\lesssim 1$ fm, although the function has a long tail that extends to longer distances as well. 
The peak of the probability function is around $r_\perp\sim 0.5$ fm such that we anticipate $h_0\sim 0.5$ fm values to be the relevant scale. 

Then, to test the relevant scale of \iccing{} we systematically vary $h_0$ for a single event and calculate the change in the multiplicity $dN^\mathrm{ch}/dy$, the mean transverse momentum $\langle p_T\rangle$, and the anisotropic flow $v_n$. 
The multiplicity is obtained from the all charged particle spectra wherein we have already integrated over the azimuthal angle $\phi$ such that our spectra varies only in $p_T$ and rapidity $y$.
Then, one can integrate over $p_T$ to obtain:
\begin{equation}
    \frac{dN^\mathrm{ch}}{dy} = \int_{0}^{p_{T}^\text{max}}dp_T\,  \frac{d^2N^\mathrm{ch}}{dp_Tdy},
\end{equation}

which is our all charged particle multiplicity. 

The $\langle p_T\rangle$ is obtained by weighting the spectra by $p_T$ and then normalizing by the multiplicity, i.e., 
\begin{equation}
    \langle p_T\rangle = \frac{\int_{p_{T}^\text{min}}^{p_{T}^\text{max}} p_T \frac{d^2N^\mathrm{ch}}{dp_Tdy}dp_T }{\int_{p_{T}^\text{min}}^{p_{T}^\text{max}}  \frac{d^2N^\mathrm{ch}}{dp_Tdy}dp_T},
\end{equation}
wherein the experiments use $p_T^{min}=0$ to $p_T=3\;[GeV]$ for the range of integration. 
Note that for $\langle p_T\rangle$ all integration must be in the same range, even if the spectra itself has been previously calculated across a wide range of $p_T$. 

Finally, we can characterize the size of the azimuthal anisotropies  by the Fourier coefficients:
\begin{equation}\label{eq:1}
\frac{dN^\mathrm{ch}}{d\phi} \propto 1 + 2\sum_{n=1}^{\infty} v_n \cos \left[n\left(\phi-\Psi_n \right)\right],
\end{equation}
where $v_n$ are the Fourier coefficients, $\phi$ is the azimuthal angle of a particle, and $\Psi_n$ defines the $n$-th order event plane.
These $v_n$ can be calculated in an individual event directly from the spectra:
\begin{equation}
    v_n^i=\frac{\int_{p_{T}^\text{min}}^{p_{T}^\text{max}} dp_T \int_0^{2\pi}d\phi\, \frac{d^3N^i}{dp_Tdyd\phi}\cos\left[n\left(\phi-\psi^i_n\right)\right] }{\int_{p_{T}^\text{min}}^{p_{T}^\text{max}}  dp_T\,\frac{d^2N^i}{dp_Tdy}},
\end{equation}
where the event-plane angle for an individual event is $\psi_n^i$. 
Further details can be found in Appendix C from \cite{Noronha-Hostler:2013gga0}.

In Fig.\ \ref{fig:initial_smoothing} the results for the percentage change of these three observables compared to the baseline of this given  \TRENTo{}+\ccake{} event are shown. 
When the $\delta \mathcal{O}\% \rightarrow 0$ for each observable $\mathcal{O}$, then this implies that there is no effect from \iccing{}. 
We can see that only if we can resolve the initial state below a scale of $h_0< 0.5$ fm do we see any significant effect from \iccing{}. 
The largest effect from \iccing{} is seen if one can resolve very small scale structure, i.e., $h_0\sim 0.1$ fm.
Our default SPH smoothing scale is $h=0.3$ fm that is equivalent to smoothing the initial scale to that same value, such that we anticipate small changes from \iccing{} at the percent level for $dN^\text{ch}/dy$ and $v_n$, but no discernible effect for $\langle p_T\rangle$.

As mentioned above, our default $h=0.3$ fm is used for the remainder of the paper.  
However, it would be interesting to study the effects of \iccing{} using even smaller $h$ to determine the influence of resolving \iccing{} at even shorter length scales. 
There is a somewhat delicate balance, though, because a smaller $h$ implies both the need for more SPH particles and large Knudsen and inverse Reynolds numbers \cite{Noronha-Hostler:2015coa}. 
Thus, we leave this study to a future work. 

%%%%%%%%%%%%%%%%%%%%%%%%%%%%%%%%%%%%%%%%%%%%%%%%%%%%%%%%%%%%%%%%%%%%%%%%%%%%%%%%%%%%%%%%%%
\section{Results: Comparisons to experimental data and predictions}
\label{sec:resultsII}
%%%%%%%%%%%%%%%%%%%%%%%%%%%%%%%%%%%%%%%%%%%%%%%%%%%%%%%%%%%%%%%%%%%%%%%%%%%%%%%%%%%%%%%%%%%%
 
In the following, we will compare to multiplicities, average transverse momentum $\langle p_T\rangle$, and flow of both all charged particles (ch) and identified particles (PID).  
Because PID results from the LHC only exists for the ALICE experiment at the moment, we focus solely on results for their momentum cuts $0.2<p_T<3$ [GeV].

For the purposes of the present paper, we are not interested in conducting a comprehensive scan of parameter space.
Instead, in order to compare the results for \TRENTo{}+\ccake{} and \TRENTo{}+\iccing{}+\ccake{}, we have only adjusted the normalization constant to reproduce the all charged particle spectra and included a reasonable value $T\eta/w=0.08$, as discussed above. 
We emphasize that these choices are made purely for illustrative purposes, and defer a Bayesian analysis to a future work. 
Additionally, we use identical \TRENTo{} events both in the \TRENTo{}+\ccake{} and \TRENTo{}+\iccing{}+\ccake{} such that direction comparisons between the two frameworks require lower statistics and one can be certain that any differences that show up are due to physics, not event-by-event fluctuations.

Our current framework differs in several minor ways from some of the recent Bayesian tunes which focus primarily on observables that correlate two particles or fewer (e.g., $v_n\left\{2\right\}$ or $dN^\mathrm{ch}/dy$). 
The most significant differences include: (i), that we omit a pre-equilibrium phase from our framework since incorporating conserved charges into this phase is still a work in progress (see initial steps in that direction \cite{Carzon:2023zfp,Dore:2023qxr}); (ii), that we do not include hadronic rescatterings; (iii), that we do not consider a finite bulk viscosity; and (iv), that we use a small value of the nucleon width $\sigma=0.3$ fm compared to $\sigma\sim 1$ fm extracted from recent Bayesian analyses \cite{JETSCAPE:2020shq,JETSCAPE:2020mzn,Bernhard:2019bmu}. 
The choice of a smaller $\sigma$ is motivated by recent studies on $\langle p_T\rangle $ correlations with $v_n$ that demonstrate a smaller $\sigma$ appears to be preferred \cite{ALICE:2021gxt,Giacalone:2021clp,Nijs:2022rme}.
Similar results are found for multiparticle cumulants in small systems \cite{Giacalone:2017uqx}.

\begin{figure}
    \centering
    \includegraphics[keepaspectratio, width=\linewidth]{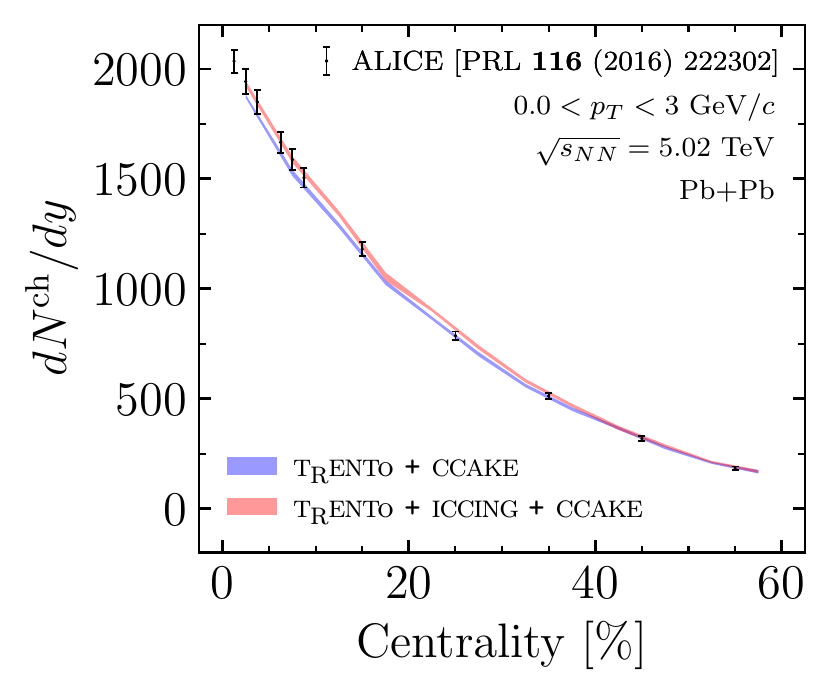}
        \caption{
        (color online) Multiplicity ($dN^\text{ch}/dy$) in Pb\texttt{+}Pb collisions at $\sqrt{s_{NN}}=5.02$  TeV for all charged particles as measured by the ALICE collaboration \cite{ALICE:2015juo} across all centralities compared to our theoretical results from \TRENTo{}+\ccake{} and \TRENTo{}+\iccing{}+\ccake{}.
        }
    \label{fig:allchg_dNdy}
\end{figure}

We begin our study by calculating the multiplicity of all charged particles at mid-rapidity $dN^\mathrm{ch}/dy$ across all centralities. 
The results of $dN^\mathrm{ch}/dy$ vs centrality compared to ALICE Pb\texttt{+}Pb data at $\sqrt{s_{NN}}=5.02$ TeV \cite{ALICE:2018vuu} are shown in Fig.\ \ref{fig:allchg_dNdy}. 
We use these results to fix our overall normalization constant for both \TRENTo{}+\ccake{} and \TRENTo{}+\iccing{}+\ccake{}. 
We found that using the same normalization constant $\,\mathcal{N}\,$

 provides results for \TRENTo{}+\ccake{} and \TRENTo{}+\iccing{}+\ccake{} that are both consistent with data, although that for \TRENTo{}+\iccing{}+\ccake{} is slightly higher.
The increase when one includes conserved charges has been discussed already in Sec.\ \ref{sec:norm}. 
Because of this small increase in $dN^\mathrm{ch}/dy$, we normalize all PID multiplicities by $dN^\mathrm{ch}/dy$ in order to ensure an apples-to-apples comparison of our framework with and without conserved charges.

%%%%%%%%%%%%%%%%%%%%%%%%%%%%%%%%%%%%%%%%%%%%%%%%%
%%%%%%%    SPECTRA AND COLLECTIVE FLOW   %%%%%%%%
%%%%%%%%%%%%%%%%%%%%%%%%%%%%%%%%%%%%%%%%%%%%%%%%%
\subsection{Identified particles in central collisions}
\label{sec:PID}
%================================================

We now compare our results directly to experimental data from central Pb\texttt{+}Pb collisions at $\sqrt{s_{NN}}=5.02$ TeV using identified particles from ALICE \cite{ALICE:2019hno,VazquezRueda:2019iuc,ALICE:2015juo,ALICE:2018rtz}.
The data is available from $0$--$5\%$ centrality for the light particles ($\pi$, $K$, $p$) and from $0$--$10\%$ centrality for the strange baryons ($\Lambda$, $\Xi$, $\Omega$). 

Identified particle yields are interesting for several reasons:
\begin{enumerate}
    \item The proton-to-pion ($p/\pi$) ratio puzzle.
    LHC particle yields have seen tension between the protons and strange baryons when comparing to thermal models, a tension which has been highlighted by poor fits to the $p/\pi$ ratio \cite{Floris:2014pta}.
    Many solutions have been proposed to this puzzle such as missing resonances \cite{Bazavov:2014xya,Noronha-Hostler:2014usa,Noronha-Hostler:2014aia}, multi-particle interactions \cite{Noronha-Hostler:2014usa,Noronha-Hostler:2014aia,Steinheimer:2015msa}, the $S$-matrix approach \cite{Andronic:2020iyg}, and two freeze-out temperatures \cite{Noronha-Hostler:2016rpd,Bellwied:2018tkc,Bluhm:2018aei,Bellwied:2019pxh,Alba:2020jir}.
    However, the solution is still an open question in the field.
    \item Strange baryons have also been notoriously hard to reproduce from hydrodynamic calculations \cite{Takeuchi:2015ana}, although an EoS with strangeness and electric charge can improve comparisons \cite{Monnai:2019hkn}.
    \item While we do not add net-strangeness or net-baryon number to the system, \iccing{} certainly leads to larger local fluctuations in BSQ, meaning there is a chance that \iccing{} calculations could produce too much of a specific type of particle.
    Comparing to identified particles thus provides an important benchmark check of the \iccing{} model.
\end{enumerate}

\begin{figure}
    \centering
    \includegraphics[keepaspectratio, width=\linewidth]{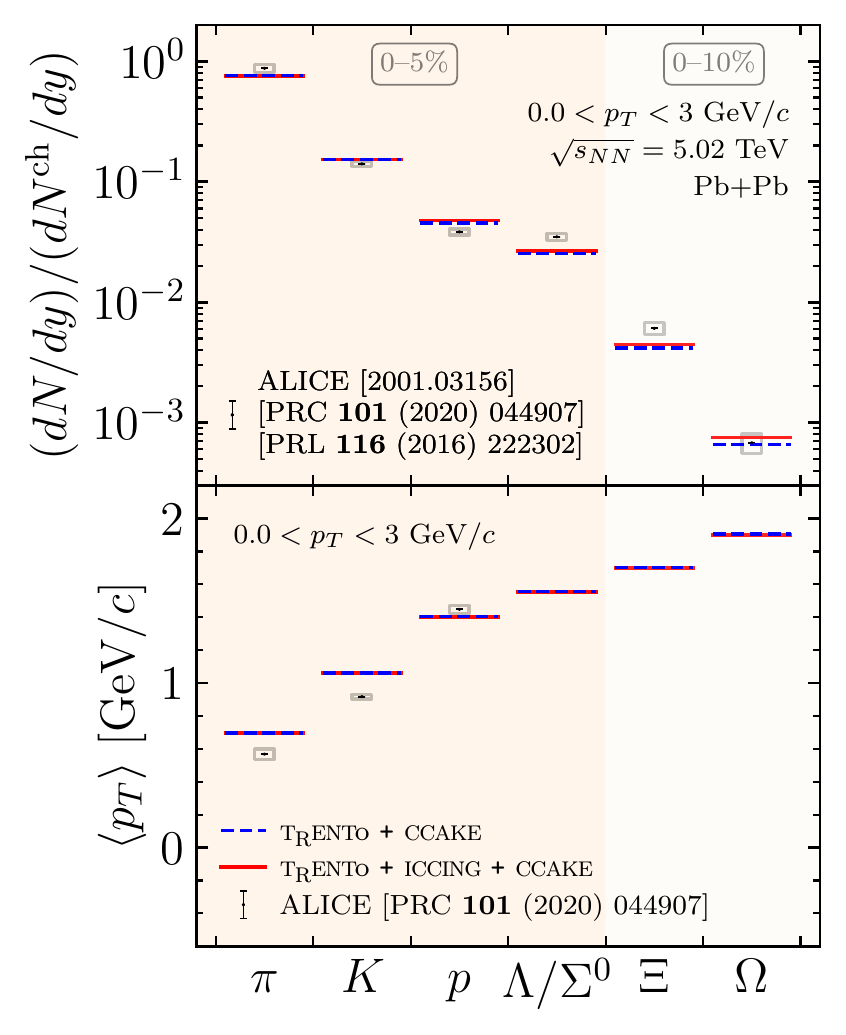}
        \caption{
        (color online) (Top) Identified particle multiplicities normalized by $dN^\text{ch}/dy$ at midrapidity and (bottom) $\langle p_T \rangle$ for central Pb\texttt{+}Pb collisions at $\sqrt{s_{NN}}=5.02$ TeV. 
        We use a shorthand notation when naming hadrons (e.g., $\pi$ refers to $\pi^+$ and $\pi^-$ and $\Xi$ refers to $\Xi^-$ and $\bar{\Xi}^+$, etc.); for multiplicities this represents the sum and for $\langle p_T \rangle$ this is the average of individual particle values.
        Solid red and dashed blue lines show the result of performing the hydrodynamic evolution with and without initial charged densities, respectively. 
        The $p_T$-cuts and the $0$--$10\%$ centrality bin used for $\Xi$ and $\Omega$ are chosen to reflect experimental reports. Experimental data points from Refs.~\cite{ALICE:2019hno,VazquezRueda:2019iuc,ALICE:2015juo}, including statistical and systematic uncertainties as error bars and boxes, respectively.
        }
    \label{fig:spec}
\end{figure}

In Fig.\ \ref{fig:spec} we compare the identified particle yields normalized by $dN^\text{ch}/dy$ (top) and the mean transverse momentum $\langle p_T\rangle$ (bottom) computed in \TRENTo{}+\ccake{} and \TRENTo{}+\iccing{}+\ccake{} with the corresponding ALICE data in central Pb+Pb collisions.
The particles yields are computed by averaging the yields for degenerate particles i.e. $\pi=(\pi^+ +\pi^-)/2$.  
For the ratio of identified particle multiplicities normalized by all charged particles i.e. $(dN^{i}/dy)/(dN^\text{ch}/dy)$ we have propagated the experimental error ourselves.  
Because there is a significant amount of correlated error between these observables, the experimental error is overestimated in our approach since we do not have access to the covariance between different uncertainties. 
The statistical error in the theoretical calculations is determined using jackknife resampling. 
However, the statistical error is extremely small and is not visible in the multiplicity plots.

We find that our comparisons to the experimental data fit quite well from  \TRENTo{}+\iccing{}+\ccake{}.  
Generally, the yields are about $10$--$20\%$. 
The contribution from \iccing{} appears to play almost no role at all for the overall yields, although there is a very tiny enhancement in $\Omega$'s. 
Thus, the overall yields by themselves do not provide any additional motivation to include quark degrees of freedom in the initial state. 

We still observe tension between the protons and strange baryons.  
The protons are slightly over-predicted, which may be a sign of needing a lower freeze-out temperature, and the strange baryons are slightly under-predicted, possibly indicating the need for a higher freeze-out temperature. 

These results are consistent with some of the studies which have addressed the previously mentioned $p/\pi$ puzzle by arguing for assigning a lower freeze-out temperature to light particles and a higher freeze-out temperature to strange particles. 
Including two freeze-out temperatures with our current parameter set would thus likely bring our model closer to the experimental data.
However, at this time we do not draw any strong conclusions because one should study the effects of varying other hydrodynamic parameters first.

Also in Fig.\ \ref{fig:spec} we compare $\langle p_T\rangle$ for identified particles in central collisions.  
Experimental results are presently available only for $\pi$, $K$, $p$.
We find that the proton $\langle p_T\rangle$ fits nearly perfectly compared to the experimental data, but that the pion and kaon theory  predictions slightly overshoot the data (around $20\%$).  This is likely a consequence of our choice to show only results without any bulk viscosity, since it has been shown previously that bulk viscosity can help with decreasing $\langle p_T\rangle$ in simulations \cite{Ryu:2015vwa}. 
Our smaller chosen nucleon width also likely  contributes to these larger $\langle p_T\rangle$ compared to the Bayesian results.
Additionally, in this work we have used the PDG2016+ list, whereas the PDG2021+ may slightly affect these results \cite{SanMartin:2023zhv} as well. 

%%%%%%%%%%%%%%%%%%%%%%%%%%%%%%%%%%%%%%%%%%%%%%%%%
%%%%%%%    SPECTRA AND COLLECTIVE FLOW   %%%%%%%%
%%%%%%%%%%%%%%%%%%%%%%%%%%%%%%%%%%%%%%%%%%%%%%%%%
\subsection{Collective flow of all charged particles}
\label{sec:AnisotropyOfBSQDensities}
%================================================

In the following,  we will compare directly to collective flow experimental data from Pb\texttt{+}Pb $\sqrt{s_{NN}}=5.02$  TeV from all charge particles $v_n\left\{2\right\}$ from ALICE in \cite{ALICE:2018rtz}.
As above, for all results we compare the \TRENTo{}\texttt{+}\ccake{} framework, in which all BSQ charge fluctuations are set to 0, with the \TRENTo{}\texttt{+}\iccing{}\texttt{+}\ccake{} set up, in which BSQ charge fluctuations are included. 
We emphasize that our \TRENTo{}\texttt{+}\ccake{} results are slightly different from past \TRENTo{}\texttt{+}\vUSPhydro{} results because we use a different tuned normalization constant here, a smaller nucleon width, and a larger $\eta/s$ at $\vect{\mu} = 0$.
Additionally, the results in this section have been obtained with a low statistics run, since our goal is primarily to understand the effects of including \iccing{} in the simulations. 
Later work will study in more detail how to better tune parameters for the inclusion of \iccing{} and the effect of different transport coefficients on fluctuations in $\vect{\mu}$.

Let us first discuss how one calculates the 2-particle collective flow observables of all charged particles, which is relevant because identified particles are calculated differently. 
The collective flow of the reference particles can be characterized by the flow vector:
\begin{equation}\label{eqn:Vn_flow}
    V_n=v_n e^{in\Psi_n},
\end{equation}
where $v_n$ is the magnitude of the flow vector and $\Psi_n$ is the event plane angle for that specific event.
For a given event, the collective flow observable should not depend on the individual event plane angle because the orientation of each collision is random.
Thus, we want rotationally invariant observables.  
The simplest possible such observable that can be created with $V_n$ is a 2-particle correlation i.e.
\begin{equation}\label{eqn:2part}
    \langle V_nV_n^*\rangle = v_n^2.
\end{equation}

The angular brackets $\langle \dots \rangle$ implies an averaging over all pairs (within specific kinematic cuts) in a single event. 
The 2-particle correlation in Eq.\ (\ref{eqn:2part}) is for a single event and must be averaged over many events to obtain a global observable. 
The averaging over many events typically includes assigning a weight $w_i$ to the $i$th event such that for an observable $\mathcal{O}$:
\begin{equation}\label{eqn:weights}
    \langle \langle \mathcal{O}\rangle_i \rangle_\text{events} =\frac{\sum_i^{N_\text{ev}}w_i \langle \mathcal{O}\rangle_i}{\sum_i^{N_\text{ev}}w_i},
\end{equation}
where an averaging over pairs occurs first within a single event and then another averaging occurs over the ensemble of events.

The most common weight is the multiplicity which we define here for a single event by $M_i=(dN^\text{ch}/dy)_i$, such that for a 2-particle correlation
\begin{equation}
    w_i=M_i(M_i-1),
\end{equation}
This skews events towards the higher multiplicity events (see \cite{Bilandzic:2012wva,Bilandzic:2010jr,Bilandzic:2013kga} for the experimental observables and \cite{Gardim:2016nrr} for the consequences).
The results in this paper include multiplicity weighting.
It is also typical in experimental analyses to include centrality re-binning, where events are originally binned into $1\%$ bins and then recombined into larger centrality classes. Due to our low statistics runs here  we will not include that effect in this work. 

The actual experimental observable calculated is then the root-mean-squared, i.e.,
\begin{equation}
    v_n\left\{2\right\}=\sqrt{\langle v_n^2\rangle },
\end{equation}
where the $\left\{2\right\}$ indicates that it is a 2-particle correlation.  
This $ v_n\left\{2\right\}$ relates the second moment, $\langle v_n^2\rangle$, of the $v_n$ distribution to the $2^\text{nd}$ cumulant of the distribution.
In \cite{Luzum:2012da} a detailed discussion on why it is important to calculate the actual cumulant and not just the mean (or the event-plane method) of the $v_n$ distribution is outlined.
Additionally, we point interested readers in the direction of \cite{Luzum:2013yya} for more discussion on the collective flow observables.
It is worth noting that our statistics are relatively low (only 1200 events across all centralities) such that it is not yet possible to calculate 4-particle cumulants.

\begin{figure}
    \centering
    \includegraphics[keepaspectratio, width=\linewidth]{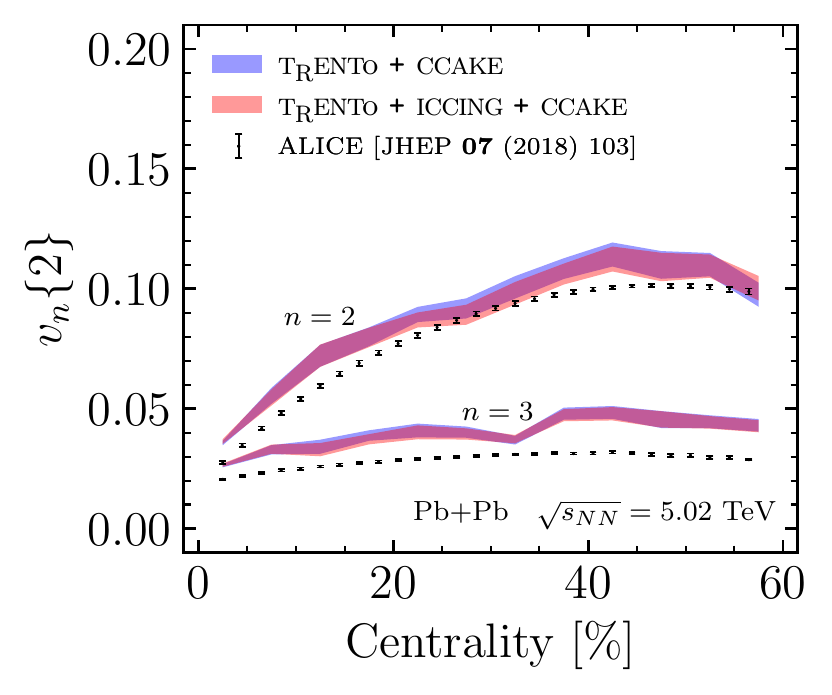}
        \caption{(color online) Elliptical and triangular flow vs centrality of integrated all charge flow $v_n\left\{2\right\}$ from $0.2\leq p_T \leq 3.0$ GeV/$c$ compared to ALICE results from Pb\texttt{+}Pb $\sqrt{s_{NN}}=5.02$  TeV collisions \cite{ALICE:2018rtz}.
        }
    \label{fig:collectiveflow}
\end{figure}

In Fig.\ \ref{fig:collectiveflow} one can see the results for all charged particles collective flow compared to the ALICE data. 
Our theoretical calculations provide a reasonably good fit to $v_2\left\{2\right\}$ and are slightly above the $v_3\left\{2\right\}$ data from ALICE.  
In principle, one could likely get closer to the data by slightly increasing $\eta/s$ at $\vect{\mu} = 0$ and/or including a finite bulk viscosity.  
For this work, we argue that these results are close enough since we are more interested in differences that appear when including \iccing{}, rather than exact fits to experimental data.
We see essentially no difference in $v_n\left\{2\right\}$ results for all charged particles with or without the inclusion of \iccing{}, which is consistent with \cite{Carzon:2019qja} wherein the energy density distribution is not altered by \iccing{}. 
Instead, we expect the effects from \iccing{} to show up in observables sensitive to charge fluctuations.

%

%%%%%%%%%%%%%%%%%%%%%%%%%%%%%%%%%%%%%%%%%%%%%%%%%
%%%%%%%    SPECTRA AND COLLECTIVE FLOW   %%%%%%%%
%%%%%%%%%%%%%%%%%%%%%%%%%%%%%%%%%%%%%%%%%%%%%%%%%
\subsection{Collective flow of identified particles (one particle of interest)}
\label{sec:flowPID}
%================================================

As a next step we study the collective flow of identified particles using 2-particle correlations. 
 Unfortunately, experimental data does not yet exist for Pb\texttt{+}Pb 5.02 TeV for  collective flow of identified particles so our results must be taken as predictions. 
 We focus solely on central collisions here since this is where we anticipate the largest effects from \iccing{} and also this centrality should have the best statistics for identified particles. 

Due to lower statistics of identified particles, the standard approach is to consider 1 reference particle (this is from all charged particles) and 1 particle of interest (POI) (this is the identified particle that we are interested in).  
Thus, our 2-particle correlation is then inherently a correlation between our background energy density and our conserved charge(s) - depending on which particle we're observing.
For instance, a $\pi^+$ only carries one conserved charge ($Q=+1$) in contrast to $\Xi^-$ that carries all 3 conserved charges ($B=+1$, $S=-1$, $Q=-1$).

For identified particles, the calculation of flow harmonics differs slightly from what was shown in Sec.\ \ref{sec:AnisotropyOfBSQDensities}.  Identified particle flow harmonics are two-particle correlations wherein one correlates a reference particle (i.e., typically a charged particle within a low $p_T$ cut-off like $0.2 \leq  p_T \leq 3$ GeV/$c$) and a particle of interest or POI (in our case, we consider identified particles like a kaon within the same $p_T $ cut-off, i.e., $0.2 \leq  p_T \leq 3$ GeV/$c$). The flow of the reference particles can be characterized by the flow vector (as discussed previously):
\begin{equation}
    V_n=v_n e^{in\Psi_n},
\end{equation}
where $v_n$ is the magnitude of the flow vector and $\Psi_n$ is the event plane angle for that specific event. 
The POI have their own flow vector that we define as:
\begin{equation}
    V_n^\prime=v_n^\prime e^{in\Psi_n^\prime},
\end{equation}
where $v_n^\prime$ is the magnitude of the POI flow vector and $\Psi_n^\prime$ is the POI event plane angle for that specific event. 
Generally, we always use $^\prime$'s to indicate the number of POI in our observables. 
For instance, an observable with $^{\prime\prime}$ indicates two POI.
Then, to calculate a two particle observable (with 1 reference particle and 1 POI) we correlate the two particles such that:
\begin{eqnarray}\label{eqn:1POI}
    v_n^\text{1POI}\left\{2\right\}&=&\frac{\langle V_n \left(V_n^\prime\right)^*\rangle }{v_n\left\{2\right\}}\\
    &=&\frac{\langle V_n \left(V_n^\prime\right)^*\rangle }{v_n\left\{2\right\}}\\
    &=&\frac{\langle v_n v_n^\prime \cos n\left(\Psi_n-\Psi_n^\prime\right)\rangle }{v_n\left\{2\right\}},
\end{eqnarray}
where $v_n\left\{2\right\}$ is the all charge particle 2-particle collective flow (the same that was calculated for the reference particles in Fig.\ \ref{fig:collectiveflow}) and the numerator contains the angular difference between the reference and the POI flow harmonics.
The reason that the 1 POI is normalized by $v_n\left\{2\right\}$ is to cancel out any effect on the overall magnitude of $v_n^\text{1POI}$ that would come from the all charge particle $v_n\left\{2\right\}$. 

The averaging over events is done in the same way as was done in Eq.\ (\ref{eqn:weights}) but the weights themselves are changed. 
For 1 reference particle and 1 POI then the weights for a given event $i$ are a mixture 
\begin{equation}
    w_i=M_i\, m_i,
\end{equation}
where $m_i=dN^i/dy$ is the multiplicity of an identified particle of interest, e.g., pions and $M_i$ is the same as before.

Recent works have begun to explore the possibility of 2 POIs.  ALICE has already been able to calculate 2POI in regards to $p_T$ dependence observables \cite{ALICE:2022dtx} in the soft sector.  
In \cite{Holtermann:2023vwr} a large number of new observables were proposed for an arbitrary number of POI for multiparticle correlations (relevant for high $p_T$, identified particles, and rapidity dependent observables etc) and in \cite{Holtermann:2024vdw} the role these 2+ POI observables play in understanding jet physics was explored. 
The motivation for these new studies with 2+ POI is that the LHC has accumulated enormous data sets such that these observables may now be (or soon be) possible.
Using these other studies as a motivation, we also explore the 2POI scenario wherein the 2-particle correlation becomes
\begin{eqnarray}
    v_n^\text{2POI}\left\{2\right\}&=&\langle V_n^\prime \left(V_n^\prime\right)^*\rangle \\
    &=&\sqrt{\langle \left(v_n^{\prime\prime}\right)^2\rangle },
\end{eqnarray}
because the event plane angles are again $100\%$ correlated such that one only takes the root-mean-squared.
Also, one does not need to normalize by the all charge particle $v_n\left\{2\right\}$ because $V_n$ does not directly enter this calculation.

Once again, the averaging over events is done in the same way as was done in Eq.\ (\ref{eqn:weights}) but the weights differ. 
For 2 POI then the weights for a given event $i$ are a mixture,
\begin{equation}
    w_i=m_i(m_i-1),
\end{equation}
such that only the multiplicity of the identified particle of interest $m_i$ enters the calculation.

\begin{figure}
    \centering
    \includegraphics[keepaspectratio, width=\linewidth]{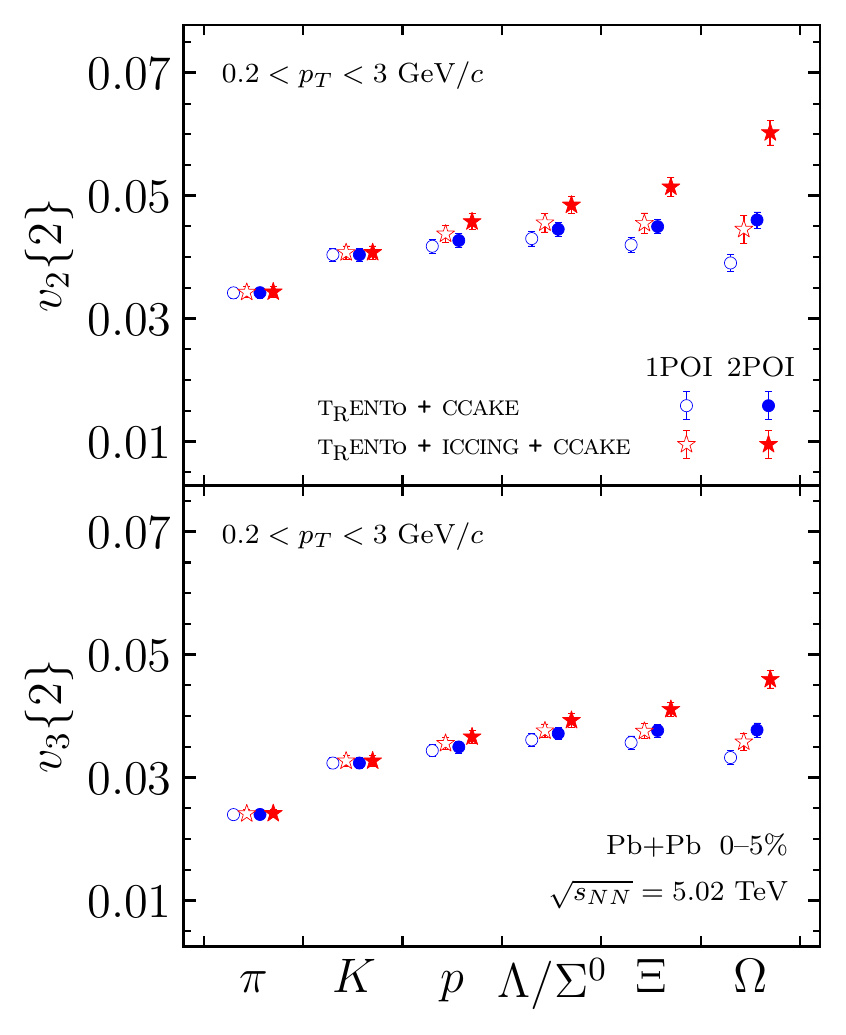}
    \caption{(color online) Elliptical (top) and triangular flow (bottom) in central collisions $0$--$5\%$ for identified particles calculated with and without \iccing{}. We compare here calculate the flow harmonics with 1 particle of interest vs 2-particles of interest.
    }
    \label{fig:vnPID}
\end{figure}

In Fig.\ \ref{fig:vnPID} the $v_n^\text{PID}\left\{2\right\}$ for Pb\texttt{+}Pb at $\sqrt{s_{NN}}=5.02$ TeV in the $0$--$5\%$ centrality class for pions, kaons, protons, lambdas and neutral sigmas, cascades, and omegas.
We only show the $0$--$5\%$ centrality class because the effect of \iccing{} is the strongest in most central collisions. 
In Fig.\ \ref{fig:vnPID} we compare both $v_n$ calculated using 1 or 2 POI, comparing \TRENTo{}+\ccake{} vs \TRENTo{}+\iccing{}+\ccake{}. 
For mesons, we have essentially no effects if we calculate our results with 1 or 2 POI, nor do we see an effect from \iccing{}.
However, for baryons (and especially strange baryons), we begin to see effects from \iccing{}. 
There is a very tiny effect that is not yet clearly statistically significant with and without \iccing{} if one considers 1POI where \iccing{} enhances $v_n$ (the effect appears larger in elliptical flow than triangular flow). 
For 1 POI the best chance of seeing this effect is for particles with the most conserved charges, i.e. $\Omega$'s, although smaller effects appear in protons, lambdas/sigmas, and cascades. 
We consistently find that \iccing{} increases the $v_n$, which agrees with what was found from the strangeness eccentricity calculations in \cite{Carzon:2019qja}.

However, a much better way to see the effect of \iccing{} is to calculate $v_n$ using 2 POI. 
In Fig.\ \ref{fig:vnPID} we find that there is a significant difference (compare the blue to brown markers) for lambdas/sigmas, cascades, and omegas that likely could be measurable. 
The difference is most pronounced for heavy strange baryons and elliptical flow but can be measured in $v_3$ as well. 

The reason that 2 POI is more sensitive to quark anti-quark fluctuations in the initial state (i.e., \iccing{}) and 1 POI is not that sensitive, is because of the nature of the scalar product in Eq.\ (\ref{eqn:1POI}). 
In Eq.\ (\ref{eqn:1POI}) if the event plane angles are decorrelated between the POI and the reference particle, then $v_n^\text{1POI}$ is suppressed even if $v_n^\prime$ is large. 
That is exactly what happens with \iccing{}. 
For \iccing{} the quark/anti-quark pairs are produced isentropically around the location of the gluon that split such that the total event plane angle may look very different for strangeness vs identified particles that are more sensitive to the energy density distribution. 
Thus, the $\cos$ term suppresses the magnitude of $v_n^\text{1POI}$, masking the effect of \iccing{}.
However, in the case of 2 POI then the strangeness event plane angle can be completely misaligned with that of the all charged particles (or rather the energy density) and the signal of $v_n^\prime$ still remains. 
Thus, flow calculations with 2 POI are important to observe effects of BSQ charge fluctuations.

%%%%%%%%%%%%%%%%%%%%%%%%%%%%%%%%%%%%%%%%%%%%%%%%%
%%%%%%%%%%%%%%%%    CONCLUSIONS   %%%%%%%%%%%%%%%
%%%%%%%%%%%%%%%%%%%%%%%%%%%%%%%%%%%%%%%%%%%%%%%%%
\section{Conclusions}
\label{sec:Conclusions}

In this work, we have presented an upgraded relativistic viscous hydrodynamic code called \ccake{} based on the SPH formalism with 3 conserved charges: baryon, strangeness, and electric charge to be used in heavy-ion collisions.  The code is then used to study the influence of BSQ charge fluctuations that arise from gluon splittings into $q\bar{q}$ pairs produced in the initial state.  The gluon distributions are described using \TRENTo{} initial conditions and are coupled to \iccing{} in order to generate BSQ charge fluctuations.

While the effect of BSQ charge fluctuations are significantly dampened throughout the hydrodynamic expansion, we find that signatures of these fluctuations still remain at freeze-out.  For instance, fluctuations in the local $\mu_B$ chemical potentials range from $\mu_B\sim \pm 400$ MeV in the initial state and are dampened out to a range of $\mu_B\sim \pm 50$ MeV.  Of course, the global $\mu_B^\text{global}=0$ MeV but we find non-trivial fluctuations around this global value.  Similar results are found for $\mu_S$ and $\mu_Q$,  Furthermore, we find non-trivial correlations between chemical potentials themselves that emphasizes the importance of developing EoS that include wide ranges of $\left\{\mu_B,\mu_S,\mu_Q\right\}$, not just along specific trajectories. 

The \TRENTo{}\texttt{+}\iccing{}\texttt{+}\ccake{} framework allows for direct comparisons to experimental data of multiplicity and collective flow of both all charged particles and identified particles.
We found two significant results in these comparisons: (i) BSQ charge fluctuations lead to an enhancement of $v_n^\text{2POI}\left\{2\right\}$ when 2-particles of interest are considered for strange baryons (ii) the remaining particle multiplicities and collective flow of all charged particles (and even those with 1 particle of interest) are not affected by BSQ charge fluctuations.
With future upgrades at the LHC and higher luminosity, it is likely possible to measure these observables with 2 POI to test the sensitive to the produce of quark anti-quark pairs in the initial state of heavy-ion collisions.

We emphasize that our results are the first study on the effect of these BSQ charge fluctuations within relativistic viscous hydrodynamics relevant for LHC energies.  Thus, these results open the door for many future explorations.  First, certain improvements to \ccake{} can be made in the future such as including second-order transport coefficients as well as the full BSQ charge diffusion matrix.  Additionally, it would be interesting to investigate out-of-equilibrium contributions to \iccing{} (coupled to something like \KOMPOST{}) and what role these play through the passage of the QCD phase diagram.  In this study, we only explored a limited number of events and centrality classes.  Future work should, of course, probe a much wider range of data, taking full advantage of the wealth of data that exists at the LHC and RHIC. While most of the fluids cells are located well within the range of the lattice QCD EoS expansion, it may be of interest to also explore EoS that include a critical point such as \cite{Parotto:2018pwx,Karthein:2021nxe}. Finally,  because we already know differences in the extracted viscous transport coefficients, a better understanding of eccentricities when BSQ fluctuations are taken into account are warranted. Overall, these BSQ charge fluctuations provide a new window into the QGP wherein we can probe not just the hydrodynamic response to the energy density but also study charge fluctuations even at LHC energies where the system is better understood and more well-controlled with a wealth of high-statistics data as compared to lower beam energies.

%%%%%%%%%%%%%%%%%%%%%%%%%%%%%%%%%%%%%%%%%%%%%%%%%
%%%%%%%%%%%%%    ACKNOWLEDGEMENTS   %%%%%%%%%%%%%
%%%%%%%%%%%%%%%%%%%%%%%%%%%%%%%%%%%%%%%%%%%%%%%%%
%================================================
\section{Acknowledgements}
%================================================
The authors would like to thank Jorge Noronha and Gabriel Denicol for providing useful discussions on the hydrodynamic analytical solutions.
Several of the authors (C.P., D.A., T.D., and J.N.H.) gratefully acknowledge the hospitality of the Kavli Institute for Theoretical Physics during “The Many
Faces of Relativistic Fluid Dynamics” Program, while this manuscript was being completed.  This research was supported in part by the National Science Foundation under Grant No. NSF PHY-1748958.
J.N.H. acknowledges support from the US-DOE Nuclear Science Grant No. DE-SC0020633 and  DE-SC00238 and from the Illinois Campus Cluster, a computing resource that is operated by the Illinois Campus Cluster Program (ICCP) in conjunction with the National Center for Supercomputing Applications (NCSA), and which is supported by funds from the University of Illinois at Urbana-Champaign.  This work was supported in part by the National Science Foundation (NSF) within the framework of the MUSES collaboration, under grant number OAC-2103680.
D.M is supported by the National Science Foundation Graduate Research Fellowship Program under Grant No. DGE – 1746047, the Illinois Center for Advanced Studies of the Universe Graduate Fellowship, and the University of Illinois Graduate College Distinguished Fellowship.
%================================================

%%%%%%%%%%%%%%%%%%%%%%%%%%%%%%%%%%%%%%%%%%%%%%%%%
%%%%%%%%%%%%%%%    BIBLIOGRAPHY   %%%%%%%%%%%%%%%
%%%%%%%%%%%%%%%%%%%%%%%%%%%%%%%%%%%%%%%%%%%%%%%%%
\bibliography{inspire,NOTinspire}

\appendix
\onecolumngrid
%%%%%%%%%%%%%%%%%%%%%%%%%%%%%%%%%%%%%%%%%%%%%%%%%
%%%%%%%%%%%%%%%%    APPENDIX A   %%%%%%%%%%%%%%%%
%%%%%%%%%%%%%%%%%%%%%%%%%%%%%%%%%%%%%%%%%%%%%%%%%
\section{Thermodynamic relations for the complete EoS}
\label{app:thermoRel}
%==============================================================================

In the following, we outline the required thermodynamic relations to construct the full EoS given the dimensionless pressure at a fixed point in temperature and chemical potentials, $p/T^4\left(T,\vect{\mu}\right)$.  A similar useful summary for the case of a single conserved charge only is given in Ref.~\cite{Floerchinger:2015efa}.

For any EoS, the starting point is always a definition of the pressure $p$ in terms of the temperature $T$ and any chemical potentials $\vect\mu$ (where the vector notation allows for multiple chemical potentials):
\begin{align*}
    p = p\l( T, \vect\mu \r)
\end{align*}
All quantities in this appendix depend explicitly on $T$ and $\vect\mu$ unless otherwise indicated.  For notational clarity below, we suppress this explicit dependence on $T$ and $\vect\mu$, and use capital letters (e.g., $X$, $Y$) to designate arbitrary conserved charges.  We also denote partial derivatives using standard subscript notation, e.g.,
\begin{equation}
    \partial^2_{A,B}f = \frac{\partial^2 f}{\partial A \partial B}
\end{equation}

Once the form of $p$ has been specified, the entropy density $s$ and charge densities $\vect\rho$ are obtained by taking suitable derivatives in the thermodynamic coordinates:
\begin{equation*}
    s = \l.\partial_T p \r|_{\mu_X} \quad \text{ and } \quad \l.\rho_X = \partial_{\mu_X} p\r|_{T,\mu_Y,\mu_Z}
\end{equation*}
Using these densities, we can determine $\varepsilon$ using the Gibbs relation:
\begin{equation*}
   \varepsilon = -P + s T + \sum_X \mu_X \rho_X
\end{equation*}

The corresponding second-order derivatives define a symmetric matrix of thermodynamic susceptibilities:
\begin{equation}
    \begin{pmatrix}
    \chi_{TT} & \chi_{TB} & \chi_{TS} & \chi_{TQ}\\
    \chi_{BT} & \chi_{BB} & \chi_{BS} & \chi_{BQ}\\
    \chi_{ST} & \chi_{SB} & \chi_{SS} & \chi_{SQ}\\
    \chi_{QT} & \chi_{QB} & \chi_{QS} & \chi_{QQ}
    \end{pmatrix}
    = \begin{pmatrix}
    \partial^2_{T,T}p & \partial^2_{T,B}p & \partial^2_{T,S}p & \partial^2_{T,Q}p\\
    \partial^2_{B,T}p & \partial^2_{B,B}p & \partial^2_{B,S}p & \partial^2_{B,Q}p\\
    \partial^2_{S,T}p & \partial^2_{S,B}p & \partial^2_{S,S}p & \partial^2_{S,Q}p\\
    \partial^2_{Q,T}p & \partial^2_{Q,B}p & \partial^2_{Q,S}p & \partial^2_{Q,Q}p\\
    \end{pmatrix}
\end{equation}

For each susceptibility ($\chi_{ab} = \partial p/ \partial a \partial b$), thermodynamic coordinates not used to take the derivatives are held constant. For instance, if $a=T$ then $\mu_{B,S,Q}$ are held constant. The generalization of these identities to include additional conserved charges is straightforward.

The final necessary ingredient in our code is the squared speed of sound $c_s^2$.  For the conformal equations of state, this is simply $1/3$.  For the non-conformal equations of state, the expression for $c_s^2$ is vastly more complicated and was first derived in \cite{Parotto:2019iro}.  It is defined by \cite{Noronha-Hostler:2019ayj}
\begin{equation}
    c_s^2 = \l. \partial_\varepsilon p \r|_{\rho_X} + \sum_X \frac{\rho_X}{\varepsilon + p} \l. \partial_{\rho_X} p \r|_{T,\rho_{Y \neq X}}
\end{equation}
This expression can be evaluated by a straightforward (albeit, tedious) expansion in terms of the other thermodynamic quantities described above.

%%%%%%%%%%%%%%%%%%%%%%%%%%%%%%%%%%%%%%%%%%%%%%%%%
%%%%%%%%%%%%%%%%    APPENDIX B   %%%%%%%%%%%%%%%%
%%%%%%%%%%%%%%%%%%%%%%%%%%%%%%%%%%%%%%%%%%%%%%%%%
\section{Derivation of energy constraint for conformal-diagonal EoS}\label{sec:emin}

Using the definition of the pressure \eqref{eqn:pressureConDiag}, one can find from the identities in Appendix \ref{app:thermoRel} that the corresponding energy density $\varepsilon$, entropy density $s$, and charge densities $\rho_X$ satisfy
\begin{align*}
   \varepsilon(T,\vect{\mu}) &=
        3 A_0 T_0^4 \left( \left( \frac{T}{T_0} \right)^4
         +\sum_{X=B,S,Q} \left( \frac{\mu_X}{\mu_{X,0}} \right)^4 \right) \\
         &= 3 p_\mathrm{cd}(T,\vect{\mu}) \\
    s(T,\vect{\mu}) &= 4 A_0 T^3 \\
    \rho_X(T,\vect{\mu}) &= \frac{4 A_0 T_0^4 \mu_X^3}{\mu_{X,0}^4}.
\end{align*}
For a generic SPH particle, we find the phase diagram coordinates by solving the set of constraints
\begin{align*}
    \varepsilon(T,\vect{\mu}) &= \varepsilon_0 \quad \text{ and } \quad \rho_X(T,\vect{\mu}) = \rho_{X,0}
\end{align*}
at the initial timestep and
\begin{align*}
    s(T,\vect{\mu}) &= s_0 \quad \text{ and } \quad \rho_X(T,\vect{\mu}) = \rho_{X,0}
\end{align*}
at subsequent timesteps for $T$ and $\vect{\mu}$.  In both cases, the solutions for the chemical potentials are trivial:
\begin{align*}
    \mu_X = \sgn\l(\rho_{X,0}\r)\l( \frac{\mu_{X,0}^4 \l|\rho_{X,0}\r|}{4 A_0 T_0^4} \r)^{1/3},
\end{align*}
where the signum function is
\begin{align*}
    \sgn\l(x\r)
    = \begin{cases}
        +1 & \text{if}\quad x > 0 \\
        0  & \text{if}\quad x = 0 \\
        -1 & \text{if}\quad x < 0.
    \end{cases}
\end{align*}
Given $s_0$, the solution for $T$ is also trivial:
\begin{equation*}
    T = \l( \frac{s_0}{4 A_0} \r)^{1/3},
\end{equation*}
If instead $\varepsilon_0$ is given, the only real and positive solution for $T$ becomes
\begin{equation}
    T = \l( \frac{\varepsilon_0 - \varepsilon_\mathrm{min}(\vect{\rho}_0)}{3 A_0} \r)^{1/4}
\end{equation}
which requires that $\varepsilon_0 \geq \varepsilon_\mathrm{min}(\vect{\rho}_0)$, where $\varepsilon_\mathrm{min}(\vect{\rho})$ is given by \eqref{emin_definition}.

%%%%%%%%%%%%%%%%%%%%%%%%%%%%%%%%%%%%%%%%%%%%%%%%%
%%%%%%%%%%%%%%%%    APPENDIX C   %%%%%%%%%%%%%%%%
%%%%%%%%%%%%%%%%%%%%%%%%%%%%%%%%%%%%%%%%%%%%%%%%%
\section{Extrapolating Taylor series EoS to \texorpdfstring{$T=0$}{T=0} MeV}
\label{sec:lowT}

The Taylor series EoS paper \cite{Noronha-Hostler:2019ayj}
 defined $T_\mathrm{min} = 30$ MeV as the lowest temperature that is reliable within that EoS.  
 Hydrodynamics, however, requires the ability to fix the thermodynamic information of fluid elements with arbitrarily small energy and charge densities, meaning that a prescription for extending the tabulated EoS to $T = 0$ MeV is necessary.  
 Notice that this problem is specific to the Taylor series EoS considered here; the ``fallback" equations of state are all analytic functions of $T$ and $\vect\mu$ which can be evaluated reliably at any value of $T>0$ MeV.  We therefore construct the necessary extrapolation only for the Taylor series EoS.

As noted above, the Taylor series EoS is constructed in terms of the coefficients $\chi_{ijk}^{BQS}(T)$, whose $T$ dependence is parameterized using either Pad\'e approximants or simple combinations of transcendental functions.  These parametrizations are adjusted in order to agree with lattice QCD calculations which have been smoothly matched to a hadron resonance gas at temperatures below $T \sim 154$ MeV \cite{Noronha-Hostler:2019ayj}.

In order to extend this EoS to $T=0$ MeV, we must extend the $\chi_{ijk}^{BQS}(T)$ to $T=0$ MeV as well.  We do this by identifying a switching temperature $T_\text{switch}$ at which we match the Taylor series EoS smoothly to the following generic form:
\begin{equation}
    \chi_{\text{low-}T}(T) = A e^{-B/T} \label{low_T_chi}
\end{equation}
The coefficients $A$ and $B$ must be determined separately for each of the $\chi_{ijk}^{BQS}(T)$.  This is done by matching the value and first derivative of each $\chi_{ijk}^{BQS}$ at $T_\text{switch}$ to those of its corresponding low-$T$ extrapolation.

Explicitly, we have the conditions
\begin{align}
    \chi_{ijk}^{BQS}\l(T_\text{switch}\r) &\equiv \chi_0
    = A e^{-B/T_\text{switch}} \\
    \frac{d\chi_{ijk}^{BQS}}{dT}\l(T_\text{switch}\r) &\equiv \chi'_0
    = \frac{A B}{T_\text{switch}^2} e^{-B/T_\text{switch}}
\end{align}
The solutions are
\begin{align}
    A &= \chi_0 \exp\l(T_\text{switch} \chi'_0/\chi_0\r) \\
    B &= T_\text{switch}^2 \l(\frac{\chi'_0}{\chi_0} \r)
\end{align}
$A$ and $B$ are thus fixed for each of the coefficients $\chi_{ijk}^{BQS}$, allowing Eq.~\eqref{low_T_chi} to be used for the extrapolation of each coefficient to $T=0$ MeV.  In this work we have set $T_\text{switch} = 70$ MeV, which is well below the transition to the hadron resonance gas and has completely negligible effects on the hydrodynamic evolution.

%%%%%%%%%%%%%%%%%%%%%%%%%%%%%%%%%%%%%%%%%%%%%%%%%
%%%%%%%%%%%%%%%%    APPENDIX D   %%%%%%%%%%%%%%%%
%%%%%%%%%%%%%%%%%%%%%%%%%%%%%%%%%%%%%%%%%%%%%%%%%
\section{Benchmark checks: Energy loss, efficiency, and convergence tests}
\label{app:Benchmark_checks}
%%%%%%%%%%%%%%%%%%%%%%%%%%%%%%%%%%%
We carried out convergence tests of three aspects of our simulations framework: (i), ensuring the conservation of energy and charge densities; (ii), optimizing the grid size used in the EoS; and (iii), stabilizing the low density regime of the fluid. 
%%%%%%%%%%%%%%%%%%%%%%%%%%%%%%%%%%%

\subsection{Energy loss in SPH}\label{app:Eloss}

Within \ccake{} any numerical issues often show up first in issues with the conservation of energy.  
The hydrodynamic code must  conserve energy and momentum i.e.
\begin{eqnarray}
     \partial_\mu T^{\mu\nu}=0. 
\end{eqnarray}
where apply the Christoffel  symbols in hyperbolic coordinates leads to 
\begin{eqnarray}
    g_{\nu\beta}\frac{1}{\tau}\partial_{\mu}\left(\tau T^{\mu\nu}\right)+g_{\nu\beta}\Gamma^{\nu}_{\lambda\mu}T^{\lambda\mu}&=&0\\
    \frac{1}{\tau}\partial^{\alpha}\left(\tau T_{\alpha\beta}\right)+g_{\beta0}\tau T^{33}&=&0
\end{eqnarray}
where $\beta=i$ provides the momentum conservation law:
\begin{equation}
    \frac{1}{\tau}\partial^{\alpha}\left(\tau T_{\alpha i}\right)=0
\end{equation}
and $\beta=0$ provides the energy conservation law:
\begin{equation}
    \frac{1}{\tau}\partial^{\mu}\left(\tau T_{\mu 0}\right)+\tau T^{33}=0.
\end{equation}

Focusing on the energy conservation term we can separate out the 0 and $i$ components in $\mu$ i.e.
\begin{eqnarray}
     \frac{1}{\tau}\partial^{\tau}\left(\tau T_{0 0}\right)+ \frac{1}{\tau}\partial^{i}\left(\tau T_{i 0}\right)+\tau T^{33}&=&0\\
     \partial^{\tau}\left(\tau T_{0 0}\right)+\tau^2 T^{33}&=&0
\end{eqnarray}
since $T^{i0}=T^{0i}$ is symmetric the middle term ($\frac{1}{\tau}\partial^{i}\left(\tau T_{i 0}\right)$) is zero due to momentum conservation.

Then we can integrate over the entire field at a specific time step $\tau$ such that
\begin{eqnarray}
    \frac{d}{d\tau}\int d\vec{x}\;\left(\tau T_{0 0}\right)+\int d\vec{x}\;\tau^2 T^{33}&=&0 \\
\end{eqnarray}
In this work, we assume Bjoerken scaling along the rapidity direction such that we can trivially relate
\begin{eqnarray}
     T^{33}&=& - g^{33} (p +\Pi) + \pi^{33} = \frac{1}{\tau^2} (p+\Pi) + \pi^{33} \\
\end{eqnarray}
Arriving to 
\begin{eqnarray}
\frac{d}{d\tau} \int \tau T^{00} + \int  (p+\Pi) + \tau^2 \pi^{33} = 0
\end{eqnarray}
\begin{figure}
    \centering
    \includegraphics[keepaspectratio, width=0.5\linewidth]{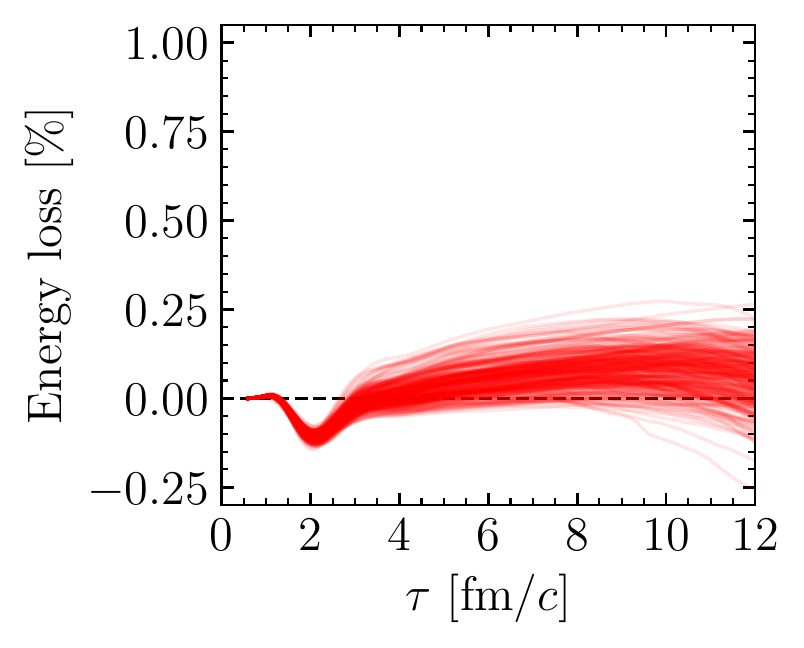}
        \caption{(color online) Percentage energy loss  as a function of $\tau$ for 300 central events.  Global energy loss is generally limited to 0.5\% or less.}
    \label{fig:energy_loss}
\end{figure}

In Fig.\ \ref{fig:energy_loss} we demonstrate the energy loss over time for 300 different events. 
Initial we see a small energy loss
Overall, the change in energy is significantly less than a percent level change, even taking account small differences event-by-event. 

\begin{figure}
    \centering
    \includegraphics[keepaspectratio, width=0.5\linewidth]{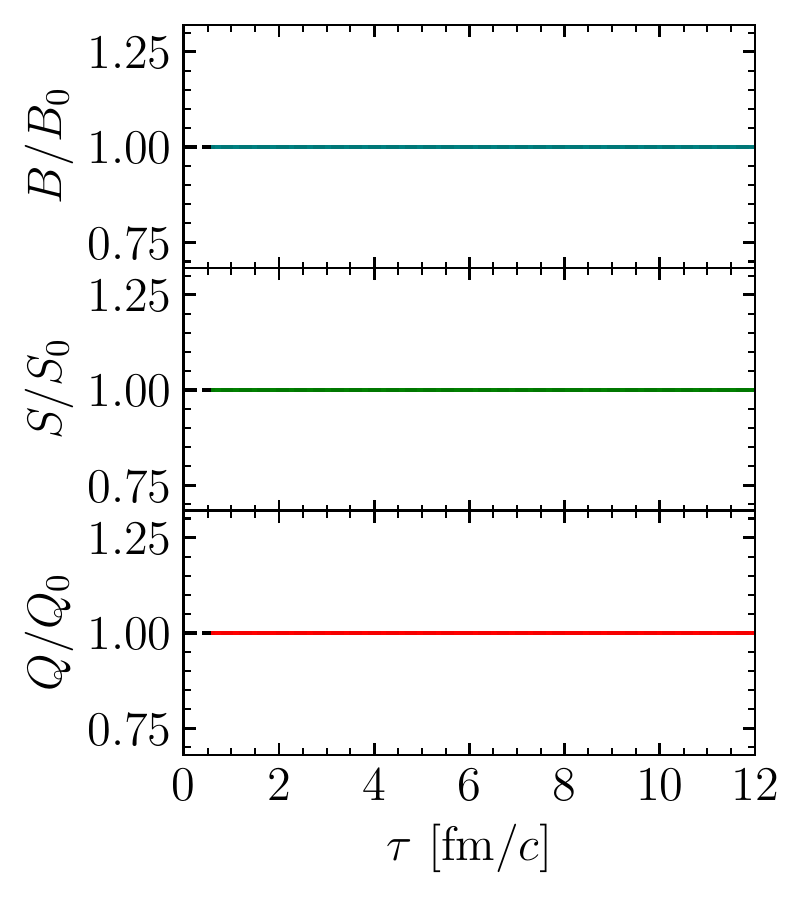}
        \caption{(color online) Change in the conservation of total Baryon number, strangeness, and electric charge over time in \ccake{}.
        }
    \label{fig:density_loss}
\end{figure}

In Fig.\ \ref{fig:density_loss} we plot the change in BSQ densities over time. 
Because the SPH formalism identically conserves BSQ charges within each SPH particle there is no loss of the total BSQ charges over time. 

\subsection{Convergence tests on the EoS grid size}\label{app:gridsize}

\begin{figure}
    \centering
    \includegraphics[keepaspectratio, width=0.5\linewidth]{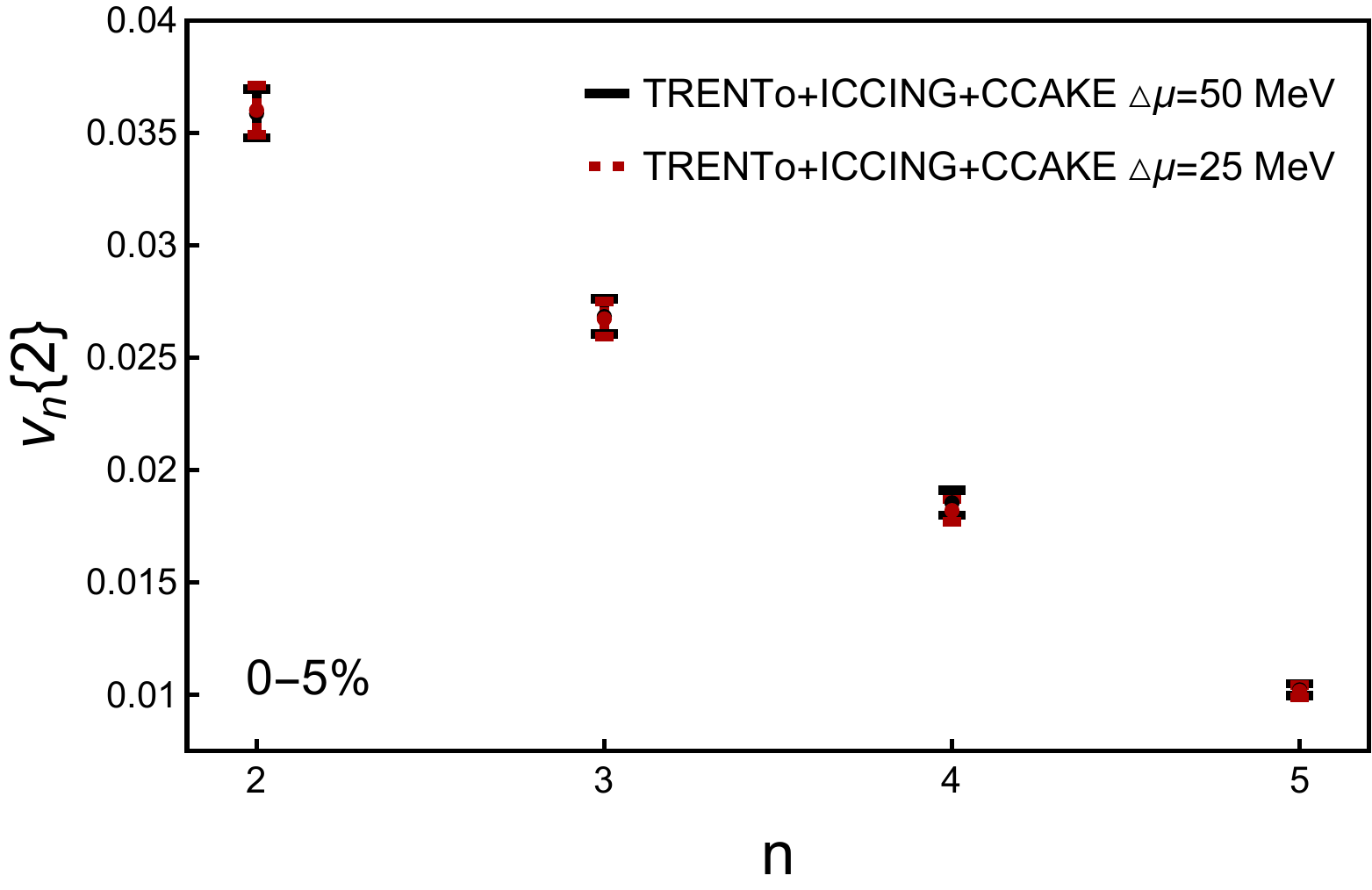}
        \caption{(color online) Sensitivity of the measurement of flow observables is within $1\%$ effect on multiplicity and $p_T$ by decreasing the grid size used in this work which confirms robustness of our calculations against the numerics of the EoS code itself.
        }
    \label{fig:EoSGrid}
\end{figure}
There is a delicate balance between including a fine grid in the EoS and ensuring an efficient run time for \ccake{}.
The finer the grid spacing, the longer the interpolation and root-finding routine will take.  
On the other hand, too coarse of a grid spacing may lead to error in the interpolation/rootfinding algorithm.
Thus, we test any changes that our grid spacing in $\Delta \mu$ has on the final flow harmonics (in past work we have tested more thoroughly the spacing in $\Delta T$). 

In this work, we have chosen $\Delta \mu=50$ MeV, which seems a reasonable assumption given that we have a cross-over phase transition.  
We have then also tested a finer grid of $\Delta \mu=25$ MeV to see if there is any discernible effect on our final flow observables. 
In Fig.\ \ref{fig:EoSGrid} we compare $v_n\left\{2\right\}$ for $n=2,3,4,5$ for $\Delta \mu=50$ MeV and $\Delta \mu=25$ MeV. 
We find that we obtain nearly identical results even with the smaller grid. 
Thus, for this EoS our results appear to converge. 
However, for an EoS with phase transitions or a critical point likely a much finer grid would be needed. 

%%%%%%%%%%%%%%%%%%%%%%%%%%%%%%%%%%%%%%%%%%%%%%%%%
%%%%%%%%%%%%%%%%    SPH buffering   %%%%%%%%%%%%%%%%
%%%%%%%%%%%%%%%%%%%%%%%%%%%%%%%%%%%%%%%%%%%%%%%%%
\subsection{SPH particle buffering}\label{app:SPHbuffer}
%%%%%%%%%%%%%%%%%%%%%%%%%%%%%%%%%%%%%%%%%%%%%%%%%%%%%%%%
%
\begin{figure}[hbt]
    \centering
    \includegraphics[keepaspectratio, width=0.5\linewidth]{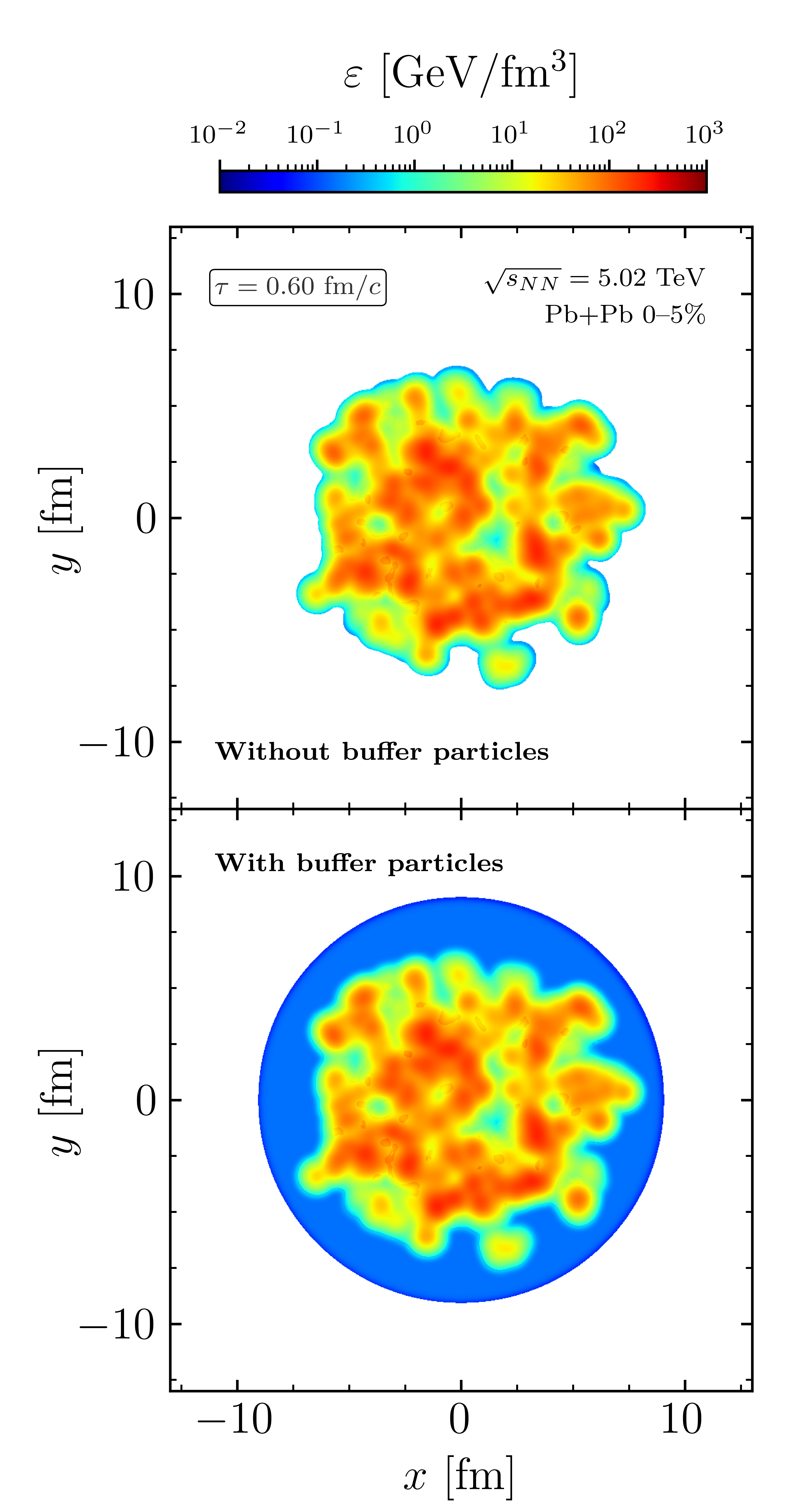}
    \caption{(color online) A central Pb\texttt{+}Pb event generated from \TRENTo{} without (top) and with (bottom) additional SPH buffer particles.  The buffer particles are each initialized with a small but non-zero energy density which has minimal effects on the overall evolution, while also minimizing potential edge effects.
    }
    \label{fig:appEbufferComparison}
\end{figure}

While in many respects SPH Lagrangian methods provide a robust alternative to standard Eulerian grid-based methods, the former nevertheless tend to encounter instabilities when SPH particles have too few neighbors within the width of the smoothing kernel to ensure a numerically stable evaluation of the locally smoothed densities in their vicinity.  
This occurs more often in regions with very low density or where local gradients are especially large, typically near the edge of the system. 
In fact, in SPH codes one cannot run fluid cells with vanishing energy densities. 
Thus, a energy density cut-off is enforced, \code{e\_cutoff}, where fluid cells below this energy density are removed from the simulation. 

The best way to deal with this issue would be to use adaptive SPH methods \cite{Monaghan:1992rr,Owen:1995qv,Shapiro:1995we,Rosswog:2009sr,Springel:2010peu,Price:2010hv}.  However, this has not yet been attempted in heavy-ion collisions so we leave that approach for a future work. 
Instead, it is occasionally useful to artificially insert additional (very low density) SPH particles, not present in the initial conditions themselves, which provide additional neighbors to those cells where instabilities tend to arise. 
We refer to these particles as ``buffer'' particles.  \ccake{} allows the user to do this by turning on the \code{buffer\_event} flag (which is set to \code{false} by default).  This can be done in one of two ways: (i), by initializing every empty cell in the initial conditions grid with a default particle; or (ii), by initializing empty cells within a fixed distance of the origin, whose scale is determined by the size of the unbuffered initial conditions.  

Each default particle is initialized with zero flow velocity, zero charge densities, and a small energy density equal to the cut-off energy density \code{e\_cutoff}.  
The default value of \code{e\_cutoff} is approximately 3 MeV/fm$^3$, roughly 100 times smaller than the freeze-out energy density, and thus ensures that the buffer particles have litte effect on the overall evolution.  ``Buffering" the event thus provides a convenient way to stabilize potentially unstable evolution without appreciably changing the geometry or physical characteristics of the system.

In Fig.\ \ref{fig:appEbufferComparison} we demonstrate the effect that the buffer particles have on a given central \TRENTo{} event.  The left figure is the energy density distribution of the transverse plan for all grid points that are above  \code{e\_cutoff}. The right figure demonstrates the addition of the buffer SPH particles where they are added in a circle around the event. Because \code{e\_cutoff} is very small compared to the center of the event, the effect of these buffer particles is minimal. Note that in the default \ccake{}, the buffer particles are turned off and they can be turned on if needed for the evolution. Important to notice that we did test the effect of their presence on the flow observables and we observed they do not affect the flow.

%%%%%%%%%%%%%%%%%%%%%%%%%%%%%%%%%%%%%%%%%%%%%%%%%
%%%%%%%%%%%%%%%%    APPENDIX F   %%%%%%%%%%%%%%%%
%%%%%%%%%%%%%%%%%%%%%%%%%%%%%%%%%%%%%%%%%%%%%%%%%
\section{Thermodynamic relations for time dependent quantities}
\label{app:thermodynamic_derivatives}
%==============================================================================

In the following, we work out the required thermodynamic relations in order to rewrite time dependent derivatives of thermodynamic quantities in order to avoid taking time derivatives in real time within the \ccake{}. This approach improves both the speed and accuracy of the code because these derivatives need only be computed once at the point when the EoS is generated.  Throughout this appendix, we assume that $X,Y,Z,W \in \l\{B,S,Q\r\}$.

%-------------------------------------------------------------------------------
%
\subsection{First Term \texorpdfstring{$\l(\frac{\partial s}{\partial T}\r)_\rho$}{(∂s/∂T)}}
\label{sec:firstderiv}
%
%-------------------------------------------------------------------------------

The desired derivative is $\l( \frac{\partial s}{\partial T} \r)_\rho$, which we want to express through the thermodynamic basis $(T, \mu_X)$:
\begin{align} \label{e:matx1}
    ds &= \l(\frac{\partial s}{\partial T}\r)_\mu dT +
    \sum_Y \l(\frac{\partial s}{\partial \mu_Y}\r)_{T, \mu'} d\mu_Y
    \notag \\
    \l(\frac{\partial s}{\partial T}\r)_\rho &= 
    \l(\frac{\partial s}{\partial T}\r)_\mu 
    + \sum_Y 
    \l(\frac{\partial s}{\partial \mu_Y}\r)_{T, \mu'} 
    \l(\frac{\partial \mu_Y}{\partial T}\r)_\rho
\end{align}
where by $\l(\frac{\partial s}{\partial \mu_Y}\r)_{T, \mu'}$ we mean that $T$ and all other chemical potentials $\mu_X \neq \mu_Y$ (other than the one being differentiated) are being held fixed.  

Suggestively, we can already reformulate \eqref{e:matx1} in terms of a matrix product:
\begin{align} \label{e:matx2}
    \l(\frac{\partial s}{\partial T}\r)_\rho &= 
    \l(\frac{\partial s}{\partial T}\r)_\mu +
    \begin{bmatrix}
        \, & \l(\frac{\partial s}{\partial \mu_Y}\r)_{T, \mu'} & \,
    \end{bmatrix}_Y
    \begin{bmatrix}
        \, \\ \l(\frac{\partial \mu_Y}{\partial T}\r)_\rho \\ \,
    \end{bmatrix}_Y ,
\end{align}
where we have explicitly indicated the matrix index $Y$ being contracted.

Now all of the nontrivial conditions on the differentials $dT, d\mu_Y$ are contained in the column vector $\l(\frac{\partial \mu_Y}{\partial T}\r)_\rho$, which means that these quantities must be obtained by simultaneously solving the 3 constraint equations:
\begin{align}
    0 = d\rho_X &= \l(\frac{\partial \rho_X}{\partial T}\r)_\mu dT +
    \sum_Y \l(\frac{\partial \rho_X}{\partial \mu_Y}\r)_{T,\mu'} d\mu_Y
\end{align}
or, equivalently,
\begin{align}
    0 &= \l(\frac{\partial \rho_X}{\partial T}\r)_\mu +
    \sum_Y \l(\frac{\partial \rho_X}{\partial \mu_Y}\r)_{T,\mu'}
    \l( \frac{\partial \mu_Y}{\partial T} \r)_\rho 
\end{align}
In matrix notation (and using the summation convention), this is
\begin{align}
    -
    \begin{bmatrix}
        \, \\ \l(\frac{\partial \rho_X}{\partial T}\r)_\mu \\ \,
    \end{bmatrix}_X
    &=
    \begin{bmatrix}
        \, & \, & \, \\
        \, & \l(\frac{\partial \rho_X}{\partial \mu_Y}\r)_{T, \mu'}  & \, \\
        \, & \, & \,
    \end{bmatrix}_{X Y}
    \begin{bmatrix}
        \, \\ \l( \frac{\partial \mu_Y}{\partial T} \r)_\rho \\ \,
    \end{bmatrix}_Y .
\end{align} 
The single matrix constraint equation is directly solved by matrix inversion to give
\begin{align}
    \begin{bmatrix}
        \, \\ \l( \frac{\partial \mu_Y}{\partial T} \r)_\rho \\ \,
    \end{bmatrix}_Y
    &=
    -
    \begin{bmatrix}
    \, & \, & \, \\
    \, & \l(\frac{\partial \rho_X}{\partial \mu_Y}\r)_{T, \mu'}  & \, \\
    \, & \, & \,
    \end{bmatrix}_{Y X}^{-1}
    \begin{bmatrix}
        \, \\ \l(\frac{\partial \rho_X}{\partial T}\r)_\mu \\ \,
    \end{bmatrix}_X
    =
    -
    \begin{bmatrix}
    \, & \, & \, \\
    \, & \chi_{X Y}  & \, \\
    \, & \, & \,
    \end{bmatrix}_{Y X}^{-1}
    \begin{bmatrix}
        \, \\ \chi_{X T} \\ \,
    \end{bmatrix}_X
\end{align}
where in the last line we have used the thermodynamic relations $\rho_X = \l(\frac{\partial p}{\partial \mu_X}\r)_{T, \mu'}$, $\l(\frac{\partial p}{\partial \mu_X\partial \mu_Y}\r)_{T,\mu'} = \chi_{X Y}$, and $\l(\frac{\partial p}{\partial \mu_X\partial T}\r)_{\mu'} = \chi_{X T}$.  Back-substituting into \eqref{e:matx2} and also using the thermodynamic relation $s = \l(\frac{\partial p}{\partial T}\r)_\mu$ allows us finally to write
\begin{align}
    \l(\frac{\partial s}{\partial T}\r)_\rho &= 
    \chi_{TT} -
    V^\dagger \, M^{-1} \, V, \hspace{0.5cm} \text{where} \hspace{0.5cm}
    V =
    \begin{bmatrix}
        \chi_{TB} \\
        \chi_{TS} \\
        \chi_{TQ}
    \end{bmatrix}
    \hspace{0.5cm} \text{and} \hspace{0.5cm}
    M =
    \begin{bmatrix}
        \chi_{BB} &
        \chi_{BS} &
        \chi_{BQ} \\
        \chi_{SB} &
        \chi_{SS} &
        \chi_{SQ} \\
        \chi_{QB} &
        \chi_{QS} &
        \chi_{QQ}
    \end{bmatrix} .
\end{align}

%-------------------------------------------------------------------------------
%
\subsection{Second Term: \texorpdfstring{$\l(\frac{\partial s}{\partial \mu_X} \r)_\rho$}{(∂s/∂μ\_X)}}
%
%-------------------------------------------------------------------------------

We now apply the same logic to the other terms.  Here we are interested in the rate of change with respect to a particular choice $i$ of chemical potential $\mu_X$, so we must distinguish between this particular $\mu_X$ and the others $\mu_{X \neq Y}$:
\begin{align}
    ds &= \l(\frac{\partial s}{\partial T}\r)_\mu dT +
    \l(\frac{\partial s}{\partial \mu_X}\r)_{T, \mu'} d\mu_X +
    \sum_{X \neq Y} \l(\frac{\partial s}{\partial \mu_Y}\r)_{T, \mu'} d\mu_Y
    \notag \\ \notag \\
    \l(\frac{\partial s}{\partial \mu_X}\r)_\rho &= 
    \l(\frac{\partial s}{\partial \mu_X}\r)_{T, \mu'} +
    \l(\frac{\partial s}{\partial T}\r)_\mu 
    \l(\frac{\partial T}{\partial \mu_X}\r)_\rho +
    \sum_{X \neq Y} \l(\frac{\partial s}{\partial \mu_Y}\r)_{T, \mu'}
    \l(\frac{\partial \mu_Y}{\partial \mu_X}\r)_\rho .
\end{align}
This equation also has the form of a matrix product, although the relevant 3-component matrix is a combination of different-looking quantities:
\begin{align}   \label{e:matx4}
    \l(\frac{\partial s}{\partial \mu_X}\r)_\rho &= 
    \l(\frac{\partial s}{\partial \mu_X}\r)_{T, \mu'} 
    +
    \begin{bmatrix}
        \l(\frac{\partial s}{\partial T}\r)_\mu & 
        \l(\frac{\partial s}{\partial \mu_{X \neq Y}}\r)_{T, \mu'} &
        \l(\frac{\partial s}{\partial \mu_{X \neq Y}}\r)_{T, \mu'}
    \end{bmatrix}
    \begin{bmatrix}
        \l(\frac{\partial T}{\partial \mu_X}\r)_\rho \\
        \l(\frac{\partial \mu_{X \neq Y}}{\partial \mu_X}\r)_\rho \\
        \l(\frac{\partial \mu_{X \neq Y}}{\partial \mu_X}\r)_\rho
    \end{bmatrix} ,
\end{align}
which mixes together $T$ and the two chemical potentials $\mu_{X \neq Y}$ which are not being fixed.  In the same way as before, we need to impose the constraints
\begin{align}
    0 = d\rho_Z &= \l(\frac{\partial \rho_Z}{\partial T}\r)_\mu dT 
    + \l(\frac{\partial \rho_Z}{\partial \mu_X}\r)_{T,\mu'} d\mu_X
    + \sum_{X \neq Y} 
    \l(\frac{\partial \rho_Z}{\partial \mu_Y}\r)_{T,\mu'} d\mu_Y
    \notag \\
    0 &=
    \l(\frac{\partial \rho_Z}{\partial T}\r)_\mu 
    \l(\frac{\partial T}{\partial \mu_X}\r)_\rho 
    + 
    \l(\frac{\partial \rho_Z}{\partial \mu_X}\r)_{T,\mu'}
    + \sum_{X \neq Y} 
    \l(\frac{\partial \rho_Z}{\partial \mu_Y}\r)_{T,\mu'} 
    \l(\frac{\partial \mu_Y}{\partial \mu_X}\r)_\rho
    \notag \\
    - \l(\frac{\partial \rho_Z}{\partial \mu_X}\r)_{T,\mu'} &=
    \l(\frac{\partial \rho_Z}{\partial T}\r)_\mu 
    \l(\frac{\partial T}{\partial \mu_X}\r)_\rho 
    + \sum_{X \neq Y} 
    \l(\frac{\partial \rho_Z}{\partial \mu_Y}\r)_{T,\mu'} 
    \l(\frac{\partial \mu_Y}{\partial \mu_X}\r)_\rho
\end{align}
which can again be expressed (and solved) as a matrix equation:
\begin{align}
    - 
    \begin{bmatrix}
        \, \\ \l(\frac{\partial \rho_Z}{\partial \mu_X}\r)_{T,\mu'} \\ \,
    \end{bmatrix}_Z
    &=
    \begin{bmatrix}
        \l(\frac{\partial \rho_Z}{\partial T}\r)_\mu &
        \l(\frac{\partial \rho_Z}{\partial \mu_{Y \neq X}}\r)_{T,\mu'} &
        \l(\frac{\partial \rho_Z}{\partial \mu_{Y \neq X}}\r)_{T,\mu'} \\
        \l(\frac{\partial \rho_Z}{\partial T}\r)_\mu &
        \l(\frac{\partial \rho_Z}{\partial \mu_{Y \neq X}}\r)_{T,\mu'} &
        \l(\frac{\partial \rho_Z}{\partial \mu_{Y \neq X}}\r)_{T,\mu'} \\
        \l(\frac{\partial \rho_Z}{\partial T}\r)_\mu &
        \l(\frac{\partial \rho_Z}{\partial \mu_{Y \neq X}}\r)_{T,\mu'} &
        \l(\frac{\partial \rho_Z}{\partial \mu_{Y \neq X}}\r)_{T,\mu'}
    \end{bmatrix}_{Z Y}
    \begin{bmatrix}
        \l(\frac{\partial T}{\partial \mu_X}\r)_\rho \\
        \l(\frac{\partial \mu_{X \neq Y}}{\partial \mu_X}\r)_\rho \\
        \l(\frac{\partial \mu_{X \neq Y}}{\partial \mu_X}\r)_\rho
    \end{bmatrix}_Y
    \notag \\ \notag \\
    \begin{bmatrix}
        \l(\frac{\partial T}{\partial \mu_X}\r)_\rho \\
        \l(\frac{\partial \mu_{X \neq Y}}{\partial \mu_X}\r)_\rho \\
        \l(\frac{\partial \mu_{X \neq Y}}{\partial \mu_X}\r)_\rho
    \end{bmatrix}_Y
    &=
    -
    \begin{bmatrix}
        \l(\frac{\partial \rho_Z}{\partial T}\r)_\mu &
        \l(\frac{\partial \rho_Z}{\partial \mu_{Y \neq X}}\r)_{T,\mu'} &
        \l(\frac{\partial \rho_Z}{\partial \mu_{Y \neq X}}\r)_{T,\mu'} \\
        \l(\frac{\partial \rho_Z}{\partial T}\r)_\mu &
        \l(\frac{\partial \rho_Z}{\partial \mu_{Y \neq X}}\r)_{T,\mu'} &
        \l(\frac{\partial \rho_Z}{\partial \mu_{Y \neq X}}\r)_{T,\mu'} \\
        \l(\frac{\partial \rho_Z}{\partial T}\r)_\mu &
        \l(\frac{\partial \rho_Z}{\partial \mu_{Y \neq X}}\r)_{T,\mu'} &
        \l(\frac{\partial \rho_Z}{\partial \mu_{Y \neq X}}\r)_{T,\mu'}
    \end{bmatrix}_{Y Z}^{-1}
    \begin{bmatrix}
        \, \\ \l(\frac{\partial \rho_Z}{\partial \mu_X}\r)_{T,\mu'} \\ \,
    \end{bmatrix}_Z
\end{align}
where we have tried to show the structure of the matrix explicitly.  Inserting this expression back into \eqref{e:matx4} yields the solution
\begin{align}
    \l(\frac{\partial s}{\partial \mu_X}\r)_\rho &=
    \l(\frac{\partial^2 p}{\partial T \, \partial \mu_X}\r)_{\mu'} -
    A^\dagger \, M^{-1} \, B
\end{align}
with
\begin{align}
    A &=
    \begin{bmatrix}
        \l(\frac{\partial^2 p}{\partial T^2}\r)_\mu \\ 
        \l(\frac{\partial^2 p}{\partial T \, \partial \mu_{X \neq Y}}\r)_{\mu'} \\
        \l(\frac{\partial^2 p}{\partial T \, \partial \mu_{X \neq Y}}\r)_{\mu'}
    \end{bmatrix}_Y
    \hspace{2cm}
    B =
    \begin{bmatrix}
        \l(\frac{\partial^2 p}{\partial \mu_X \, \partial \mu_Z}\r)_{T,\mu'} \\ 
        \l(\frac{\partial^2 p}{\partial \mu_X \, \partial \mu_Z}\r)_{T,\mu'} \\ 
        \l(\frac{\partial^2 p}{\partial \mu_X \, \partial \mu_Z}\r)_{T,\mu'}
    \end{bmatrix}_Z
    \notag \\ \notag \\
    M &=
    \begin{bmatrix}
        \l(\frac{\partial^2 p}{\partial \mu_Z \, \partial T}\r)_{\mu'} &
        \l(\frac{\partial^2 p}{\partial \mu_Z \, \partial \mu_{Y \neq X}}\r)_{T,\mu'} &
        \l(\frac{\partial^2 p}{\partial \mu_Z \, \partial \mu_{Y \neq X}}\r)_{T,\mu'} \\
        \l(\frac{\partial^2 p}{\partial \mu_Z \, \partial T}\r)_{\mu'} &
        \l(\frac{\partial^2 p}{\partial \mu_Z \, \partial \mu_{Y \neq X}}\r)_{T,\mu'} &
        \l(\frac{\partial^2 p}{\partial \mu_Z \, \partial \mu_{Y \neq X}}\r)_{T,\mu'} \\
        \l(\frac{\partial^2 p}{\partial \mu_Z \, \partial T}\r)_{\mu'} &
        \l(\frac{\partial^2 p}{\partial \mu_Z \, \partial \mu_{Y \neq X}}\r)_{T,\mu'} &
        \l(\frac{\partial^2 p}{\partial \mu_Z \, \partial \mu_{Y \neq X}}\r)_{T,\mu'}
    \end{bmatrix}_{Z Y} .
\end{align}
Finally, since the notation and matrix organization may be a little opaque, let us switch to the notation of Appendix \ref{app:thermoRel} and write out the result explicitly for the case of $\mu_X = \mu_B$:
\begin{align}
    \l(\frac{\partial s}{\partial \mu_B}\r)_\rho &=
    \chi_{TB} -
    A^\dagger \, M^{-1} \, B
\end{align}
with
\begin{align}
    A &=
    \begin{bmatrix}
        \chi_{TT} \\ 
        \chi_{TS} \\
        \chi_{TQ}
    \end{bmatrix}_Y
    \hspace{1cm}
    B =
    \begin{bmatrix}
        \chi_{BB} \\ 
        \chi_{BS} \\ 
        \chi_{BQ}
    \end{bmatrix}_Z
    \hspace{1cm}
    M =
    \begin{bmatrix}
        \chi_{BT} &
        \chi_{BS} &
        \chi_{BQ} \\
        \chi_{ST} &
        \chi_{SS} &
        \chi_{SQ} \\
        \chi_{QT} &
        \chi_{QS} &
        \chi_{QQ}
    \end{bmatrix}_{Z Y} .
\end{align}
One can then obtain the other equations by cyclic permutation of $B/S/Q$ in the matrices.

%-------------------------------------------------------------------------------
%
\subsection{Third Term: \texorpdfstring{$\l(\frac{\partial \rho_X}{\partial T} \r)_{s, \rho_X \neq \rho_Y}$}{(∂ρ\_X/∂T)}}
%
%-------------------------------------------------------------------------------

Starting with
\begin{align}   \label{e:matx5}
    d\rho_X &= \l(\frac{\partial \rho_X}{\partial T}\r)_\mu dT +
    \sum_Y \l(\frac{\partial \rho_X}{\partial \mu_Y}\r)_{T, \mu'} d\mu_Y
    \notag \\ \notag \\
    \l(\frac{\partial \rho_X}{\partial T}\r)_{s, \rho_X \neq \rho_X}
    \hspace{-0.5cm} &= \l(\frac{\partial \rho_X}{\partial T}\r)_\mu +
    \sum_Y \l(\frac{\partial \rho_X}{\partial \mu_Y}\r)_{T, \mu'} 
    \l(\frac{\partial \mu_Y}{\partial T}\r)_{s, \rho_X \neq \rho_X}
    \notag \\ \notag \\
    \l(\frac{\partial \rho_X}{\partial T}\r)_{s, \rho_X \neq \rho_X}
    \hspace{-0.5cm} &=
    \l(\frac{\partial \rho_X}{\partial T}\r)_\mu 
    +
    \begin{bmatrix}
        \l(\frac{\partial \rho_X}{\partial \mu_B}\r)_{T, \mu_S, \mu_Q} & 
        \l(\frac{\partial \rho_X}{\partial \mu_S}\r)_{T, \mu_B, \mu_Q} & 
        \l(\frac{\partial \rho_X}{\partial \mu_Q}\r)_{T, \mu_B, \mu_S}
    \end{bmatrix}
    \begin{bmatrix}
        \l(\frac{\partial \mu_B}{\partial T}\r)_{s, \rho_X \neq \rho_X} \\ 
        \l(\frac{\partial \mu_S}{\partial T}\r)_{s, \rho_X \neq \rho_X} \\ 
        \l(\frac{\partial \mu_Q}{\partial T}\r)_{s, \rho_X \neq \rho_X}
    \end{bmatrix}
\end{align}
we now have the constraints $ds = 0$ and $d \rho_{X \neq Y} = 0$:
\begin{subequations}
\begin{align}
    ds = 0
    &\Longrightarrow 
    \l(\frac{\partial s}{\partial T}\r)_\mu  
    + \sum_Z
    \l(\frac{\partial s}{\partial \mu_Z}\r)_{T, \mu'} 
    \l(\frac{\partial \mu_Z}{\partial T}\r)_{s, \rho_X \neq \rho_X} = 0 ,
    \\ \notag \\
    d\rho_{X \neq Y} = 0
    &\Longrightarrow 
    \l(\frac{\partial \rho_{X \neq Y}}{\partial T}\r)_\mu
    + \sum_Z
    \l(\frac{\partial \rho_{X \neq Y}}{\partial \mu_Z}\r)_{T, \mu'} 
    \l(\frac{\partial \mu_Z}{\partial T}\r)_{s, \rho_X \neq \rho_X} = 0 .
\end{align}
\end{subequations}
While these three constraint equations look different, they can again be combined and solved as a single matrix equation:
\begin{align}
    -
    \begin{bmatrix}
        \l(\frac{\partial s}{\partial T}\r)_\mu \\
        \l(\frac{\partial \rho_{X \neq Y}}{\partial T}\r)_\mu \\
        \l(\frac{\partial \rho_{X \neq Y}}{\partial T}\r)_\mu
    \end{bmatrix}_Y
    &=
    \begin{bmatrix}
        \l(\frac{\partial s}{\partial \mu_B}\r)_{T, \mu_S, \mu_Q} &
        \l(\frac{\partial s}{\partial \mu_S}\r)_{T, \mu_B, \mu_Q} &
        \l(\frac{\partial s}{\partial \mu_Q}\r)_{T, \mu_B, \mu_S} \\
        \l(\frac{\partial \rho_{X \neq Y}}{\partial \mu_B}\r)_{T, \mu_S, \mu_Q} &
        \l(\frac{\partial \rho_{X \neq Y}}{\partial \mu_S}\r)_{T, \mu_B, \mu_Q} &
        \l(\frac{\partial \rho_{X \neq Y}}{\partial \mu_Q}\r)_{T, \mu_B, \mu_S} \\
        \l(\frac{\partial \rho_{X \neq Y}}{\partial \mu_B}\r)_{T, \mu_S, \mu_Q} &
        \l(\frac{\partial \rho_{X \neq Y}}{\partial \mu_S}\r)_{T, \mu_B, \mu_Q} &
        \l(\frac{\partial \rho_{X \neq Y}}{\partial \mu_Q}\r)_{T, \mu_B, \mu_S}
    \end{bmatrix}_{Y Z}
    \begin{bmatrix}
        \l(\frac{\partial \mu_B}{\partial T}\r)_{s, \rho_X \neq \rho_X} \\
        \l(\frac{\partial \mu_S}{\partial T}\r)_{s, \rho_X \neq \rho_X} \\
        \l(\frac{\partial \mu_Q}{\partial T}\r)_{s, \rho_X \neq \rho_X}
    \end{bmatrix}_Z
    \notag \\ \notag \\
    \begin{bmatrix}
    \l(\frac{\partial \mu_B}{\partial T}\r)_{s, \rho_X \neq \rho_X} \\
    \l(\frac{\partial \mu_S}{\partial T}\r)_{s, \rho_X \neq \rho_X} \\
    \l(\frac{\partial \mu_Q}{\partial T}\r)_{s, \rho_X \neq \rho_X}
    \end{bmatrix}_Z
    &= 
    -
    \begin{bmatrix}
    \l(\frac{\partial s}{\partial \mu_B}\r)_{T, \mu_S, \mu_Q} &
    \l(\frac{\partial s}{\partial \mu_S}\r)_{T, \mu_B, \mu_Q} &
    \l(\frac{\partial s}{\partial \mu_Q}\r)_{T, \mu_B, \mu_S} \\
    \l(\frac{\partial \rho_{X \neq Y}}{\partial \mu_B}\r)_{T, \mu_S, \mu_Q} &
    \l(\frac{\partial \rho_{X \neq Y}}{\partial \mu_S}\r)_{T, \mu_B, \mu_Q} &
    \l(\frac{\partial \rho_{X \neq Y}}{\partial \mu_Q}\r)_{T, \mu_B, \mu_S} \\
    \l(\frac{\partial \rho_{X \neq Y}}{\partial \mu_B}\r)_{T, \mu_S, \mu_Q} &
    \l(\frac{\partial \rho_{X \neq Y}}{\partial \mu_S}\r)_{T, \mu_B, \mu_Q} &
    \l(\frac{\partial \rho_{X \neq Y}}{\partial \mu_Q}\r)_{T, \mu_B, \mu_S}
    \end{bmatrix}_{Z Y}^{-1}
    \begin{bmatrix}
    \l(\frac{\partial s}{\partial T}\r)_\mu \\
    \l(\frac{\partial \rho_{X \neq Y}}{\partial T}\r)_\mu \\
    \l(\frac{\partial \rho_{X \neq Y}}{\partial T}\r)_\mu
    \end{bmatrix}_Y
\end{align}
Inserting this result back into \eqref{e:matx5} gives
\begin{align}
    \l(\frac{\partial \rho_X}{\partial T}\r)_{s, \rho_X \neq \rho_X}
    \hspace{-0.5cm} &=
    \l(\frac{\partial^2 p}{\partial T \, \partial \mu_X}\r)_{\mu'}
    - A^\dagger \, M^{-1} \, B
\end{align}
with
\begin{align}
    A &=
    \begin{bmatrix}
        \l(\frac{\partial^2 p}{\partial \mu_X \, \partial \mu_B}\r)_{T, \mu'} \\ 
        \l(\frac{\partial^2 p}{\partial \mu_X \, \partial \mu_S}\r)_{T, \mu'} \\
        \l(\frac{\partial^2 p}{\partial \mu_X \, \partial \mu_Q}\r)_{T, \mu'}
    \end{bmatrix}_Y
    \hspace{2cm}
    B =
    \begin{bmatrix}
        \l(\frac{\partial^2 p}{\partial T^2}\r)_\mu \\
        \l(\frac{\partial^2 p}{\partial T \, \partial \mu_{X \neq Y} }\r)_{\mu'} \\
        \l(\frac{\partial^2 p}{\partial T \, \partial \mu_{X \neq Y} }\r)_{\mu'}
    \end{bmatrix}_Z
    \notag \\ \notag \\
    M &=
    \begin{bmatrix}
    \l(\frac{\partial^2 p}{\partial T \, \partial \mu_B}\r)_{\mu_S, \mu_Q} &
    \l(\frac{\partial^2 p}{\partial T \, \partial \mu_S}\r)_{\mu_B, \mu_Q} &
    \l(\frac{\partial^2 p}{\partial T \, \partial \mu_Q}\r)_{\mu_B, \mu_S} \\
    \l(\frac{\partial^2 p}{\partial \mu_{X \neq Y} \, \partial \mu_B}\r)_{T, \mu'} &
    \l(\frac{\partial^2 p}{\partial \mu_{X \neq Y} \, \partial \mu_S}\r)_{T, \mu'} &
    \l(\frac{\partial^2 p}{\partial \mu_{X \neq Y} \partial \mu_Q}\r)_{T, \mu'} \\
    \l(\frac{\partial^2 p}{\partial \mu_{X \neq Y} \, \partial \mu_B}\r)_{T, \mu'} &
    \l(\frac{\partial^2 p}{\partial \mu_{X \neq Y} \, \partial \mu_S}\r)_{T, \mu'} &
    \l(\frac{\partial^2 p}{\partial \mu_{X \neq Y} \partial \mu_Q}\r)_{T, \mu'}
    \end{bmatrix}_{Z Y} ,
\end{align}
and again it is instructive to write explicitly the case for $\rho_X = \rho_B$:
\begin{align}
    \l(\frac{\partial \rho_B}{\partial T}\r)_{s, \rho_S, \rho_Q}
    &=
    \chi_{TB}
    - A^\dagger \, M^{-1} \, B
\end{align}
with
\begin{align}
    A &=
    \begin{bmatrix}
        \chi_{BB} \\ 
        \chi_{BS} \\
        \chi_{BQ}
    \end{bmatrix}_Y
    \hspace{1cm}
    B =
    \begin{bmatrix}
        \chi_{TT} \\
        \chi_{TS} \\
        \chi_{TQ}
    \end{bmatrix}_Z
    \hspace{1cm}
    M =
    \begin{bmatrix}
    \chi_{TB} &
    \chi_{TS} &
    \chi_{TQ} \\
    \chi_{SB} &
    \chi_{SS} &
    \chi_{SQ} \\
    \chi_{QB} &
    \chi_{QS} &
    \chi_{QQ}
    \end{bmatrix}_{Z Y} ,
\end{align}
and the other equations can be obtained by cyclic permutation.
%-------------------------------------------------------------------------------
%
\subsection{Fourth Term: \texorpdfstring{$\l(\frac{\partial \rho_Y}{\partial \mu_X} \r)_{s, \rho_X \neq \rho_Y}$}{(∂ρ\_Y/∂μ\_X)}}
%
%-------------------------------------------------------------------------------
Starting with
\begin{align}   \label{e:matx6}
    d\rho_Y &= \l(\frac{\partial \rho_Y}{\partial T}\r)_\mu dT 
    +
    \l(\frac{\partial \rho_Y}{\partial \mu_X}\r)_{T, \mu'} d\mu_X
    + \sum_{W \neq X} 
    \l(\frac{\partial \rho_Y}{\partial \mu_\ell}\r)_{T, \mu'} d\mu_\ell
    \notag \\ \notag \\
    \l(\frac{\partial \rho_Y}{\partial \mu_X}\r)_{s, \rho_X \neq \rho_Y} &= \l(\frac{\partial \rho_Y}{\partial T}\r)_\mu 
    \l(\frac{\partial T}{\partial \mu_X}\r)_{s, \rho_X \neq \rho_Y}
    +
    \l(\frac{\partial \rho_Y}{\partial \mu_X}\r)_{T, \mu'} 
    + \sum_{W \neq X} 
    \l(\frac{\partial \rho_Y}{\partial \mu_\ell}\r)_{T, \mu'} 
    \l(\frac{\partial \mu_\ell}{\partial \mu_X}\r)_{s, \rho_X \neq \rho_Y}
    \notag \\ \notag \\
    \l(\frac{\partial \rho_Y}{\partial \mu_X}\r)_{s, \rho_X \neq \rho_Y} &=
    \l(\frac{\partial \rho_Y}{\partial \mu_X}\r)_{T, \mu'} 
    +
    \begin{bmatrix}
        \l(\frac{\partial \rho_Y}{\partial T}\r)_\mu &
        \l(\frac{\partial \rho_Y}{\partial \mu_{k \neq i}}\r)_{T, \mu'} &
        \l(\frac{\partial \rho_Y}{\partial \mu_{k \neq i}}\r)_{T, \mu'} 
    \end{bmatrix}_W
    \begin{bmatrix}
        \l(\frac{\partial T}{\partial \mu_X}\r)_{s, \rho_X \neq \rho_Y} \\
        \l(\frac{\partial \mu_{k \neq i}}{\partial \mu_X}\r)_{s, \rho_X \neq \rho_Y} \\
        \l(\frac{\partial \mu_{k \neq i}}{\partial \mu_X}\r)_{s, \rho_X \neq \rho_Y}
    \end{bmatrix}_W 
\end{align}
we have the constraints $ds = 0$ and $d \rho_{Z \neq Y} = 0$:
\begin{subequations}
\begin{align}
    ds = 0
    &\Longrightarrow
    0 = \l(\frac{\partial s}{\partial T}\r)_\mu
    \l(\frac{\partial T}{\partial \mu_X}\r)_{s, \rho_X \neq \rho_Y}
    + \l(\frac{\partial s}{\partial \mu_X}\r)_{T, \mu'}
    + \sum_{W \neq X}
    \l(\frac{\partial s}{\partial \mu_\ell}\r)_{T, \mu'} 
    \l(\frac{\partial\mu_\ell}{\partial \mu_X}\r)_{s, \rho_X \neq \rho_Y}
    ,
    \\ \notag \\
    d\rho_{Z \neq Y} = 0
    &\Longrightarrow
    0 = 
    \l(\frac{\partial \rho_{Z \neq Y}}{\partial T}\r)_\mu 
    \l(\frac{\partial T}{\partial \mu_X}\r)_{s, \rho_X \neq \rho_X}
    +
    \l(\frac{\partial \rho_{Z \neq Y}}{\partial \mu_X}\r)_{T, \mu'} 
    + \sum_{W \neq X}
    \l(\frac{\partial \rho_{Z \neq Y}}{\partial \mu_\ell}\r)_{T, \mu'} \l(\frac{\partial \mu_\ell}{\partial \mu_X}\r)_{s, \rho_X \neq \rho_X}
    .
\end{align}
\end{subequations}
We cast the constraints into matrix form and solve them:
\begin{align}
    -
    \begin{bmatrix}
        \l(\frac{\partial s}{\partial \mu_X}\r)_{T, \mu'} \\
        \l(\frac{\partial \rho_{Z \neq Y}}{\partial \mu_X}\r)_{T, \mu'} \\
        \l(\frac{\partial \rho_{Z \neq Y}}{\partial \mu_X}\r)_{T, \mu'} 
    \end{bmatrix}_Z
    &=
    \begin{bmatrix}
         \l(\frac{\partial s}{\partial T}\r)_\mu &
         \l(\frac{\partial s}{\partial \mu_{W \neq X}}\r)_{T, \mu'} &
         \l(\frac{\partial s}{\partial \mu_{W \neq X}}\r)_{T, \mu'} \\
         \l(\frac{\partial \rho_{Z \neq Y}}{\partial T}\r)_\mu &
         \l(\frac{\partial \rho_{Z \neq Y}}{\partial \mu_{W \neq X}}\r)_{T, \mu'} &
         \l(\frac{\partial \rho_{Z \neq Y}}{\partial \mu_{W \neq X}}\r)_{T, \mu'} \\
         \l(\frac{\partial \rho_{Z \neq Y}}{\partial T}\r)_\mu &
         \l(\frac{\partial \rho_{Z \neq Y}}{\partial \mu_{W \neq X}}\r)_{T, \mu'} &
         \l(\frac{\partial \rho_{Z \neq Y}}{\partial \mu_{W \neq X}}\r)_{T, \mu'}
    \end{bmatrix}_{Z W}
    \begin{bmatrix}
        \l(\frac{\partial T}{\partial \mu_X}\r)_{s, \rho_X \neq \rho_Y} \\
        \l(\frac{\partial \mu_{W \neq X}}{\partial \mu_X}\r)_{s, \rho_X \neq \rho_X} \\
        \l(\frac{\partial \mu_{W \neq X}}{\partial \mu_X}\r)_{s, \rho_X \neq \rho_X}
    \end{bmatrix}_W
    \notag \\ \notag \\ 
    \begin{bmatrix}
        \l(\frac{\partial T}{\partial \mu_X}\r)_{s, \rho_X \neq \rho_Y} \\
        \l(\frac{\partial \mu_{W \neq X}}{\partial \mu_X}\r)_{s, \rho_X \neq \rho_X} \\
        \l(\frac{\partial \mu_{W \neq X}}{\partial \mu_X}\r)_{s, \rho_X \neq \rho_X}
    \end{bmatrix}_W
    &=
    -
    \begin{bmatrix}
         \l(\frac{\partial s}{\partial T}\r)_\mu &
         \l(\frac{\partial s}{\partial \mu_{W \neq X}}\r)_{T, \mu'} &
         \l(\frac{\partial s}{\partial \mu_{W \neq X}}\r)_{T, \mu'} \\
         \l(\frac{\partial \rho_{Z \neq Y}}{\partial T}\r)_\mu &
         \l(\frac{\partial \rho_{Z \neq Y}}{\partial \mu_{W \neq X}}\r)_{T, \mu'} &
         \l(\frac{\partial \rho_{Z \neq Y}}{\partial \mu_{W \neq X}}\r)_{T, \mu'} \\
         \l(\frac{\partial \rho_{Z \neq Y}}{\partial T}\r)_\mu &
         \l(\frac{\partial \rho_{Z \neq Y}}{\partial \mu_{W \neq X}}\r)_{T, \mu'} &
         \l(\frac{\partial \rho_{Z \neq Y}}{\partial \mu_{W \neq X}}\r)_{T, \mu'}
    \end{bmatrix}_{W Z}^{-1}
    \begin{bmatrix}
        \l(\frac{\partial s}{\partial \mu_X}\r)_{T, \mu'} \\
        \l(\frac{\partial \rho_{Z \neq Y}}{\partial \mu_X}\r)_{T, \mu'} \\
        \l(\frac{\partial \rho_{Z \neq Y}}{\partial \mu_X}\r)_{T, \mu'} 
    \end{bmatrix}_Z .
\end{align}
Back-substituting into \eqref{e:matx6} gives the solution
\begin{align}
    \l(\frac{\partial \rho_Y}{\partial \mu_X}\r)_{s, \rho_X \neq \rho_Y} &=
    \l(\frac{\partial^2 p}{\partial \mu_Y \, \partial \mu_X}\r)_{T, \mu'}
    - A^\dagger \, M^{-1} \, B    
\end{align}
with
\begin{align}
    A &=
    \begin{bmatrix}
        \l(\frac{\partial^2 p}{\partial \mu_Y \, \partial T}\r)_{\mu'} \\
        \l(\frac{\partial^2 p}{\partial \mu_Y \, \partial \mu_{W \neq X}}\r)_{T, \mu'} \\
        \l(\frac{\partial^2 p}{\partial \mu_Y \, \partial \mu_{W \neq X}}\r)_{T, \mu'} 
    \end{bmatrix}_W
    \hspace{2cm}
    B =
    \begin{bmatrix}
        \l(\frac{\partial^2 p}{\partial T \, \partial \mu_X}\r)_{\mu'} \\
        \l(\frac{\partial^2 p}{\partial \mu_{Z \neq Y} \, \partial \mu_X}\r)_{T, \mu'} \\
        \l(\frac{\partial^2 p}{\partial \mu_{Z \neq Y} \, \partial \mu_X}\r)_{T, \mu'}
    \end{bmatrix}_Z
    \notag \\ \notag \\
    M &=
    \begin{bmatrix}
         \l(\frac{\partial^2 p}{\partial T^2}\r)_{\mu_B, \mu_S, \mu_Q} &
         \l(\frac{\partial^2 p}{\partial T \, \partial \mu_{W \neq X}}\r)_{\mu'} &
         \l(\frac{\partial^2 p}{\partial T \, \partial \mu_{W \neq X}}\r)_{\mu'} \\
         \l(\frac{\partial^2 p}{\partial \mu_{Z \neq Y} \, \partial T}\r)_{\mu'} &
         \l(\frac{\partial^2 p}{\partial \mu_{Z \neq Y} \, \partial \mu_{W \neq X}}\r)_{T, \mu'} &
         \l(\frac{\partial^2 p}{\partial \mu_{Z \neq Y} \, \partial \mu_{W \neq X}}\r)_{T, \mu'} \\
         \l(\frac{\partial^2 p}{\partial \mu_{Z \neq Y} \, \partial T}\r)_{\mu'} &
         \l(\frac{\partial^2 p}{\partial \mu_{Z \neq Y} \, \partial \mu_{W \neq X}}\r)_{T, \mu'} &
         \l(\frac{\partial^2 p}{\partial \mu_{Z \neq Y} \, \partial \mu_{W \neq X}}\r)_{T, \mu'}
    \end{bmatrix}_{Z W} .
\end{align}

For the concrete substitutions of $B/S/Q$, there are a few cases to consider.  First consider the diagonal cases $\mu_X = \mu_B$ and $\rho_Y = \rho_B$:
\begin{align}
    \l(\frac{\partial \rho_B}{\partial \mu_B}\r)_{s, \rho_S, \rho_Q} &=
    \chi_{BB}
    - A^\dagger \, M^{-1} \, B    
\end{align}
with
\begin{align}
    A &=
    \begin{bmatrix}
        \chi_{BT} \\
        \chi_{BS} \\
        \chi_{BQ}
    \end{bmatrix}_W
    \hspace{1cm}
    B =
    \begin{bmatrix}
        \chi_{TB} \\
        \chi_{SB} \\
        \chi_{QB}
    \end{bmatrix}_Z
    \hspace{1cm}
    M =
    \begin{bmatrix}
         \chi_{TT} &
         \chi_{TS} &
         \chi_{TQ} \\
         \chi_{ST} &
         \chi_{SS} &
         \chi_{SQ} \\
         \chi_{QT} &
         \chi_{QS} &
         \chi_{QQ}
    \end{bmatrix}_{Z W} .
\end{align}

Similarly for $\mu_X = \mu_B$ and $\rho_Y = \rho_S$ we have
\begin{align}
    \l(\frac{\partial \rho_S}{\partial \mu_B}\r)_{s, \rho_B, \rho_Q} &=
    \chi_{SB}
    - A^\dagger \, M^{-1} \, B    
\end{align}
with
\begin{align}
    A &=
    \begin{bmatrix}
        \chi_{ST} \\
        \chi_{SS} \\
        \chi_{SQ}
    \end{bmatrix}_W
    \hspace{1cm}
    B =
    \begin{bmatrix}
        \chi_{TB} \\
        \chi_{BB} \\
        \chi_{QB}
    \end{bmatrix}_Z
    \hspace{1cm}
    M =
    \begin{bmatrix}
         \chi_{TT} &
         \chi_{TS} &
         \chi_{TQ} \\
         \chi_{BT} &
         \chi_{BS} &
         \chi_{BQ} \\
         \chi_{QT} &
         \chi_{QS} &
         \chi_{QQ}
    \end{bmatrix}_{Z W} .
\end{align}
As before, all other necessary derivatives of this form can be obtained from suitable permutations of these examples.

%%%%%%%%%%%%%%%%%%%%%%%%%%%%%%%%%%%%%%%%%%%%%%%%%
%%%%%%%%%%%%%%%%    APPENDIX G   %%%%%%%%%%%%%%%%
%%%%%%%%%%%%%%%%%%%%%%%%%%%%%%%%%%%%%%%%%%%%%%%%%
\section{Shear-stress tensor evolution}
\label{sec:shear-rel-sph}
%%%%%%%%%%%%%%%%%%%%%%%%%%%%%%%%%%%%%%%%%%%%%%%%%

Several dynamical quantities in the SPH equations of motion are defined and evolved in relation to the SPH reference density $\sigma$, e.g., $\Pi/\sigma$ and $\rho_X/\sigma$ (cf. Eqs.\ \eqref{eq:bulk_evolution_in_SPH} - \eqref{eq:charge_evolution_in_SPH}).  The shear viscous tensor $\pi_{\mu \nu}$, on the other hand, is evolved directly, as indicated below:

%%%%%%%%%%%%%%%%%%%%%%%%%%%%%%%%%%%%%%%%%%%%%%%%%%%%%%%%%%%%%%%%%%%%%%%%%%%%%%
\begin{align}
\frac{d \pi_{\mu \nu}}{d \tau}&=\nonumber\\
&\frac{\eta}{2 \gamma \tau_\pi}\left[\partial_\mu u_\nu+\partial_\nu u_\mu\right]-\frac{\eta}{2 \tau_\pi}\left[u_\mu \frac{d u_\nu}{d \tau}+u_\nu \frac{d u_\mu}{d \tau}\right]-\frac{\eta}{3 \gamma \tau_\pi} \Delta_{\mu \nu}\left(\partial_\beta u^\beta+\frac{\gamma}{\tau}\right) \nonumber\\
&-\frac{\eta}{\gamma \tau_\pi}\left[\frac{u_3}{\tau} g_\nu^3 g_\mu^0+\frac{u_3}{\tau} g_\nu^0 g_\mu^3+\gamma \tau g_\nu^3 g_\mu^3\right]\\
&-\frac{\eta\left(u_3\right)^2}{2 \tau^3 \gamma \tau_\pi}\left(u_\mu g_\nu^0+u_\nu g_\mu^0\right) -\frac{1}{\gamma \tau_\pi} \pi_{\mu \nu}-\frac{1}{\gamma}\left(\gamma u_\mu \pi_\nu^j+\gamma u_\nu \pi_\mu^j-u_\mu \pi_\nu^0 u^j-u_\nu \pi_\mu^0 u^j\right) \frac{d u_j}{d \tau} \\
& -\frac{\pi_{\mu \nu}}{\gamma}\left(\frac{\gamma}{\tau}-\frac{\left(u_3\right)^2}{\gamma \tau^3}-\frac{\gamma}{\sigma^*} \frac{d \sigma^*}{d \tau}\right) \\
& -\frac{1}{\tau^3 \gamma} u_3\left(g_{\mu 0} \pi_{\nu 3}+g_{\nu 0} \pi_{\mu 3}\right)-\frac{1}{\tau \gamma} u_3\left(g_{\mu 3} \pi_{\nu 0}+g_{\nu 3} \pi_{\mu 0}\right)-\frac{1}{\tau^3}\left(g_{\mu 3} \pi_{\nu 3}+g_{\nu 3} \pi_{\mu 3}\right)
\label{eq:shear-stress-sph}
\end{align}
%%%%%%%%%%%%%%%%%%%%%%%%%%%%%%%%%%%%%%%%%%%%%%%%%%%%%%%%%%%%%%%%%%%%%%%%%%%%%%

As noted in the main text, the effects of propagating $\pi_{\mu \nu}$ directly (as opposed to $\pi_{\mu \nu}/\sigma$) are negligible, and is documented here only in the interest of completeness.

\end{document}